\documentstyle[sprocl]{article}
\input{psfig}
\catcode`\@=11
\def\@oldcitex[#1]#2{\if@filesw\immediate\write\@auxout
	{\string\citation{#2}}\fi
\def\@citea{}\@oldcite{\@for\@citeb:=#2\do
	{\@citea\def\@citea{,}\@ifundefined
	{b@\@citeb}{{\bf ?}\@warning
	{Citation `\@citeb' on page \thepage \space undefined}}
	{\csname b@\@citeb\endcsname}}}{#1}}
\def\oldcite{\@ifnextchar [{\@tempswatrue
	\@oldcitex}{\@tempswafalse\@oldcitex[]}}
\def\@oldcite#1#2{{[{#1}]\if@tempswa\typeout
	{IJCGA warning: optional citation argument 
	ignored: `#2'} \fi}}
\catcode`\@=12
\newcommand{\lesssim}{\stackrel{\scriptscriptstyle<}{\scriptscriptstyle\sim}}

\def\detadphi{$\Delta\eta\times\Delta\phi$} 
\def\ppbar{$p\overline{p} $}            
\def\pbarp{$\overline{p}p $}            
            \def\ttbar{$t\overline{t}$}             
\def\bbbar{$b\overline{b}$}             
\def\epm{$e^+e^-$}                      
\def\ino{\widetilde \chi}               
\def\winog{\widetilde \chi^\pm}               
\def\zinog{\widetilde \chi^0}               
\def\winol{\widetilde \chi^\pm_1}             
\def\winoop{\widetilde \chi^+_1}              
\def\winoom{\widetilde \chi^-_1}              
\def\zinol{\widetilde \chi^0_1}               
\def\winoh{\widetilde \chi^\pm_2}             
\def\zinoh{\widetilde \chi^0_2}               
\def\zinot{\widetilde \chi^0_3}               
\def\zinof{\widetilde \chi^0_4}               
\def\w3{\widetilde W_3}               
\def\Bino{\widetilde B}               
\def\Wino{\widetilde W}               
\def\Zino{\widetilde Z}               
\def\photino{\widetilde \gamma}               
\def\higgsino{\widetilde H}               
\def\sbottom{\tilde b}             
\def\stop{\tilde t}             
\def\stau{\tilde\tau}             
\def\slepton{\tilde\ell}             
\def\selectron{\tilde e}           
\def\squark{\widetilde Q}             
\def\sneutrino{\tilde \nu}          
\def\gluino{\tilde g}             
\def\gravitino{\widetilde G}          
\def\LSP{LSP}                         
\def\tanb{$\tan{\beta}$}                           
\def\m0{$m_0$}                         
\def\mhalf{$m_{1/2}$}                     
\def\a0{$A_0$}                         
\def\pt{$p_T$}                          
\def\et{$E_T$}                          
\def\met{\mbox{${\hbox{$E$\kern-0.6em\lower-.1ex\hbox{/}}}_T$}} 
\def\htran{$H_T$}                       
\def\ipb{pb$^{-1}$}                     
\def\gevc{GeV/c}                        
\def\gev{GeV}                           
\def\deg{$^\circ$}                      
\def\lumint{$\int\! {\cal{L}} dt$}      
\def\D0{D\O}                            
\def \intlum {\int {\cal L} dt}
\def\goes{\rightarrow}

\def\winoi{\widetilde \chi^\pm_i}             
\def\zinoi{\widetilde \chi^0_i}               
\def\sQsQ   	{\widetilde Q \widetilde Q^*}  		
\def\sQgl   	{\widetilde Q \tilde g}           	
\def\glgl  	{\tilde g \tilde g}          	
\def\sEsE {\selectron\selectron}			
\def\sEsN {\selectron\sneutrino}		
\def\sWsZ    {\winol\zinol}			
\def\sWsZh   {\winol\zinoh}			
\def\sWsW    {\winol\winol}			
\def\sZlsZl    {\zinol\zinol}			
\def\sZsZ    {\zinoh\zinoh}			
\def\sWsZt    {\winol\widetilde{\chi}^0_3}		
\def\sWpsWm    {\widetilde{\chi}^\pm_1\widetilde{\chi}^\mp_1}  
\def\sQsZ  {\squark\zinol}			
\def\sQjone   	{\squark \to q \zinol}
\def\sQjthree   {\squark \to q\winol \to q (q\bar{q}\zinol)}
\def\sQsjthree   {\squark^* \to q\winol \to q (q\bar{q}\zinol)}
\def\gljtwo  	{\gluino \to q\bar{q}\zinol}
\def\gljtwoh  	{\gluino \to q\bar{q}\zinoh}
\def\gljfour  	{\gluino \to q\squark \to 
           q (q\winol \to q (q\bar{q}\zinol))}
\def\gljsix   {\gluino \to t\stop \to (Wb)(\winol b) \to (jjb)(jjb\zinol)}
\def\glqqwino  	{\gluino \to qq'\winol}
\def\glqqwinoe  	{\gluino \to qq'\winol(\to e\nu\zinol)}

\def\glqqzinoh  {\gluino \to q\bar{q}\zinoh}

\def\sBbzinol   	{\sbottom \to b \zinol} 	
\def\sBbgluino   	{\sbottom \to b \gluino}	
\def\sBstW              {\sbottom \to \stop W }       
\def\sTbwinol		{\stop \to b \winol}      	
\def\sTczinol		{\stop \to c \zinol}      	
\def\glbbbar  		{\gluino \to b\bar{b}\zinol}	

\def\sEenzinol {\selectron \to \nu \winol(\to e\nu\zinol)}	
\def\sEezinol {\selectron \to e \zinol}	
\def\sEezinoh {\selectron \to e \zinoh}	
\def\sNnzinol {\sneutrino \to \nu \zinol}	
\def\sNewinol {\sneutrino \to e \winol}	

\def\sWenlsp  {\winol \to  e\nu\zinol}	
\def\sWenZh   {\winol \to  e\nu\zinoh}	
\def\sWqq     {\winol \to  qq'\zinol}	
\def\sWwsZ    {\winol \to W \zinol}
\def\sWstb    {\winol \to \stop b, \stop \to c \zinol} 
\def\sZnn    {\zinoh \to  \nu\nu\zinol}	%
\def\sZee     {\zinoh \to  ee\zinol}	%
\def\sZqq     {\zinoh \to  q\bar{q}\zinol}	%
\def\sZtqq    {\tilde{\chi}^0_3\to q\bar{q}\zinoh(\to ee\zinol)}
\def\sZgG     {\zinol \to  \gamma\gravitino}	
\def\sZgZ     {\zinoh \to  \gamma\zinol}		
\def\sZgsz    {\zinoh \to \gamma \zinol}
\def\lstop{$m_{\stop}<m_b+M_{\winol}$}
\def\yratio  {{Y_t \over Y_t^{ir}}}
\def\yratlo  {{Y_t/Y_t^{ir}}}
\begin{document}
\begin{rightline}
{\today}
\end{rightline}
\title{ The Search for Supersymmetry at the Tevatron Collider}
\author{
M. Carena,$^{1}$
R.L. Culbertson,$^{2}$
S. Eno,$^{3}$
H.J. Frisch,$^{2}$ and
S. Mrenna$^{4}$}
\address{
\centerline{$^{1}$~Fermi National Accelerator Laboratory}
\centerline{$^{2}$~University of Chicago}
\centerline{$^{3}$~University of Maryland}
\centerline{$^{4}$~Argonne National Laboratory}
}
\maketitle
\begin{abstract}
We review the status of searches for Supersymmetry at the Tevatron Collider.
After discussing the theoretical aspects relevant to the 
production and decay of supersymmetric particles at the Tevatron, we present
the current results for Runs Ia and Ib as of the summer of 1997.
\end{abstract}
\clearpage  
\setcounter{footnote}{0}
\section{Introduction}
\label{sec:intro}

The Tevatron is a 1--kilometer radius superconducting accelerator ring, located
at the Fermi National Accelerator Laboratory (Fermilab),  in Batavia,  Illinois,
U.S.A. The ring is used in two modes:  as a source of high energy beams for
fixed target experiments, and, in conjunction with the Antiproton Source, as a
proton--antiproton collider, operating with a \ppbar~center--of--mass  energy of
$\sqrt{s}=1.8$~TeV. The Tevatron Collider has been the world's 
accelerator--based high
energy frontier since it first began taking data in 1987, and has thus been a
prime location to search for the final pieces of the 
Standard Model\,\cite{GlashowSM} (SM) and new
phenomena  beyond. 

With the discovery of the top quark at the Tevatron,\cite{topdiscovery} the SM
particle spectrum is almost complete, with only the Higgs boson (and, arguably,
the tau neutrino) lacking direct experimental confirmation. 
If the interactions of the 
leptons, quarks and gauge bosons of the SM remain perturbative up to very high
energies  (as appears to be the case from the measured running of the 
gauge couplings), 
then the mechanism responsible for electroweak symmetry breaking
(EWSB) 
is expected to contain
one or more fundamental scalar Higgs bosons that are light, {\it i.e.} with
masses of the order of the symmetry--breaking scale.

The Higgs mechanism\,\cite{Higgs}
plays a crucial role in the SM.
The neutral component of the Higgs boson
acquires a vacuum expectation value to give mass to the $W$ and $Z$  
gauge bosons as well as to the SM fermions.  In addition, 
the couplings of the Higgs boson to
the gauge bosons  and fermions 
are such as to cancel the infinities in the electroweak
radiative  corrections  and  prevent  unitarity violation in the longitudinal
scattering of the gauge bosons. 
However, if the SM is valid up to an energy cutoff
of  $\Lambda_{\rm cutoff} \approx M_{\rm Planck}$   
($M_{\rm Planck} = 1.22\times 10^{19}$ GeV),  
the model has a fundamental problem, the so--called 
{\it naturalness problem}.\cite{naturalness}  

The radiative corrections to the Higgs boson mass--squared calculated in the SM
are quadratically divergent (proportional to $\Lambda_{\rm cutoff}^2$). A
physical Higgs mass of the order of the electroweak scale requires a
cancellation of one part in $10^{16}$ between these radiative corrections,
which
come from the interactions of the  Higgs bosons with all other particles in the
theory, and the bare Higgs mass at the Planck scale:
\begin{equation}
m_H^2 \simeq m_H^2(\Lambda_{\rm cutoff}) - \alpha \Lambda^2_{\rm
cutoff}.
\end{equation}
Either there is an extreme ``fine tuning'' necessary to have  a cancellation of 
two independent effects (the naturalness problem), or there must be  some new
principle at work.

Supersymmetry (SUSY)\,\cite{SUSY,SUSY2,Dawson_lectures} 
is a new symmetry which provides a 
well--motivated extension of the SM  with an elegant solution to the naturalness
problem.  Supersymmetric transformations relate
fermionic and bosonic degrees of  freedom. Each left--handed and right--handed
fermion of the SM is postulated to have its own bosonic  superpartner with equal
mass and coupling strengths. Similarly each SM boson would have its own
fermionic  superpartner, again with equal mass and couplings.   Because bosons and
fermions induce radiative effects of opposite signs, SUSY naturally provides
an exact cancellation of the otherwise quadratically--divergent  radiative
corrections to the Higgs boson mass.\cite{canceldiv}

Given that no superparticles  have been observed so far, it is assumed
that SUSY is broken, and that in general the sparticles must be heavier
than their partners.   
In order
to break SUSY without spoiling the necessary
cancellation of quadratic divergences, the 
splittings between the masses of the SM particles and their
SUSY partners should not be much larger than a few TeV. 
If SUSY is a consistent description of Nature, then the
lower range of sparticle masses  can be within the reach of the 
Tevatron,\cite{naturalness2}
motivating a wide range of searches in a large number of 
channels.\cite{tevchannels} The mass
of the lightest neutral
Higgs boson is strongly constrained within SUSY,\cite{two_loop}
and could be
within the reach of the upgraded Tevatron.\cite{higgssim,mrenna_higgs}
In addition, the decay of the heavy
top quark, pair 
produced in strong interactions with a cross section of $\simeq 6$
pb at the Tevatron,  gives a unique mechanism for producing lighter
supersymmetric particles which might not otherwise be
produced at a large rate in proton--antiproton collisions.

Two experimental collaborations, CDF and \D0, have large, general--purpose
detectors at Fermilab. The Tevatron had initial running periods in 1985 and 1987
with low luminosity.\footnote{The 1985 run produced the first detected
luminosity, with 20 events recorded by CDF.} In $1988-1989$ (the
`89 Run'), the Tevatron operated at $\sqrt{s}=1.8$ TeV with an average
instantaneous luminosity of $1.6 \times 10^{30}$ cm$^{-2}$s$^{-1}$,  and the
CDF detector collected approximately 4.4 \ipb~of data.   In $1992-93$ (Run Ia),
CDF and \D0~accumulated approximately 20 and 15 \ipb, respectively, and in
$1994-95$ (Run Ib) 90 and 108 \ipb, for a total Run I integrated luminosity of
more than 100 \ipb~per detector.\footnote{The difference in integrated 
luminosities in Run Ib comes partly from the fact that \D0~uses the 
11.3~\ipb~from  
Run Ic while CDF ignores it, and the fact that the \D0~experiment
normalizes its luminosity (and hence all cross sections) 
to an inelastic cross section that is 2.4\% smaller
than that used by CDF. For the actual luminosities used in each analysis see 
Table~\ref{tab:summary}.
}  The average instantaneous luminosity during Run
I was approximately $1\times 10^{31}$~cm$^{-2}$s$^{-1}$.  A new run (Run II)
utilizing the new Main Injector and Recycler rings and other major accelerator
improvements is scheduled to begin around the year 2000, reaching an average
instantaneous luminosity of $1\times 10^{32}$ cm$^{-2}$s$^{-1}$ and an 
expected integrated luminosity of 1000 \ipb~per year.\cite{peoples} 
In addition, the
energy of the machine will be increased to $\sqrt{s}=2$ TeV, substantially increasing the
cross section for producing heavy particles (the top quark pair production cross section,
for example, increases by 40\% from $\sqrt{s}=1.8$ TeV to 2 TeV).  Run II, with
an upgraded Collider and detectors, holds a great deal of promise for Higgs
boson and sparticle searches.

\setcounter{footnote}{0}
\section{The MSSM}
\label{section:mssm}

In the past two decades, a detailed picture of the 
Minimal Supersymmetric extension of the Standard Model (MSSM),
has emerged.\,\cite{R13A,R13B,CPW1}
In the MSSM, the particle spectrum is doubled by SUSY.
Moreover,
to generate masses for up-- and down--type fermions while
preserving SUSY and gauge invariance, 
the Higgs sector must contain two doublets.\cite{R14}
After EWSB, there is a quintet of physical Higgs boson
states: two CP--even scalar $(h,H)$, one CP--odd pseudoscalar
$(A)$, and a pair of charged $(H^\pm)$ Higgs bosons.\cite{hhg}
All the Higgs bosons and other SM particles have superpartners
with the same quantum numbers under
the SM gauge groups
$SU(3)_C \times SU(2)_L \times U(1)_Y$, but with different spin.\cite{SUSY2}
The spin--1/2 partners of the gauge bosons (gauginos) are denoted as 
winos $\Wino^\pm$, zinos $\Zino$, 
photinos \footnote{The superpartners of the $U(1)_Y$ and $SU(2)_L$ gauge
bosons (before EWSB) are the Bino $\Bino$, the unmixed neutral
Wino $\w3$, and the unmixed charged Winos $\Wino_1$
and $\Wino_2$.} $\photino$,
and gluinos $\gluino$.   
The spin--1/2 partners
of the Higgs bosons (Higgsinos) are $\higgsino_1, \higgsino_2$ and
$\higgsino^\pm$.
Because of EWSB, 
the Higgsinos and $SU(2)_L \times U(1)_Y$ gauginos
mix to give physical mass eigenstates consisting of
two Dirac fermions of electric charge one, the charginos 
$\winog_{1,2}$,
and four neutral Majorana fermions, the neutralinos $\zinog_{1-4}$.
The spin--0 partners of the fermions 
(sfermions)~\footnote{Charge
conjugate scalars are denoted by $^{*}$, {\it e.g.} $\squark^*$.}   
are squarks $\squark$, \footnote{To allow easier reading,
we use the non--standard symbol $\squark$ instead of $\tilde q$.
This has no special significance.}
sleptons $\slepton$ and  sneutrinos $\sneutrino$.
Each charged lepton or quark has two scalar partners, one associated with each 
chirality.  These are   
named left--handed squarks 
and sleptons,  
which belong to  $SU(2)_L$ doublets,  and right--handed squarks  
and sleptons,
which are $SU(2)_L$ singlets. 
The neutrinos have only left--handed
superpartners $\sneutrino$, which belong to  $SU(2)_L$ doublets.
The gluino $\gluino$ and squarks $\squark$ carry color indices and
are $SU(3)_C$ octets and triplets, respectively.

The MSSM Lagrangian contains interactions between particles
and sparticles, fixed by SUSY. There are also a number of
soft SUSY--breaking mass parameters.  ``Soft'' means
that they break the
mass degeneracy between SM particles and their SUSY partners
without reintroducing quadratic divergences while respecting
the gauge invariance of the theory.
The soft SUSY--breaking parameters are extra mass  terms for
gauginos and scalar fermions, and trilinear scalar couplings.
The exact number of extra parameters
depends on the exact mechanism of SUSY breaking.
In the remainder of this section, the MSSM particle
spectrum and properties will be described in general, and 
also with reference to specific SUSY--breaking
scenarios and variations.  

\subsection{Sparticle Spectrum}

The chargino and neutralino masses and 
their mixing angles (that is, their gaugino and Higgsino composition)
are determined by the SM gauge boson masses ($M_W$ and $M_Z$),
$\tan \beta$,\,\footnote{One Higgs doublet, $H_2$, couples
to $u, c,$ and $t$, while the other, $H_1$, couples to $d, s, b, e, \mu$, and $\tau$.
The parameter $\tan\beta$ is the ratio of vacuum expectation values $\langle H_2 \rangle/
\langle H_1 \rangle$ $\equiv$ $v_2/v_1$, and $v^2=v_1^2+v_2^2$, where $v$ is
the order parameter of EWSB.}
two soft SUSY--breaking
parameters (the $SU(2)_L$ gaugino mass $M_2$ and the $U(1)_Y$ gaugino 
mass $M_1$), and the SUSY Higgsino mass 
parameter $\mu$, 
all evaluated at the electroweak scale $M_{EW}$.\footnote{The electroweak
scale $M_{EW}$ is roughly the scale of the sparticle
masses themselves.  Usually, in the literature, for simplicity,
$M_{EW}\simeq M_Z$.}
Explicit solutions are found by considering the $2\times 2$ chargino 
$\mathbf{M_C}$ and
$4\times 4$ neutralino $\mathbf{M_N}$ mass matrices: \footnote{Beware of different sign conventions
for $\mu$ in the literature.  Both 
{\tt PYTHIA} and {\tt ISAJET} use the convention used here.} 
\begin{eqnarray}
\label{eqn:inomatrices}
&\mathbf{M_{C}} =  \left( \begin{array}{cc}
 M_2 &  \sqrt{2}M_W s\beta \\
\sqrt{2}M_W c\beta & \mu \\ \end{array} \right); 
\mathbf{M_{N}} =  \left( \begin{array}{cc}
 \mathbf{M}_i &  \mathbf{Z} \\
 \mathbf{Z^T} & \mathbf{M}_\mu \\ \end{array} \right) &  \\
 & \nonumber \\
&\mathbf{M}_i =  \left( \begin{array}{cc}
 M_1 &  0 \\
 0   &  M_2 \\ \end{array} \right);
\mathbf{M}_\mu =  \left( \begin{array}{cc}
 0 &  -\mu \\
 -\mu   &  0 \\ \end{array} \right);
\mathbf{Z} =  \left( \begin{array}{cc}
 -M_Z c\beta s_W & M_Z s\beta s_W   \\
 M_Z c\beta c_W & -M_Z s\beta c_W  \\ \end{array} \right) & \nonumber
\end{eqnarray}
$\mathbf{M_C}$ is written in the $\Wino^+-\higgsino^+$ basis,
$\mathbf{M_N}$ in the $\Bino-\Wino^3-\higgsino_1-\higgsino_2$ basis,
with the notation $s\beta=\sin\beta,c\beta=\cos\beta,s_W=\sin\theta_W$
and $c_W=\cos\theta_W$.
In general, the mass eigenstates are admixtures of the interaction
states, but, for large values 
of $|\mu|$ or $M_1$ and $M_2$, the limit is reached where the mass eigenstates
are mostly pure gaugino or Higgsino states (independent of $\tan\beta$).
In particular, 
if $|\mu| \gg M_Z$ and $M_1,M_2 \simeq M_Z$, with $M_1 < M_2$, 
the lightest eigenstates are gaugino--like and the
heaviest are Higgsino--like, 
leading to the spectrum:
\begin{eqnarray}
\label{eq:mularge}
M_{\winol} \simeq M_2 ; \;\;\;\; 
M_{\winoh} \simeq |\mu| 
\\
 M_{\zinol} \simeq M_1 ; \;\;\;\; 
 M_{\zinoh} \simeq M_2 ; \;\;\;\;
 M_{\zinot} \simeq M_{\zinof} \simeq |\mu|. \nonumber
\end{eqnarray}
Similarly, if  $M_1,M_2 \gg M_Z$ and $|\mu| \simeq M_Z$,
the lightest eigenstates are Higgsino--like and the heaviest are
gaugino--like: 
\begin{eqnarray}
\label{eq:M2large}
M_{\winol} \simeq |\mu|; \;\;\;\; 
M_{\winoh} \simeq M_2 
 \\ M_{\zinol} \simeq M_{\zinoh} \simeq |\mu|; 
\;\;\;\; 
M_{\zinot} \simeq M_1 ; \;\;\;\; 
M_{\zinof} \simeq M_2. \nonumber
\end{eqnarray}
Another interesting example 
where $\zinol$ is Higgsino--like, $\zinoh$ is photino--like,
and all other charginos and neutralinos are mixtures,
occurs for
$M_1 = M_2 \simeq |\mu| \simeq M_Z$ 
and $\tan\beta\simeq 1$:
\begin{eqnarray}
\label{eq:M2eqmu}
M_{\winog_{1,2}} = {1\over 2}|M_2+\mu\mp\sqrt{(M_2-\mu)^2+4M_W^2}| 
\\ \nonumber
M_{\zinol} = |\mu|;\;\;\; M_{\zinoh} = M_1;\;\;\;
M_{\zinog_{3,4}} =  {1\over 2}|M_2+\mu\mp\sqrt{(M_2-\mu)^2+4M_Z^2}|. 
\nonumber 
\end{eqnarray}
Since the $SU(3)_C$ symmetry of the SM is not broken, the
gluinos have masses determined by the $SU(3)_C$ gaugino mass parameter
$M_3$.\footnote{The physical gluino mass is shifted from the value
of the gluino mass parameter $M_3$ because of radiative corrections.
As a result, there is an indirect dependence on the squark masses.}
The neutralinos and the gluinos are Majorana particles, and do not
distinguish between states and their charged conjugate.
Depending on their Higgsino or gaugino composition,
the $\ino$ couplings to gauge bosons, and left-- and right--handed 
sfermions will differ substantially, and the production 
and decay processes will strongly depend on that composition
 (see the discussion below).

The mass eigenstates of squarks and sleptons are, in principle,
 mixtures of their left-- and right--handed components, given
for the first generation by:
\begin{eqnarray}
m_{\tilde{u}_L}^2 
\simeq m_{Q_{1}}^2 + m^2_u + D_{\tilde{u}_L} &
m_{\tilde{u}_R}^2 
\simeq m_{U_{1}}^2 + m^2_u + D_{\tilde{u}_R}
\nonumber \\
m_{\tilde{d}_L}^2 
\simeq  m_{Q_{1}}^2 + m^2_d + D_{\tilde{d}_L} & 
m_{\tilde{d}_R}^2 
\simeq m_{D_{1}}^2 + m^2_d + D_{\tilde{d}_R}
\nonumber \\
m_{\selectron_L}^2 
\simeq  m_{L_{1}}^2 + m^2_e + D_{\selectron_L} &
m_{\selectron_R}^2 
\simeq m_{E_{1}}^2 + m^2_e + D_{\selectron_R}
\nonumber \\
m_{\sneutrino_e}^2  \simeq  m_{L_{1}}^2 + D_{\sneutrino_e}, &
\label{eq:LRmasseigen}  
\end{eqnarray}
where
$ m_{Q_{1}}^2$, $ m_{L_{1}}^2$, $ m_{U_{1}}^2$, $ m_{D_{1}}^2$,
and $ m_{E_{1}}^2$ are soft SUSY--breaking parameters
and 
$D_{\tilde{f}_L}= M_Z^2 \cos (2 \beta) (T_{3_{f}} -
Q_f \sin^2\theta_W)$, 
$D_{\tilde{f}_R}= M_Z^2 \cos( 2 \beta) Q_f\sin^2\theta_W$ are 
$D$--terms~\footnote{$D$--terms are terms in the scalar potential which are
quartic in the fields and are proportional to the gauge couplings squared.}
associated with EWSB
 ($T_{3_f}$ is the weak isospin eigenvalue of the fermion, $Q_f$ the electric charge).
A similar expression holds for the second (third) generation with the
substitutions $u\to c(t), d\to s(b), e\to\mu(\tau), 1\to 2(3).$
In most high--energy models, the soft SUSY--breaking sfermion mass parameters are
taken to be equal at the high--energy scale, but, in principle, they can be different for
each generation or even within a generation.
However, the sfermion flavor dependence can have important effects on
low--energy observables, and it is often strongly constrained.
The
suppression of flavor changing neutral currents (FCNC's), such as
$K_L\to\pi^\circ\nu\bar\nu$, requires that either $(i)$ the squark soft--SUSY
breaking mass matrix is diagonal and degenerate, or $(ii)$ the masses of the
first-- and second--generation sfermions are very large.\,\cite{stefan}

The magnitude of left--right 
sfermion mixing is always proportional to the mass of
the corresponding fermion.
The left--right mixing of squarks and sleptons of 
the first and second generation is thus negligible,
and $\squark_{L,R}$, with $\squark = \tilde u,\tilde d,\tilde c,\tilde s$, and 
$\slepton_{L,R},\sneutrino_\ell$,
with $\ell = e, \mu$, are the real mass eigenstates with
masses $m_{\squark_{L,R}}$ and $m_{\slepton_{L,R}}, m_{\sneutrino_{\ell}},$
respectively.
For the third generation  sfermions, the left--right
mixing can be  nontrivial.
The mass matrix for the top squarks (stops) in the ($\stop_L,  \stop_R$) basis is
 given by
\begin{equation}
M^2_{\stop} =  \left( \begin{array}{cc}
 m_{Q_3}^2 + m_t^2 + D_{\stop_L} &  m_t ( A_t - \mu/ \tan \beta) \\
 m_t ( A_t - \mu/ \tan \beta) &  m_{U_3}^2 + m_t^2 + D_{\stop_R} \\
 \end{array} \right), 
\label{stop_matrix}
\end{equation}
where $A_t$ is a soft SUSY--breaking parameter.\footnote{Beware 
also of different sign conventions for $A_t$.
Both {\tt PYTHIA} and {\tt ISAJET} use the convention used here.}    
Unless there is
a cancellation between $A_t$ and $\mu/\tan\beta$, left--right mixing
occurs for the stop squarks because of the large top quark mass.
The stop  mass eigenstates are then given by  
\begin{eqnarray}
\stop_1 & = & 
\cos \theta_{\stop} \;\;\stop_L + \sin \theta_{\stop} \;\;
 \stop_R
\nonumber \\
\stop_2 &= & 
 - \sin \theta_{\stop} \;\;\stop_L + \cos \theta_{\stop} \;\;
\stop_R,
\end{eqnarray}
where the masses and mixing angle $\theta_{\stop}$ are fixed by
diagonalizing the squared--mass matrix Eq.~(\ref{stop_matrix}).
Because of the large mixing, the lightest stop $\stop_1$ can be
one of the lightest sparticles.
For the sbottom, an
analogous formula for the mass matrix holds with $m_{U_3}\to m_{D_3}$, 
$A_t\to A_b$, $D_{\stop_{L,R}}\to D_{\sbottom_{L,R}}$, $m_t\to m_b$, 
and $\tan\beta\to 1/\tan\beta$.  For the stau, substitute
$m_{Q_3}\to m_{L_3}$, 
$m_{U_3}\to m_{E_3}$, $A_t\to A_\tau$, $D_{\stop_{L,R}}\to
D_{\stau_{L,R}}$,
$m_t\to m_\tau$ and $\tan\beta\to$ 1/$\tan\beta$.
The parameters
$A_t$, $A_b$, and $A_\tau$ can be independent soft SUSY--breaking
parameters, or they might be related by some underlying principle.
When $m_b\tan\beta$ or $m_\tau\tan\beta$ is large (${\cal O}(m_t))$, 
left--right
mixing can also become relevant for the sbottom and stau.
It will become clear below that $A_b$ and $A_\tau$ do not contribute in
a major way to left--right mixing, since they do not have a
\tanb~enhancement.

The Higgs boson spectrum at tree level
can be expressed in terms of the weak gauge boson masses, the
CP--odd Higgs boson mass $M_A$ and $\tan\beta$: \footnote{In the MSSM, $M_A$ is
related to the values of the Higgs soft SUSY--breaking 
parameters $m_{H_1}$ and $m_{H_2}$ and the Higgsino mass
parameter $\mu$ through 
$M_A^2=m_{H_1}^2+m_{H_2}^2+2\mu^2$.}
\begin{eqnarray}
  M_{h,H}^2 & = & 1/2 \left[ M_A^2 + M_Z^2 \mp
\sqrt{(M_A^2+M_Z^2)^2-4M_Z^2M_A^2\cos^2 2\beta} \right]  \nonumber \\
  M_{H^\pm}^2 &=& M_A^2 + M_W^2. 
\end{eqnarray}
These relations yield $M_h \lesssim M_Z$, but this
result is strongly modified by radiative corrections
that depend on other MSSM parameters.\cite{Higgsrad}
The dominant radiative corrections to $M_h$ grow as
$m_t^4$ and are logarithmically dependent on the third--generation
squark masses.  
The heavy CP--even and charged Higgs boson masses,
$M_H$ and $M_{H^\pm}$, respectively, are directly controlled by $M_A$.
If all SUSY particles were
heavy, but $M_A$ were small, then
the low--energy theory would look like a two--Higgs--doublet model.  
For sufficiently large $M_A$, 
the heavy Higgs doublet decouples, and the effective low--energy 
theory has only one light 
Higgs doublet with SM--like couplings to gauge bosons and fermions.  

Within the MSSM, a $\:${\it general}$\:$ {\bf upper}$\:$ bound 
on $\: M_h \:$ can be
determined by
a careful evaluation of the one--loop and
dominant two--loop radiative corrections.\cite{two_loop}
For $m_t=175$ GeV and an extremely conservative set of
assumptions,\,\footnote{To produce this bound, the masses of all SUSY particles and
$M_A$ are chosen to be around a TeV, $\tan\beta>20$, and 
the stop mixing parameters are varied to give the largest possible effect.}
the upper bound on the lightest Higgs mass is maximized, yielding 
$M_h \lesssim 130$ GeV. 
For more moderate values of the MSSM parameters,
the upper bound on $M_h$ becomes smaller.
Most importantly, 
given the general upper bound on $M_h$ of
about 130 GeV,
the upgraded Tevatron has the potential to provide a crucial test
of the minimal supersymmetric extension
of the SM.\cite{higgssim,mrenna_higgs}

R--parity, defined as \cite{FarrarFayet}
R=$(-1)^{2S+3B+L}$, is a discrete multiplicative symmetry 
where $S$ is the particle spin, $B$ is the
baryon number, and $L$ is the lepton number. 
All SM particles have R=1, while all superpartners have R=$-1$, so
a single SUSY particle cannot decay into just SM particles if
R--parity is conserved.
In this case, the lightest superpartner (LSP)
is absolutely stable.
Astrophysical considerations
imply that a stable
LSP should be color-- and charge--neutral.  The best candidates,
then, are the lightest neutralino $\zinol$ and the sneutrino
$\sneutrino$, or alternatively (see below) the gravitino $\gravitino$.  
Since the LSP can carry away energy without interacting
in a detector, the apparent violation of momentum conservation
is an important
part of SUSY phenomenology.\cite{FarrarFayet}
Also, when R--parity is conserved, superpartners must be produced
in pairs from a SM initial state.
The breaking of the R--parity symmetry would result in lepton and/or
baryon number violating processes.  While there are strong experimental
constraints on some classes of R--parity violating interactions, 
others are hardly
constrained at all.  Unless it is explicitly stated otherwise, R--parity 
conservation is assumed below.

Quite generally, the dependence of the SUSY spectrum on $\tan \beta$
 can be very strong, and it is necessary to determine the possible 
range of  values for this essential parameter of the theory.
The fermion masses, which are not fixed by SUSY, are a function of $\tan \beta$ and the 
 SM Yukawa couplings. 
For the up-- and down--quark and lepton masses, it follows that
$m_u = h_u v \sin \beta$, 
$m_d = h_d v \cos \beta$, and 
$m_\ell = h_\ell v \cos \beta$,
where $h_{f=u,d,\ell}$ 
is the corresponding Yukawa coupling and  $v$= 174 GeV is the 
order parameter of EWSB.  Equivalently, $m_u = h_u v_2, m_d = h_d v_1$
and $m_\ell = h_\ell v_1$, where $v_2=v\sin\beta$ and $v_1=v\cos\beta$.
The value of \tanb~and the Yukawa couplings can vary in a range
consistent with the experimental values of the fermion masses.
However, for the theory to remain
perturbatively well defined up to a given cutoff scale, the  
Yukawa couplings should  remain finite up to this cutoff scale. 
Whether this is the case can be determined by studying the renormalization
group evolution of each Yukawa coupling from low to high energy scales.
In particular, the large value of the top 
quark mass is associated with a large value of the top Yukawa coupling at 
low energies, which, depending on  $\tan \beta$, 
may become too large to be compatible with a 
perturbative description of the theory.\cite{A-GPW,CCBSW,CPW1,BBO-LP-BCPW}
The measured value of the top quark mass, 
$m_t \simeq 175$ GeV, defines a lower bound on $\tan \beta$ of about 1.2,
provided that the top Yukawa coupling remains finite
up to a scale  of the order of $10^{16}$ GeV.
If, instead, the top Yukawa coupling should remain finite 
only up to scales 
of order of a TeV, values of $\tan \beta$    
as low as .5 are still possible.\,\footnote{This implies that a perturbative
description of the MSSM would 
only be valid up to the weak scale, which  is, of course, not a very   
interesting possibility.}
A similar situation occurs when $\tan \beta$ is large, but now 
the crucial role is played by the bottom Yukawa coupling. 
If $\tan \beta$ becomes too large, large values of the bottom Yukawa 
coupling are necessary  to obtain values 
of the bottom mass compatible with experiment. 
The exact bound  on $\tan \beta$ depends on the SUSY spectrum, since there are 
radiative corrections to the bottom mass coming from sparticle 
exchange loops.\cite{bmcorrec,COPW2}
Generically, it can be shown
that values of $\tan \beta \geq 60$  are difficult to obtain if the MSSM is 
expected to remain a valid theory up to scales 
of order $10^{16}$ GeV.

\subsection{Supergravity}  

At present, the exact mechanism of SUSY breaking is unknown. Supergravity (SUGRA)
models assume the existence of extra superfields (the so--called 
``hidden sector'') which couple
to the MSSM particles only through gravitational--like interactions. When SUSY
is spontaneously or dynamically broken in the  hidden sector,
some of the components of the hidden sector acquire vacuum expectation values. 
Interaction
terms between those components of the hidden sector and
the MSSM superfields give rise to the effective soft SUSY--breaking terms
of the MSSM, which are proportional vacuum
expectation values of the hidden sector 
divided by powers of $M_{\rm Planck}$.
The low--energy Lagrangian then looks like  a SUSY--conserving MSSM with a
number of extra terms that break SUSY.   Although low--energy 
gravitational interactions depend only  on mass and spin, the
general supergravity Lagrangian may contain higher--dimensional
interactions between the hidden sector and MSSM superfields that
are flavor dependent.

The number  of SUSY breaking terms is over one hundred. In
the minimal SUGRA scenario, however, the MSSM sparticles couple universally to
the hidden sector, and the number of terms is greatly reduced.
Using this guiding principle, at a scale of order $M_{\rm Planck}$ (or,
approximately, $M_{GUT}$, the scale where the gauge couplings unify),
all scalars (Higgs bosons, sleptons, and squarks) are assumed to have a
common squared--mass $m_0^2$, all gauginos (Bino, Wino, and gluino) have
a common mass \mhalf, and all trilinear couplings have the value $A_0$.  
After specifying $\tan\beta$, 
all that remains is to relate the values of the soft
SUSY--breaking parameters specified at $M_{GUT}$ to
their values at $M_{EW}$.  This is accomplished using
renormalization group equations (RGE's).
The physical sparticle masses are then determined using 
relations like Eq.~(\ref{eq:LRmasseigen}), or by diagonalizing
mass matrices like those in Eqs. (\ref{eqn:inomatrices})
and (\ref{stop_matrix}).
Finally,
the Higgsino mass parameter is determined (up to a sign)
by demanding the correct radiative EWSB. At the tree level,
this requires
\begin{equation}
\label{mufix}
\tan^2\beta = \frac{\mu^2+m_{H_1}^2+M_Z^2/2}{\mu^2+m_{H_2}^2+M_Z^2/2},
\end{equation}
where $m_{H_1}$ and $m_{H_2}$ are soft SUSY--breaking mass parameters for the two
Higgs doublets evaluated at $M_{EW}$.

Not surprisingly, the masses 
of the gluinos, charginos and neutralinos are strongly correlated.
The gaugino mass parameters $M_i$ at the electroweak scale depend mainly
on the running of the gauge couplings between $M_{GUT}$ and $M_{EW}$:
\begin{eqnarray}
M_3 = \frac{\alpha_3(M_3)}{\alpha_{GUT}} m_{1/2} \simeq 2.6 m_{1/2}
\nonumber \\
M_2 = \frac{\alpha_2(M_2)}{\alpha_{GUT}} m_{1/2} \simeq 0.8  m_{1/2}
\nonumber \\
M_1 = \frac{\alpha_1(M_1)}{\alpha_{GUT}} m_{1/2} \simeq 0.4 m_{1/2},
\label{eq:gauginounif}
\end{eqnarray}
(where we consider $m_{1/2}\sim M_{EW}$).
As will be shown below, once the RGE evolution of the 
Higgs mass parameters is included in
Eq.~(\ref{mufix}), it follows that 
the Higgsino mass term $\mu$  tends to be larger than
the Bino and Wino masses $M_1, M_2$, becoming the largest for values of 
$\tan \beta$ closest to 1.  
As a result,
the lightest two neutralinos and
the lightest chargino tend to be gaugino--like.
This is similar to the example
presented in Eq.~(\ref{eq:mularge}) with the approximate mass hierarchy:
\begin{equation}
M_{\zinoh}\simeq 2 M_{\zinol} \simeq M_{\winol} \simeq 1/3 M_{\gluino}.
\label{eq:sugra_mass}
\end{equation}
The lightest neutralino $\zinol$ can be the LSP.

Because sleptons have only EW quantum numbers and the 
lepton Yukawa couplings
are small, the slepton mass parameters
do not evolve much from $M_{GUT}$ to $M_{EW}$.  The left-- and 
right--handed soft SUSY--breaking 
parameters at the scale $M_{EW}$ which determine the mass eigenstates
through Eq. (\ref{eq:LRmasseigen}) are given by
\begin{equation}
m_{L_{1,2}}^2 \simeq  m_{L_3}^2 \simeq m_0^2 + 0.5 m^2_{1/2} 
; \;\;\;\;\;
m_{E_{1,2}}^2 \simeq  m_{E_3}^2 \simeq m_0^2 + 0.15 m^2_{1/2}.
\label{eq:slep_mass}
\end{equation}
For $\tan\beta \ge 40$, 
the third generation
mass parameters also receive non--negligible
contributions from the $\tau$ Yukawa coupling 
in their running which can modify these expressions.
The different coefficients of \mhalf~in Eq.~(\ref{eq:slep_mass})
arise from the different
EW quantum numbers for sleptons in $SU(2)_L$ doublets and singlets.
If \m0~and \mhalf~are of the same order of magnitude, 
physical slepton masses are dominated by \m0.  
When \m0~is small, the sneutrino can be the LSP instead of 
the $\zinol$.  The $\sneutrino$ mass is
fixed by a sum rule 
$m_{\sneutrino_{\ell}}^2 = m_{\slepton_L}^2 + M_W^2\cos 2\beta$.

The squark mass parameters
evolve mainly through the strong coupling to the gluino, 
so their dependence on
the common gaugino mass is stronger than for sleptons. 
For the first and second generation,
the left-- and right--handed soft SUSY--breaking parameters at $M_{EW}$
are given approximately by
\begin{equation}
\label{eq:approxmQU}
m_{Q_{1,2}}^2 \simeq  m_0^2 + 6.3 m^2_{1/2} 
; \;\;\;\;\;
m_{U_{1,2}}^2 \simeq  m_{D_{1,2}}^2 \simeq m_0^2 + 5.8 m^2_{1/2}.
\end{equation}
In general, the squarks are heavier than
the sleptons or the lightest neutralino and chargino.
Since first and second generation squark soft SUSY--breaking parameters
are the same for squarks with the same quantum numbers, FCNC's are
suppressed.

For the third--generation squarks, the large
top and bottom Yukawa couplings
play a crucial role in the RGE evolution.
As mentioned generically in section $2.1$, the top Yukawa 
coupling $h_t$ is related to the
top quark mass by $m_t = (246/\sqrt{2}) h_t\sin\beta$ GeV and 
the bottom Yukawa coupling $h_b$ is
given by $m_b = (246/\sqrt{2}) h_b\cos\beta$ GeV, so that
$h_b$ is large (of order $h_t$) when
 $\tan \beta$ is about 40 or larger.
When $h_t$ at $M_{GUT}$ is
sufficiently large, it turns out
that its low--energy value is independent of
its exact value at $M_{GUT}$.  This
behavior is known as the {\it infrared fixed
point} solution of the top quark mass.\cite{A-GPW,CCBSW,BBO-LP-BCPW,IRFP}
With the definition $\displaystyle Y_t\equiv h_t^2 / (4\pi)$,
the infrared fixed point value of $Y_t$ at the scale $m_t$ is
$Y_t^{ir}\simeq 8\alpha_3/9$.
Within the one--loop approximation, the effects
of the top Yukawa coupling on the RGE evolution can
be parameterized in terms of the ratio $\yratlo$.
For small and moderate values of $\tan\beta$,
the left-- and right--handed soft SUSY--breaking parameters which determine 
the stop and sbottom masses are then given by\,\cite{COPW-C,CW,Drees}
\begin{eqnarray}
m_{Q_3}^2 \simeq {m_0^2}\left(1-\frac{1}{2}\yratio\right) +
  m^2_{1/2}\left(6.3+\yratio\left(-\frac{7}{3} + \yratio\right)\right) 
\nonumber \\
m_{U_3}^2 \simeq {m_0^2}\left(1-\yratio\right) +
               m^2_{1/2}\left(5.8+\yratio\left(-\frac{14}{3}+
2\yratio\right)\right),
\label{eq:soft_3gen}
\end{eqnarray}
and $m_{D_3} \simeq  m_{D_{1,2}}$.
For large $\tan\beta$, assuming $t-b$ Yukawa coupling unification 
at high energies ($Y_b=Y_t$ at $M_{GUT}$, which is a generic
prediction of $SO(10)$ $GUT$ models), 
the expressions for the third generation
soft SUSY--breaking parameters 
are:\,\cite{COPW2}
\begin{eqnarray}
m_{Q_3}^2 \simeq {m_0^2}\left(1-\frac{6}{7}\yratio\right) +
m^2_{1/2}\left(6.3+\yratio\left(-4 + \frac{12}{7}\yratio\right)\right) 
\nonumber \\
m_{U_3}^2 \simeq m_{D_3}^2 \simeq {m_0^2}\left(1-\frac{6}{7}\yratio\right) + 
m^2_{1/2}\left(5.8-\yratio\left(-4+\frac{12}{7}\yratio\right)\right).
\label{eq:soft_3gen_tanb}
\end{eqnarray}
Contributions proportional to $A_0^2$ and $A_0 m_{1/2}$ with a 
prefactor proportional to $(1-\yratlo)$
are also present in Eqs.~(\ref{eq:soft_3gen}) and (\ref{eq:soft_3gen_tanb}).
For $m_t \simeq 175$ GeV, the value of 
the ratio $\yratlo$ varies from $3/4$ to $1$ depending on
$\tan\beta$, with $ \yratlo \to 1$ as $\tan\beta\to 1$, and 
$\yratlo \simeq $ 0.85 for $\tan\beta=40$.
The value of $A_t$ is governed by $m_{1/2}$,
and, for large values of
the top Yukawa coupling, depends weakly on its 
initial value and $\tan\beta$,\cite{COPW-C}
\begin{equation}
A_t \simeq \left(1-\yratio\right)A_0 - 2 m_{1/2}.
\end{equation}
The exact values of $A_b$ and $A_{\tau}$ are not important, since
the mixing in the stau and sbottom sectors is governed by the terms 
$m_b \mu\tan\beta$ and $m_\tau \mu\tan\beta$, respectively.
In SUGRA models,
the above relations between the mass parameters 
leads to the general prediction, $m_{\squark} \ge
0.85 M_{\gluino}$ (for the five lightest squarks and small or moderate
$\tan\beta$).

The soft--SUSY breaking parameters in the Higgs sector also have
simple expressions.
For small and moderate $\tan\beta$, \cite{COPW-C,CW,Drees}
\begin{eqnarray}
m_{H_1}^2 &\simeq& m_0^2 + 0.5 m^2_{1/2} \nonumber \\
m_{H_2}^2 &\simeq& {m_0^2}\left(1-\frac{3}{2}\yratio\right) +
m^2_{1/2}\left(0.5+\yratio\left(-7+3\yratio\right)\right).
\label{eq:higgs_parameters}
\end{eqnarray}
Substituting these relations back into Eq.~(\ref{mufix}) yields
the result:
\begin{eqnarray}
\mu^2 + M_Z^2/2 = m_0^2 \left(1 + \left(\frac{3}{2} \yratio -1\right) \tan^2
\beta\right)\frac{1}{ \tan^2
\beta -1} + \nonumber \\
        m_{1/2}^2 \left(0.5 -
\left(0.5+\yratio\left(-7+3\yratio\right)\right)\tan^2 \beta\right)
\frac{1}{ \tan^2 \beta -1}.
\label{eq:EWSB}
\end{eqnarray}
Note in Eq.~(\ref{eq:higgs_parameters}) that $m_{H_2}^2<0$, which
is usually a sufficient condition to induce EWSB.
For large $\tan\beta$, the Higgs mass parameters
are more complicated.  In the limit of $t-b$ Yukawa unification,
they simplify to\,\cite{COPW2}
\begin{equation}
m_{H_1}^2 \simeq 
m_{H_2}^2 \simeq {m_0^2}\left(1-\frac{9}{7}\yratio\right) + m^2_{1/2}
\left(0.5+\yratio\left(-6+\frac{18}{7}\yratio\right)\right),
\end{equation}
and Eq.~(\ref{eq:EWSB}) must be modified accordingly.
All of these relations are only approximate:  the coefficients of
\mhalf~depend on the exact values of $\alpha_s$ and the scale of
the sparticle masses; the coefficients of \m0~and $A_0$ depend
mainly on $\tan\beta$. 

The SUGRA model presented here is minimal (mSUGRA) 
in the sense that it is
defined in terms of only five parameters at a high scale: 
\m0, \mhalf, \a0, \tanb, and 
the sign of $\mu$.  It is natural to question exact universality of 
the soft SUSY--breaking parameters.\,\cite{nonuniv}
For example, in
a $SU(5)$ SUSY GUT model, the left--handed sleptons and right--handed
down--type squarks reside in 
the same 5--multiplet of $SU(5)$, and naturally have the common mass
parameter $m^{(5)}_0$ at the GUT scale.  Similarly, $\tilde u_L, \tilde d_L$, 
$\tilde u_R$, and $\selectron_R$, which reside in the same
10--multiplet, have a common mass $m^{(10)}_0$.  
The two Higgs bosons doublets reside
in different 5-- and $\bar 5$--multiplets, with masses $m^{(5')}_0$
and $m^{(\bar 5')}_0$.  There is no symmetry principle that
demands that all these mass parameters should be the same.  The most
naive breakdown of exact universality is to consider different
values for $m^{(5')}_0$ and $m^{(\bar 5')}_0$, taking 
$m_0$ as the common mass for sleptons and squarks.

Depending on the exact mechanism of SUSY breaking, it may occur
that $m_{1/2}\simeq 0$. Low--energy gaugino
masses are then dominated by contributions of  
stop--top and Higgs--Higgsino loops.\cite{Barbieri}
In this case the gluino could be the LSP with a mass of 
$M_{\tilde{g}}\lesssim$~a~few~GeV
and the lightest neutralino 
may be somewhat heavier due to contributions from 
electroweak loops.\,\cite{Farrar_loop}  Light gluino scenarios are
being explored by many experiments.\cite{Farrar_rev}

\subsection{Gauge--Mediated Supersymmetry Breaking}

In SUGRA, gravitational interactions 
generate the soft SUSY--breaking terms in the MSSM Lagrangian.
Alternatively, soft SUSY--breaking terms
can be generated through gauge interactions.
This has the feature that mass degeneracies between sfermions with
the same quantum numbers (and, hence, the same gauge couplings)
occur naturally, which suppresses FCNC's.  
Also, in gauge--mediated models, the
scale of the SUSY breaking is much smaller than
the scale where gravity becomes relevant, so there is no possibility 
of Planck--scale corrections to these degeneracies (as there
can be between the GUT and Planck scales in SUGRA).\cite{kolda_rev}
In simple models,\cite{dine_nelson}
the existence of heavy messenger superfields $\psi$ with SM
quantum numbers is postulated.
SUSY is broken in a hidden sector which also couples to the messengers,
so that the $\psi$ fermion components have mass $M$, while
the scalar components have masses $M\sqrt{1\pm x}$,
where $x$ is a dimensionless parameter that controls the size
of SUSY breaking.
The MSSM gauginos and sfermions 
acquire masses different from their SM partners because of the radiative
effects generated by the messenger fields.
It is more convenient to define $\Lambda\equiv xM$, 
which fixes $\Lambda$ as the overall mass scale of the MSSM sparticles.
The gaugino masses at a low energy scale $\mu$ are
\begin{equation}
M_i(\mu)=\frac{\alpha_i(\mu)}{4\pi}\,g(x)(b_i-b'_i)\Lambda,
\label{eq:gaugino}
\end{equation}
where $i$ specifies the gauge group, $b_i$ is the MSSM coefficient
of the beta function for the running of $\alpha_i$, $b'_i$ 
includes the additional effect of the messenger fields in the running,
and $g(x)\simeq 1+x^2/6$.
The mass--squared of the MSSM scalars acquire values
\begin{eqnarray}
m^2_{i}(\mu)=2\Lambda^2\sum_i\left(\frac{\alpha_i(M)}{4\pi}\right)^2
C_i \left[ f(x)(b_i-b'_i) + \nonumber \right. \nonumber \\
\left. g(x)^2\left((b_i-b'_i)^2/b_i\right)
\left(\alpha_i^2(\mu)/\alpha_i^2(M)-1\right) \right],
\label{eq:scalar}
\end{eqnarray}
where $C_1=5/3Y^2, C_2=3/4,$ and $C_3=4/3$ ($Y$ is the weak hypercharge)
and $f(x)\simeq 1+x^2/36$.
The mass formula in Eq.~(\ref{eq:scalar}) ignores Yukawa couplings, and will be modified for
the stop and possibly sbottom and stau.
By comparing the previous two equations, it is clear that
the gaugino and scalar masses are roughly of the same order of magnitude.
Even after evolving these mass parameters to $M_{EW}$ (ignoring the effects
of Yukawa couplings),
sfermions with the same quantum numbers acquire the same masses,
yielding a
natural mass hierarchy between weakly and strongly interacting
sfermions; the mass
hierarchy of the gauginos is fixed by the gauge couplings (as in SUGRA models).
If the superfields $\psi$ reside in a complete representation
of $SU(5)$ or $SO(10)$, then unification of the gauge couplings
at a high scale is not compromised, though the unification will occur
at stronger values of the couplings and at a slightly different scale 
from the naive GUT scale.
One distinctive feature of these models is that the spin--3/2 superpartner
of the graviton, the gravitino $\gravitino$, 
can play a crucial role in the phenomenology.
Since the gravitino mass is given by the relation
$M_{\gravitino} = M\Lambda/M_{\rm Planck}$, the gravitino
can be very light depending on the value of $M$,
unlike SUGRA, where the gravitino has a mass on the order
of $M_W$.
As a result, in gauge--mediated Supersymmetry breaking, the gravitino
can be the LSP.  

In the above discussion, it is assumed that the messengers $\psi$ form a complete GUT 
multiplet.  However, if the messengers were neutral under some gauge group,
then the associated gauginos would be massless at one--loop because
of gauge invariance.
In particular, it is possible to construct a model where the gluino is 
a stable LSP with a mass of a few tens of GeV.\cite{Raby}
In this case, the missing energy signal for SUSY
disappears, since a 
stable LSP gluino will form stable hadrons.  

\subsection{R--Parity Violation}
\label{sec:r_parity}

One simple extension of the MSSM is to break the multiplicative
R--parity symmetry.
Presently, neither experiment nor any theoretical argument 
demand R--parity conservation, so it is natural to consider the
most general case of R--parity breaking.
It is convenient to introduce a function of superfields
called the superpotential, from which the Feynman rules for
R--parity violating processes can be derived.
The 
R--parity violating (RPV) terms 
which can contribute to the superpotential are~\footnote{In Eq. 
(\ref{eq:superpot}) bilinear terms are ignored.
A discussion of the phenomenological implications of such terms can be found
in the literature.\cite{Diaz}}
\begin{equation}
W_{RPV} =  \lambda_{ijk} L^i L^j \bar{E}^k +
           \lambda^{'}_{ijk} L^i Q^j \bar{D}^k +
           \lambda^{''}_{ijk} \bar{U}^i \bar{D}^j \bar{D}^k  
\label{eq:superpot}
\end{equation}
where $i,j,k$  are generation indices (1,2,3),
$L^i_1 \equiv \nu^i_L$, $L^i_2=\ell^i_L$ and $Q^i_1=u^i_{L}$, $Q^i_2=d^i_{L}$
are lepton and quark components of $SU(2)_L$ doublet superfields,
and
$E^i=e^i_{R}$, $D^i=d^i_{R}$ and $U^i=u^i_R$ 
are lepton, down and up-- quark $SU(2)_L$ singlet superfields, respectively.
The unwritten SU(2)$_L$ and SU(3)$_C$ indices imply that the first 
term is antisymmetric under $i \leftrightarrow j$, 
and the third term is antisymmetric under $j \leftrightarrow k$. Therefore,
$i \neq j$ in   $L^i L^j \bar{E}^k$ and $j \neq k$ in  
$\bar{U}^i \bar{D}^j \bar{D}^k$. 
The coefficients $\lambda_{ijk}$, $\lambda^{'}_{ijk}$ and 
$\lambda^{''}_{ijk}$ are Yukawa couplings, and there is no {\it a priori}
generic prediction for their values.
In principle, $W_{RPV}$ contains 45 extra parameters over the
R--parity--conserving MSSM case.

Expanding Eq.~(\ref{eq:superpot})
as a function of the superfield components, the interaction
Lagrangian derived from the first term is
\begin{equation}
{\cal{L}}_{LLE} = \lambda_{ijk} \left\{ \tilde{\nu}_L^i e_L^j \bar{e}^k_R + 
                \tilde{e}_L^i \nu_L^j \bar{e}^k_R + 
                (\tilde{e}_R^k)^* \nu_L^i e^j_L + h.c. \right\} 
\label{eq:RPVLLE}
\end{equation}
and from the second term,
\begin{eqnarray}
{\cal{L}}_{LQD} = \lambda^{'}_{ijk} \left\{
 \tilde{\nu}_L^i d_L^j \bar{d}^k_R -
                \tilde{e}_L^i u_L^j \bar{d}^k_R +
                \tilde{d}_L^j \nu_L^i \bar{d}^k_R -
                \tilde{u}_L^j e_L^i \bar{d}^k_R + \right. \nonumber \\
                \left. (\tilde{d}_R^k)^* \nu_L^i d^j_L - 
                (\tilde{d}_R^k)^* e_L^i u^j_L + h.c. \right\}
\end{eqnarray}
Both of these sets of interactions violate lepton number. 
The $\bar U\bar D\bar D$ term, instead, violates 
baryon number.
In principle, all types of R--parity violating terms may co--exist,
but this can lead to a proton 
with a lifetime shorter than the present experimental limits. 
The simplest way to avoid this
 is to allow only operators which  conserve baryon--number but  violate
 lepton--number or vice versa.

There are several effects on the SUSY phenomenology due to these
new couplings: (1) lepton or baryon number violating processes
are allowed, including the
production of single sparticles (instead of pair production),
(2) the LSP is no longer stable, but
can decay to SM particles within a collider detector,
and 
(3) because it is unstable, 
the LSP need not be the neutralino or sneutrino, but
can be charged or colored.

Present data are in remarkable agreement with the SM predictions, and very 
strong bounds on the R--parity--breaking operators can be derived
from the following 
processes: ${(a)}$ charged--current universality, ${(b)}$ $
\Gamma(\tau\to e\nu{\bar\nu})/\Gamma(\tau\to\mu\nu{\bar\nu})$, $
{(c)}$ the bound on the mass of $\nu_e$, ${(d)}$
neutrino--less double--beta decay, ${(e)}$ atomic parity
violation, ${(f)}$ $D^0-{\bar D}^0$ mixing, 
${(g) }$ $R_\ell =\Gamma_{had}(Z^0)/\Gamma_\ell (
Z^0)$, ${( h)}$ $\Gamma(\pi\to e{\bar\nu})/\Gamma(\pi\to\mu
{\bar\nu})$, ${(i) }$ $BR(D^+\to{\bar K}^{0*}\mu^+\nu_\mu)/BR(D^+
\to {\bar K}^{ 0*}e^+\nu_\mu)$, ${(j)}$ $\nu_\mu$
deep--inelastic scattering, ${(k)}$ $BR(\tau\to\pi\nu_\tau)$, 
${(l)}$ heavy nucleon decay, and ${(m)}$ $n-{\bar n}$
oscillations.  Additional limits can be derived from deep inelastic 
experiments at HERA.\cite{rplimits}
On the other hand, within the allowed values of the R--parity--violating
couplings, $\lambda_{ijk}, \lambda'_{ijk}$, a whole new world opens
up for SUSY searches.

\subsection{Run Ia Parameter Sets (RIPS) }
\label{sugra_inspired}

Some CDF and \D0~SUSY searches 
are analyzed in the framework of so--called ``SUGRA--inspired
models.''
To understand the limits that 
appear in many published analyses, it is necessary to
state explicitly the framework behind RIPS.

First, there are five main input parameters: $M_{\gluino}, m_{\squark},
M_A, \tan\beta$ and the magnitude and sign of $\mu$.  
The gluino mass $M_{\gluino}$ is defined to be $M_3$, which is
equivalent to
specifying \mhalf~and, hence, $M_1$ and $M_2$ using the unification 
relations Eq.~(\ref{eq:gauginounif}).  The
chargino and neutralino properties are then fixed by $M_1, M_2, \tan\beta$
and $\mu$.  In practice, the value of $\mu$ is set much larger than 
$M_1$ and $M_2$, 
so the properties of the neutralinos, charginos, and gluino are similar to 
those in a pure SUGRA model. 

Next, 
all squark soft SUSY--breaking mass parameters are set to $m_{\squark}$,
and the $D$--terms are neglected.
The result is that the first 5 squarks are degenerate in mass.
This may be unrealistic if $\tan\beta$ is large, since the
sbottom mass can be naturally lighter because of non--negligible
off--diagonal elements in the sbottom mass matrix.
The stop squarks are made heavier than the other squarks
by fixing $A_t = \mu/\tan\beta$ (see
Eq.~(\ref{stop_matrix})), which tunes away the mixing between
$\stop_L$ and $\stop_R$.
The resulting stop masses are 
$m_{\stop_1}=m_{\stop_2}=\sqrt{m_{\squark}^2+m_t^2}$. 
Therefore, experimental limits placed on RIPS
show no sensitivity to the stop squarks.
Note that, in SUGRA models, the stop squared--mass soft SUSY--breaking
parameters $m^2_{Q_3}$ and
$m^2_{U_3}$ are generally not equal and are smaller than the other
squark parameters at $M_{EW}$, so that one stop squark is lighter than the
other squarks. 

Giving the other five squarks a common value at the weak scale
ignores the details of running from the GUT
scale (see Eq.~(\ref{eq:approxmQU})) and the different $D$--terms.  
However, using an average of the two formulae in
Eq.~(\ref{eq:approxmQU}), a specific \mhalf~and $m_{\squark}$
roughly determine a value of $m_0^2$.  Whenever there is a solution with
$m_0^2>0$ (which implies $m_{\squark} > 0.85 M_{\gluino}$), 
RIPS has many features of a SUGRA model.
Indeed, when $m_{\squark} > M_{\gluino}$, 
the approximate
SUGRA relations
$m^2_{\slepton_L}=m^2_{\squark}-0.73M^2_{\gluino}-0.27M^2_Z 
\cos 2\beta$,
$m^2_{\slepton_R}=m^2_{\squark}-0.78M^2_{\gluino}+0.23M^2_Z 
\cos 2\beta$, and
$m^2_{\sneutrino}=m^2_{\squark}-0.73M^2_{\gluino}+0.5M^2_Z \cos 2\beta$
are used to fix the slepton masses.\footnote{Observe that the $D$--terms
for the sleptons,
although correct, are negligible in comparison with the approximation
made in defining a common $m_{\squark}$.
However, $D$--terms are included to
assure the correct splittings between the $\slepton_L$ 
and $\sneutrino$ masses.}
The region $m_{\squark} < M_{\gluino}$
is very hard to realize in SUGRA models,
but is also worth investigating.
In this case, for some analyses, a constant value of 350 GeV is
set by hand for
$m_{\slepton_L}$, $m_{\slepton_R}$, and $m_{\sneutrino}$.
Accordingly, experimental limits placed on RIPS when
$m_{\squark} < M_{\gluino}$
show little sensitivity to the sleptons.

Finally, the Higgs mass $M_A$ is used to determine the Higgs boson
sector.  This is equivalent to considering partial non--universality
for the scalar sfermion and Higgs boson soft
SUSY--breaking mass parameters at high energies,
{\it i.e.} $m_0 \ne m^{(5')}_{0} \ne m^{(\bar 5')}_{0}$
(see the discussion near the end of Sec. 2.2).
In practice,
the CP--odd Higgs boson mass $M_A$ is 
set to a large value, so that the
lightest neutral Higgs boson $h$ has SM--like couplings to 
gauge bosons and fermions, and
all other Higgs bosons are 
so heavy they are 
not kinematically accessible at the Tevatron.
\setcounter{footnote}{0}
\section{The CDF and \D0~Detectors}

The CDF~\cite{CDFnim}
and \D0~\cite{D0nim} detectors are located at the interaction regions
B0 and D0 in the accelerator ring.\footnote{The Tevatron ring has 
six--fold symmetry, with the centers
of the straight sections labelled as A0, B0, C0, D0, E0, and F0.}
 Both detectors feature particle
tracking detectors close to the interaction region, surrounded by  
quasi--hermetic calorimetry covering the region of 
pseudorapidity\,\footnote{The pseudorapidity $\eta$ is defined as
 $-\ln(\tan\frac{\theta}{2})$.
 In the CDF and \D0~ coordinate systems, 
 $\theta$ and $\phi$ are
 the polar and azimuthal angles, respectively, with respect to the proton beam
 direction $z$. 
}~of approximately $|\eta|< 4$. Muon detection
systems are located outside the calorimeters for both detectors.

\subsection{The CDF Detector}
The CDF detector is distinguished by its magnetic spectrometer:  a \mbox{3--m}
diameter, 5--m long superconducting solenoidal magnet, which creates a 
1.4~T field uniform at the 0.1\% level 
and contains the particle tracking detectors.  A four--layer
silicon microstrip vertex detector (SVX),\cite{svxnim} located directly
outside the beampipe, tracks charged particles in the $r-\phi$ plane.
The SVX measures the impact parameter of tracks coming from secondary vertices
of bottom and charm decays with a typical resolution of 30
$\mu$m, providing heavy--flavor tagging for jets.
A set of vertex time projection chambers (VTX) surrounding
the SVX  provides tracking in the radial and $z$ directions and
is used to find the $z$ position
of the \pbarp~interaction. Outside the VTX, between 
from a radius of 30 to 150 cm, the 3.2--m long central tracking 
chamber (CTC) is used to measure the momentum of
charged particles, with up to 84 measurements per track. 

The calorimeter
is divided into a
central barrel (\mbox{$|\eta| < 1.1$}), end--plugs
\footnote{``End--plugs'' because they plug into the ends of
the solenoid and central calorimeter.} (\mbox{$1.1 < |\eta| < 2.4$}) 
and forward/backward modules (\mbox{$2.4 < |\eta| < 4.2$}).  Each of these
is segmented into projective \footnote{Projective means pointing  
approximately at the interaction region.} electromagnetic and hadronic towers 
subtending 0.1 in $\eta$ by 15\deg~in $\phi$ in the central
calorimeter and 5\deg~elsewhere. Wire chambers with
cathode strip readout give information on electromagnetic shower profiles in the
central and plug calorimeters (CES and PES systems, respectively).  A system of
drift chambers (CPR) outside the magnet coil and in front of the electromagnetic
calorimeters serves as a ``preradiator,''  allowing additional photon/$\pi^0$
discrimination on a statistical basis. Muons are identified with the central
muon chambers, situated outside the calorimeters in the region   \mbox{$|\eta| <
1.1$}. 

The magnetic spectrometer measures muon and other  charged particle
transverse momenta with a resolution ${\sigma_{p_T}}/{p_T} < 0.001 p_T$
($p_T$ in
GeV) and allows
a precision calibration of the electromagnetic calorimeters by comparing the
measured calorimeter response to the measured momentum from high--energy
electrons from $W$ decays.\cite{CDF_Wmass_PRD}
Electron energies are measured 
with a resolution
${\sigma_E}/{E} =
{.135}/{\sqrt{E_{T}}}\oplus .01$ ($E_T$ in GeV).\footnote{The symbol $\oplus$
denotes addition in quadrature, {\it e.g.} $a\oplus b=\sqrt{a^2+b^2}$.
The total resolution can be parameterized this way when there are two or more
independent components of resolution.}
Jets are reconstructed as
energy clusters in the calorimeter, using a cone algorithm \cite{cone} with a
cone radius of either $R = 0.7$~\cite{CDF_QCD_jets} 
or $R = 0.4$~\cite{CDF_top_jets}
in \detadphi~space.
The jet energy resolution in the central region is approximately
${\sigma_E}/{E} = .80/\sqrt{E} \oplus .04$ ($E$ in GeV).

Missing transverse energy,\footnote{The transverse momentum of a 
particle with  momentum $p$ is
 $p_T = p \sin \theta$. The analogous quantity using energy, defined as
$E_T = E\sin\theta$, is called transverse energy.} (\met), 
a key quantity in SUSY searches,
is calculated as $\sum E^{\mbox{tower}} \ ({\bf {\hat
n}}\cdot{\bf {\hat r}}) \ {\bf {\hat r}}$, where   the sum is over both
electromagnetic and hadronic calorimeter towers \footnote{A tower is
a cell in $\eta-\phi$ space.}
in $|\eta|<3.6$, $E^{\mbox{tower}}$  is the
energy measured in the  
tower, ${\bf {\hat n}}$ is the unit vector pointing in the direction of the
center of the tower from the event vertex, and ${\bf {\hat r}}$ is the unit
vector in the radial direction.  The \met~is always
corrected for the momentum of muons; for many SUSY analyses, it
is also corrected for the calorimeter response to jets.
The typical resolution on a component of \met~is 5.7 \gev~in 
$Z^0 \rightarrow e^+e^-$ events.

\subsection{The \D0\ Detector}
The \D0\ detector 
consists of three major components: a non--magnetic central tracking
system,                 
central and forward Liquid Argon sampling calorimeters, 
and a toroidal muon spectrometer.
The central tracking system consists of four detector subsystems:
a vertex drift chamber, a transition radiation detector, 
a central drift chamber, and two forward drift chambers. 
Its outer radius is 76 cm.  The system
provides identification of charged tracks in the pseudorapidity
range $|\eta| \leq 3.5$.  It measures the trajectories of charged
particles with a resolution of 2.5 mrad in $\phi$ and 28 mrad in
$\theta$.  Using the reconstructed charged tracks, the position of
the primary interaction along the beamline direction is reconstructed with
a resolution of 8 mm.  The central tracking system also
measures the ionization of tracks to allow single charged
particles to be distinguished from \epm~pairs from photon conversion.
The transition radiation detector aids in
distinguishing electrons 
from charged pions.

The calorimeter  is transversely segmented into pseudoprojective
towers with \detadphi~$= 0.1\times 0.1$  and 
provides full coverage to
$|\eta| \leq 4.2$.
The calorimeter is divided into three parts, a central calorimeter
and two end calorimeters.  These are further segmented into 
an inner electromagnetic section, followed by a fine hadronic section,
and then a coarse hadronic section.  
Between the central and end--cap 
calorimeters, a set of scintillator tiles provides improved
energy resolution for jets that straddle the two detectors.
The electromagnetic (EM) calorimeter is divided into 32
modules in $\phi$, each of which has
22  layers, each approximately 1 radiation length~\footnote{The radiation
length is the mean distance traversed by an electron in a given material
during which it radiates a fraction $1-e^{-1}$ of its energy via bremsstrahlung.}
thick, with Liquid Argon as the active element and 
$^{238}U$ plates as the passive element.  These layers are arranged
into four longitudinal segments per tower, called cells.
The first cell contains 2 layers, the second cell contains 2 more layers, 
the third cell is finely segmented, with  
\detadphi~ = 0.05 x 0.05 and contains 7 layers, and the last cell contains 
10 layers.  The fine hadronic calorimeter uses a Uranium--Niobium alloy as its
passive element, and the coarse hadronic uses copper.
The electron energy resolutions, as measured in the EM
calorimeter, are  
${\sigma_E/E} = 0.130/\sqrt{E_T} \oplus 0.0115 \oplus 0.4/E$
for $|\eta|<1.1$, 
and 
${\sigma_E/E} = 0.157/\sqrt{E} \oplus 0.010 \oplus 0.4/E$
for $1.4 \le |\eta| \le 3.0$.\cite{qzhu}
The azimuthal position resolution for electrons above 
50 \gev~as measured by the calorimeter is 2.5 mm.
The muon spectrometer provides muon
detection in the range $|\eta| \leq 3.3$. 
The total thickness of the calorimeter plus the toroid varies
from 13 to 19 interaction lengths, making 
hadronic punch--through backgrounds
negligible.  
The muon momentum resolution is
${\sigma_p/p} = {{0.18(p-p_0)}/p}\oplus 0.008p$ ($p$ in \gevc, $p_0=2$ GeV/c).
The \D0~detector has a compact tracking volume which 
helps control backgrounds to prompt muons from in--flight decays of $\pi$
and $K$ mesons.
Jets are reconstructed as
energy clusters in the calorimeter, using a cone algorithm with a
cone radius of $R = 0.5$ or $R=0.7$ in \detadphi.\cite{jetshape}
The jet energy resolution is approximately
${\sigma_E}/{E}=0.8/\sqrt{E}$ ($E$ in \gev).\cite{d0jetres}

\met~is calculated using the vector sum of energy
deposited in all calorimeter cells, over the full calorimeter coverage
for $|\eta| \leq 4.2$,
with corrections applied to
clustered cells to take account of the jet energy scale, and to
unclustered cells as determined from studies of $E_T$ balance in 
$Z^0\to e^+e^-$ events that do not contain hadronic calorimeter
clusters.   The resolution on a component of \met~in 
``minimum--bias'' events~\footnote{These events are defined by the 
requirement that a beam--beam collision took place as measured by arrays of
scintillation counters forward and backward near the beampipe, 
and therefore have 
a smaller selection bias than events selected with more selective triggers.}
is 1.1 \gev + 0.02 $\sum_{}^{} E_T$, where  $\sum_{}^{} E_T$ is the scalar
sum of transverse energies in all calorimeter cells.
For some analyses, the \met~is corrected for the
presence of muons, which only leave a small fraction of their energy
in the calorimeter.    

\subsection{Experimental Realities}
\label{subsec:realities}

There are potentially two types of backgrounds to any experimental
signature, physics and instrumental.  Physics backgrounds mimic the
event signature even in an ideal detector, while instrumental backgrounds
arise because of detector flaws.
Experimental signatures -- SUSY or otherwise -- are identified from 
``objects'' -- the building blocks of the event. 
Examples
of objects that CDF and/or \D0~use are electrons, muons, tau's, \met,
so--called ``generic jets'' (presumably from quarks and/or gluons, but without
flavor identification), $c$'s, $b$'s, photons, and, using another level of 
kinematic reconstruction, $W$'s, $Z$'s, and $t$'s.  
The selection of
each of these objects carries with it an efficiency and 
also a ``fake rate,''
a probability that the object is actually a different object which has been
misidentified.
The description of an object as an ``electron,'' for example, 
more precisely
means an ``electron candidate that passes the electron cuts,'' and no more
and no less.

For the majority of
searches, the signal and its physics backgrounds can be estimated using
Monte Carlo simulation.
The output of an event generator
such as {\tt ISAJET}~\cite{isajet} or {\tt PYTHIA}~\cite{spythia}
can be folded with relatively simple parameterizations of the
detector response to give a good description of the data.
A typical simple simulation transforms the final state partons
from a Monte Carlo into jets, using a clustering algorithm 
similar to the one used for the data.
It then convolutes the momenta of the electrons, photons, muons, and jets with
the appropriate experimental resolutions, generating
``smeared'' momenta.  \met~is calculated
by first summing the smeared visible momenta, and then adding the effects
of additional minimum--bias events in the same beam crossing.
When calculating the geometric acceptance of the detector, 
it is necessary to include 
the distribution for the interaction vertex position in $z$, 
which is Gaussian with an RMS of approximately 30 cm. 
Finally, the detection
efficiencies for electrons, photons, $b$--quarks, $\tau$'s,
and muons (jets above 20 \gev~are 
usually found with good efficiency) are applied. 
Initial and final state gluon radiation need to be included, since they
can affect the efficiency by adding extra jets which can modify the event 
signature or ``promote'' backgrounds into the signature.

In order to make even rough predictions of instrumental backgrounds,
 imperfections in the 
detectors must be taken into account.
Two effects make these difficult to estimate: 
fakes, and tails on jet energy resolution distributions.  
Because of the very large 
multi--jet production rate at the Tevatron, there can be significant fake 
backgrounds, even if the fake probability is very small.    
Fakes are very complicated, and the fake rate must be
evaluated for each analysis using the appropriate data. 
In general, the efficiency for properly identifying an object
and the probability that another object fakes it are complementary.  
For signatures dominated by instrumental backgrounds,
tighter selection
criteria (``cuts'') make for a purer sample, but reduce the efficiency.
For signatures dominated by physics backgrounds, 
looser cuts are preferred because they produce a higher efficiency for the
same ratio of signal to background.
Fakes are an especially serious problem for signatures involving photons, 
tau's,  $b$-- and $c$--quark tagging, and \met~in events with jets.
Although jet energy resolutions are roughly
Gaussian, even small non--Gaussian tails, convoluted
with the large jet cross section,  can lead to significant numbers of
events with large fake \met. 
In addition, there are some other factors which contribute to fakes and  which
are unique to working at the Tevatron, such as the long interaction region, the
existence of multiple collisions in a single  event, the presence of the Main
Ring in the same tunnel as the beam, 
and  larger cosmic ray backgrounds than found at
detectors that are deeper underground (such as the LEP experiments).
Even a full detector simulation cannot correctly model all
detector imperfections and these other effects.

\section{ The Present Status of Sparticle Searches }

Most previous Tevatron searches have been made
under very specific assumptions.
Several of the classic signatures, such as ``jets+\met,''\,\cite{ref1}
``trileptons,''\,\cite{ref2,trilepton_golden}
and ``same--sign dileptons''\,\cite{ref3}
are likely to be fruitful in many models; others
may be specific to a certain model.
We advocate a signature--based approach, in which 
a broad range of channels are studied for departures from SM expectations,
without engineering the analysis for a specific class of models
(see Appendices A and B).
While this may sound obvious, it is a large task with no 
well--defined beginning.  With the experience of several years of data
taking, however, experimentalists now have an idea of what they can
do well.  Signatures involving high--\pt~isolated electrons, muons, and
photons, $b$--quarks, $c$--quarks, $\tau$ leptons, and/or \met~can be measured
accurately and have understandable backgrounds.  Such signatures then
are a practical starting point for the new generation of 
searches.
In addition, there are motivations that are more general than the predictions of
specific models for studying samples of 
$(i)$ high--mass, high--\met~events to probe gluino and 
squark production, 
$(ii)$ inclusive leptons, lepton pairs, and 
gauge bosons ($\gamma$, $W$, and $Z$) 
to probe both direct and cascade production of charginos and neutralinos, 
and $(iii)$  
third--generation fermions ($t,b$, and $\tau$) to probe decays of 
light squarks and
sleptons, as well as decays of Higgs bosons and Higgsinos.   
When setting
limits, it is convenient to use specific models which reduce the number of free
parameters (such as SUGRA or RIPS), but the quoted limits are valid only 
in that context and are not general limits. 
New physics may appear where we do not expect it.

\begin{table}[!htb]
\centering
\caption{A compilation of results from Run I Tevatron SUSY searches
as of the summer of 1997.
The symbol $b$ denotes an additional $b$--tagged jet. Also listed are the
references and the section of this chapter where each analysis is
discussed. More information is available for \D0~at 
$~~http://www-d0.fnal.gov/public/new/new\_public.html$, and for
CDF at $~~http://www-cdf.fnal.gov/$}
\begin{tabular}{|l|l|l|l|l|c|c|}
\hline \hline 
Sparticle & Signature & Expt. & Run & $\intlum$(pb$^{-1})$  & Ref. &
Sec. \\ \hline
Charginos   & \met+trilepton & CDF & Ia  &  19 & \oldcite{CDF_trilepton_1a}
& 4.1 \\ \cline{2-7} 
 and& \met+trilepton & CDF & Iab & 107 & \oldcite{CDF_trilepton_1ab} & " \\ \cline{2-7}
Neutralinos         & \met+trilepton & \D0  & Ia  & 12.5 
&\oldcite{d0_wino1a} & " \\  \cline{2-7} 
& \met+trilepton & \D0  & Ib  &  95   & \oldcite{d0_wino1b} & " \\  \cline{2-7}
& $\gamma\gamma+$\met or jets & CDF & Ib & 85 & \oldcite{CDF_stop_rlc} 
                             & 4.10 \\ \cline{2-7} 
& $\gamma\gamma+$\met  & \D0 & Iab & 106 & \oldcite{d0_diphoton} & " \\ \cline{2-7} 
\hline
Squarks & \met$+\ge$3,4 jets  & CDF & Ia & 19   & \oldcite{CDF_metjets_1a}
& 4.2 \\ \cline{2-7} 
and     & \met$+\ge$3,4 jets    & \D0  & Ia & 13.5 & \oldcite{D0_metjets_1a}  & " \\ \cline{2-7} 
Gluinos & \met$+\ge$3 jets    & \D0  & Ib & 79.2 & \oldcite{D0_metjets_1b} & " \\ \cline{2-7} 
        & dilepton$+\ge$2 jets & CDF & Ia & 19   & \oldcite{CDF_metdilepton_1a} & " \\ \cline{2-7} 
        & \met$+$dilepton$+\ge$2 jets  & CDF & Ib & 81   & \oldcite{CDF_metdilepton_1b} & " \\ \cline{2-7} 
        & \met$+$dilepton     & \D0  & Ib & 92.9   & \oldcite{D0_metdilepton_1b} & " \\ \hline
Stop    & \met+$\ell+\ge$2 jets+$b$    & CDF & Ib & 90 &
\oldcite{CDF_stop_gold} & 4.3 \\ \cline{2-7}
        & \met+$\ell+\ge$3 jets+$b$    & CDF & Iab & 110 & \oldcite{CDF_stop_carmine} & " \\ \cline{2-7}
        & dilepton+jets     & \D0  & Ib & 74.5 & \oldcite{D0_stopee} & " \\ \cline{2-7}
        & \met+2 jets        & \D0  & Ia & 7.4 &  \oldcite{D0_stopjj}& " \\ \cline{2-7}
        & \met+$\gamma$+$b$  & CDF & Ib & 85 & \oldcite{CDF_stop_rlc} & 4.11 \\ \hline
Sleptons  &$  \gamma\gamma$ \met& \D0~&Iab & 106 & \oldcite{d0_diphoton} & 
4.10\\ \cline{2-7}\hline
Charged & dilepton + \met & CDF & Ia & 19 
&\oldcite{CDF_ch_higgs_jinsong}  & 4.5 \\ \cline{2-7} 
Higgs & $\tau$+2 jets+\met & CDF & Ia & 19 & \oldcite{CDF_ch_higgs_couyoumt}  & " \\  \cline{2-7} 
& $\tau$+$b$+\met+($\ell$,$\tau$,jet) & CDF & Iab & 91
&\oldcite{CDF_ch_higgs_rutgers}  & " \\ \cline{2-7} 
& $\tau$+$b$+\met+($\ell$,$\tau$,jet) & \D0 & Iab & 125
&\oldcite{D0_chhiggs}  & " \\ \hline
Neutral & $WH \goes \ell+\met$+$b$+jet & CDF& Iab &109& 
\oldcite{CDF_neu_higgs_weiming} & 4.6 \\ \cline{2-7} 
 Higgs& $WH \goes \ell+\met$+$b$+jet & \D0& Ib &100& 
\oldcite{d0_neu_higgs} & " \\ \cline{2-7} 
 & $WH,ZH \goes \gamma\gamma$+2 jets & \D0& Ib &101.2& 
\oldcite{d0_gg_higgs} & " \\ \cline{2-7} 
 & $ZH \goes b$+jet+\met & \D0& Ib &20& 
\oldcite{d0_hedin_higgs} & " \\ \cline{2-7} 
   & $WH,ZH \goes$ 2 jets+2 $b$'s      & CDF& Ib & 91  & 
\oldcite{CDF_neu_higgs_valls} & " \\ \hline  
$R$ violating & dilepton+$\ge2$ jets & CDF & Iab & 105 &
     \oldcite{CDF_Rparity} & 4.8 \\ \hline
Charged LSP & slow, long--lived particle & CDF & Ib & 90 & 
     \oldcite{CDF_stables} & 4.7 \\ \cline{2-7} 
\hline
\end{tabular}
\label{tab:summary}
\end{table}

In the sections below we discuss the present status of searches at the
Tevatron. We also discuss the phenomenology behind these
searches and comment on possible improvements.  
Table \ref{tab:summary} summarizes those 
CDF and \D0 analyses that
have been published or presented at conferences.
The pace and scope of supersymmetry searches at the Tevatron, as well as the
sophistication, have grown enormously in the last several years as the
emphasis has shifted beyond the top quark and more data have
become available; there are many analyses currently in
progress.  A much broader picture of the Tevatron's capabilities 
should emerge as these results become available.

\setcounter{footnote}{0}
\subsection{Charginos and Neutralinos}
\label{sec:gauginos}
In SUGRA models, the light neutralinos and charginos
are much lighter than the gluino or squarks, and may be the only
sparticles directly accessible at the Tevatron.
In general, the lightest neutralino is a good LSP candidate, so,
assuming all charginos and neutralinos are relatively light, a
discussion of their phenomenology is a good starting point for
an overview of Tevatron searches.

Chargino and neutralino pairs would be 
produced  directly\,\footnote{They may also be produced 
indirectly in the  decays of heavier sparticles.} at hadron colliders 
through their electroweak couplings to squarks and the vector bosons $\gamma$,
$W$, and $Z$.   The production cross sections  are not a simple function of
chargino and neutralino masses, but depend also on their (model--dependent)
mixings and the squark
masses. Quite generally, there are three contributions to $\ino\ino$
production:
$(i)$ s--channel gauge boson production,
$(ii)$ t--channel squark exchange, and $(iii)$ interference.
For type $(i)$, the reactions
$q\bar q^{'}\to W^*\to\ino^\pm_i \ino^0_j$ 
and $q\bar q\to \gamma^*/Z^*\to\ino^+_i \ino^-_j$ 
occur
through Wino and Higgsino components, and 
$q\bar q\to Z^*\to\zinog_i \zinog_j$ through Higgsino 
components of the neutralinos and charginos,
respectively.\footnote{An
asterisk superscript $^*$ on a gauge boson refers to a resonance
off the mass shell.}
In type $(ii)$, the scattering quarks and antiquarks exchange squarks
subject to the constraint that
$\squark_R$ couples only to Bino component of neutralinos, and $\squark_L$
to Bino  and
Wino components of the neutralinos  and to charginos.
Any $\squark_R$ or $\squark_L$ 
coupling to Higgsino--components is proportional to the corresponding
quark
mass and can be ignored for chargino and neutralino pair production
(even though $\sbottom$ and $\stop$ have large couplings, the
contribution of $b$ and $t$ initial states from hadrons is small).
Hence, in chargino pair production ({\it e.g.} $\winoop\winoom$), only $\squark_L$ 
exchange is important, since $\squark_R$ only couples to the charged
Higgsino.
In the case when all squarks are heavy, type $(i)$ contributions dominate.
When the squarks are sufficiently light, both type $(ii)$ and $(iii)$ can be
important.
For example, if $|\mu|\gg M_1,M_2$ (see Eq.~(\ref{eq:mularge})),
then $\zinol$, $\zinoh$, and $\winol$ are mostly gaugino--like, and both
$\squark_L$ and $\squark_R$ can have electroweak strength couplings to them.
If, on the other hand, $|\mu|\ll M_1,M_2$ (see Eq.~(\ref{eq:M2large})),
$\squark$ exchange is only important for the heavier states, which
might not be kinematically accessible.

\begin{figure}[!ht]
\centerline{
\hbox{
\hbox{\psfig{figure=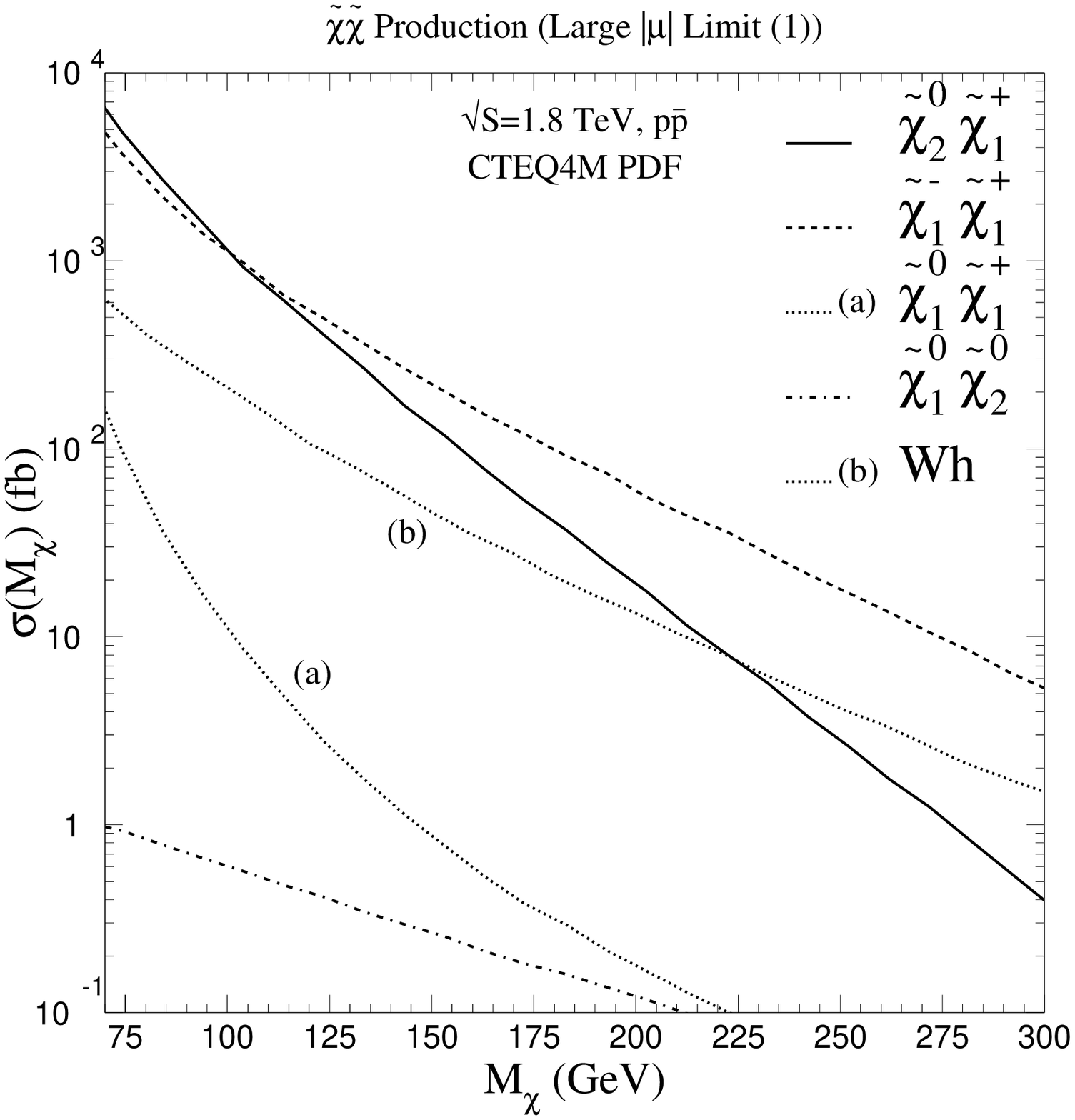,height=70mm}}
\hbox{\psfig{figure=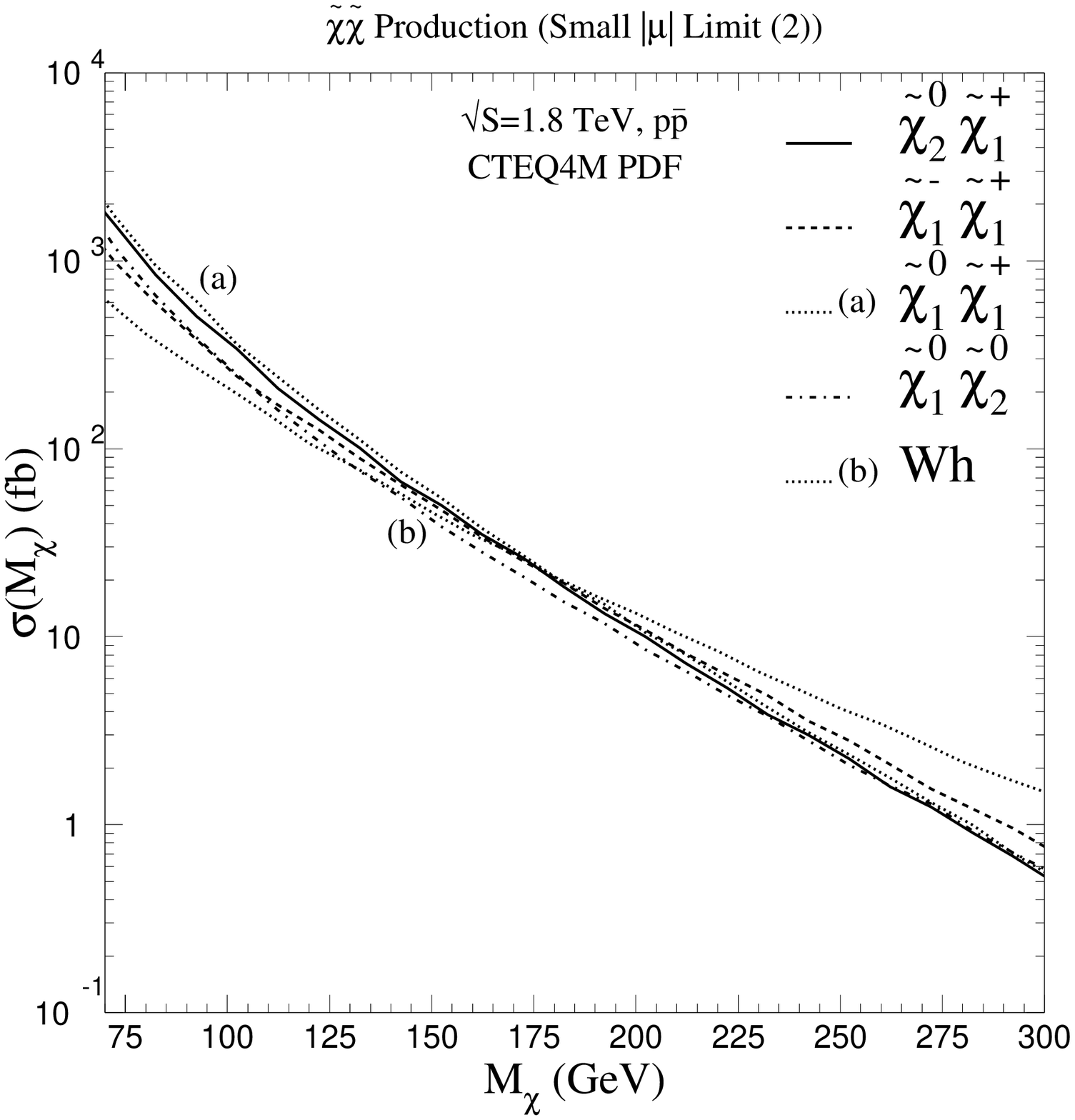,height=70mm}}
}
}
\caption{Production cross sections at the Tevatron 
for chargino and neutralino pair production
versus the lightest chargino mass
for two limiting models discussed in the text: large $|\mu|$ (Left) and
small $|\mu|$ (Right).  
The $Wh$ cross section (curve $b$) is shown for 
reference as a function of $M_h$.
}
\label{fig:susyxsecgauge}
\end{figure}

Figure~\ref{fig:susyxsecgauge} shows the production cross sections
of various chargino and neutralino pairs at the Tevatron for the limiting
cases considered earlier in Eq.~(\ref{eq:mularge}) 
and Eq.~(\ref{eq:M2large}).
In this figure, $\tan\beta=2$, $m_{\squark}=500$ GeV, and 
the gauginos obey the unification relations (see Eq.~(\ref{eq:gauginounif})).
The left figure is generated by fixing $\mu=-1$ TeV and varying
\mhalf, the right figure by fixing \mhalf=1 TeV and varying $\mu$.
For reference, the SM $W^\pm h$ production cross section is shown
as a function of $M_h$.  The example of Eq.~(\ref{eq:mularge}) is most
like a pure SUGRA model, in which  the couplings of $W \winog_1\zinoh$ and $Z
\winoop\winoom$ are large, so the $\winog_1\zinoh$ and $\winoop\winoom$
production cross sections are the largest.  

Charginos and neutralinos can also be produced in associated production: 
$\squark\ino$ and $\gluino\ino$.  This is discussed further in 
Section \ref{sec:squark_gluinos}.

The decay patterns for the charginos and neutralinos are also very
model dependent. When kinematically allowed,
a tree--level 2--body decay dominates over a tree--level 3--body decay, 
because the latter has an extra factor of $g^2$/(4$\pi^2$) in
the decay rate.
Possible 2--body decays of the chargino are to $W^\pm\zinol$, 
$H^\pm\zinol$,
$\tilde{\ell}_{L,R}\nu$, $\tilde{\nu} \ell$ and $\squark q'$. 
The heavier chargino $\winoh$ can also
decay to $Z\winol$ and $h\winol$
(when Higgs bosons are part of the event signature,
the final states can contain heavy flavor quarks or tau leptons).
When no 2--body final states are kinematically allowed,
the chargino will decay to a 3--body final state
with contributions similar
to those for chargino production (described previously): $(i)$
virtual gauge boson decays, $(ii)$ virtual sfermion decays, and
$(iii)$ interference.
A common decay is $\winog\rightarrow\zinog f\bar{f}'$.
If sfermions are much heavier than a chargino, then type $(i)$
dominates, and the decays proceed through $W^*$
with branching ratios similar to those of the on--shell $W$ boson.
Virtual squarks and sleptons can significantly alter the branching
ratios to a specific $f\bar f'$ final state depending on the squark
and slepton masses, so that a 100\% branching ratio to $\ell\nu\zinol$ or
$jj\zinol$ final states is possible.
A 3--body decay $\winog_j\to\gluino q\bar q'$ can occur
through virtual squark decays, provided the gluino is light enough
(see Appendix A).  

Similarly, for neutralinos, the 2--body decays to 
$Z\zinog$, $h\zinog$, $W^\mp\winog$, $H^\mp\winog$, $\tilde{\ell}_{L,R}\ell$,
$\tilde{\nu} \nu$
or $\squark q$ will dominate when kinematically allowed.
When this is not the case, the 3--body decay
$\zinog_i\to\zinog_j f\bar f$ (or $\zinog\to\gluino q\bar q$) can occur
through virtual $Z$ bosons, squarks, or sleptons.
Three--body decays can also be in competition
with a loop decay, so that $\zinoh\to\zinol\gamma$,
since the same factor of $g^2/(4\pi)^2$ coming from a loop integral 
for a 2--body decay also
appears in the 3--body decay rate.  Such a decay is important when
the $\zinol$~is Higgsino--like and $\zinoh$~is gaugino--like, or
{\it vice versa}.  

In general,
for  $\zinoh\winol$ production, 
the final states are $(i)$ four leptons and \met, or 
$(ii)$  two leptons, two jets and \met, or 
$(iii)$ four jets and \met.
Some of the leptons can be neutrinos.
For $\winoop\winoom$~production, the final states are
$(i)$ two acollinear charged leptons and \met,
$(ii)$ one charged lepton, 2 jets, and \met, or 
$(iii)$ four jets and \met.  
A wide variety of signatures
is possible from the production of other chargino and neutralino
combinations.  

\subsubsection{Trileptons}
\label{sssec:trileptons}
The production of $\winog_1\zinoh$, followed by the decays $\winog_1\to\zinol
\ell\nu$ and $\zinoh\to\ell^+\ell^-\zinol$, is a source of three charged leptons
($e$ or $\mu$) and \met, called trilepton events (the \met~is silent).   The
trilepton signal has small SM backgrounds, and is consequently one of the 
``golden'' SUSY signatures.\cite{ref2,hbaerslepton,trilepton_golden,multilepton}
 
The overall efficiency for $\winog_1\zinoh$ production with decays into three detected
leptons is set mainly by the branching ratio for the trilepton final state,
which is highly model--dependent. The efficiency depends on mass
splittings between the $\winol$ and $\zinoh$ and the $\zinol$. 
For example, if the
2--body decay chain 
$\zinoh\to\slepton\ell$ and $\slepton\to\ell\zinol$ occurs, and 
the mass splitting between $\zinoh$ and $\slepton$ or between $\slepton$ and
$\zinol$ is small, one of the leptons can be too soft to detect.
In addition, if the mass splitting between $\zinol$ and $\zinoh$ is large,
decays to real $Z$ or $h$ bosons are possible.  Real $Z$ bosons
have a small branching ratio to $e^+e^-$ and $\mu^+\mu^-$ and are
a SM background, and $h$
will decay mainly to $b\bar b$, decreasing the trilepton rate.  A similar
discussion holds for the decays of the $\winol$, especially in SUGRA since
$M_{\winol}\simeq M_{\zinoh}$.
For example, in SUGRA models, the branching ratios can be
$ BR(\winol\to\ell\nu\zinol)=0.22, BR(\zinoh\to\ell^+\ell^-\zinol)=0.32$
when the sleptons are off--shell but lighter than the squarks, or 
$BR(\winol\to\ell\nu\zinol)=0.66, BR(\zinoh\to\ell^+\ell^-\zinol)\simeq 0$
when the sneutrinos are light enough to allow 
$\zinoh\to\tilde{\nu}\nu$ on--shell.
The branching fractions also depend on \tanb, and, for large \tanb,
decays to $b$'s and $\tau$'s are enhanced.
The decays of $\zinoh$~to $e^+e^-$ and $\mu^+\mu^-$ are
strongly suppressed for large \tanb, falling a factor of about 5
between \tanb=2 and \tanb=20 if the
squarks are much heavier than the sleptons.\cite{baertanb}

\begin{figure}[!ht]
\centerline{
   \hbox{
      \hbox{\psfig{figure=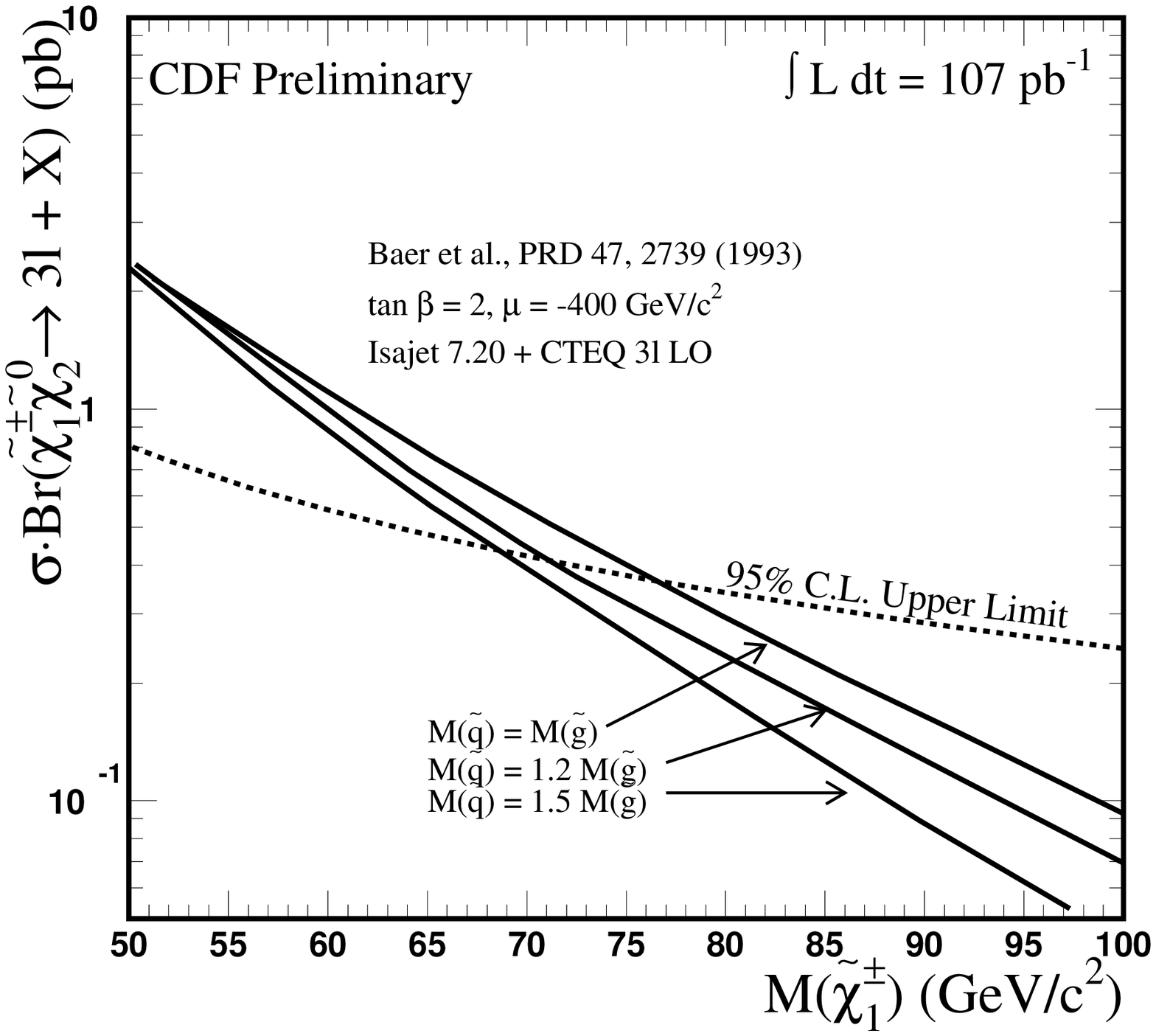,height=75mm}}
      \hbox{\psfig{figure=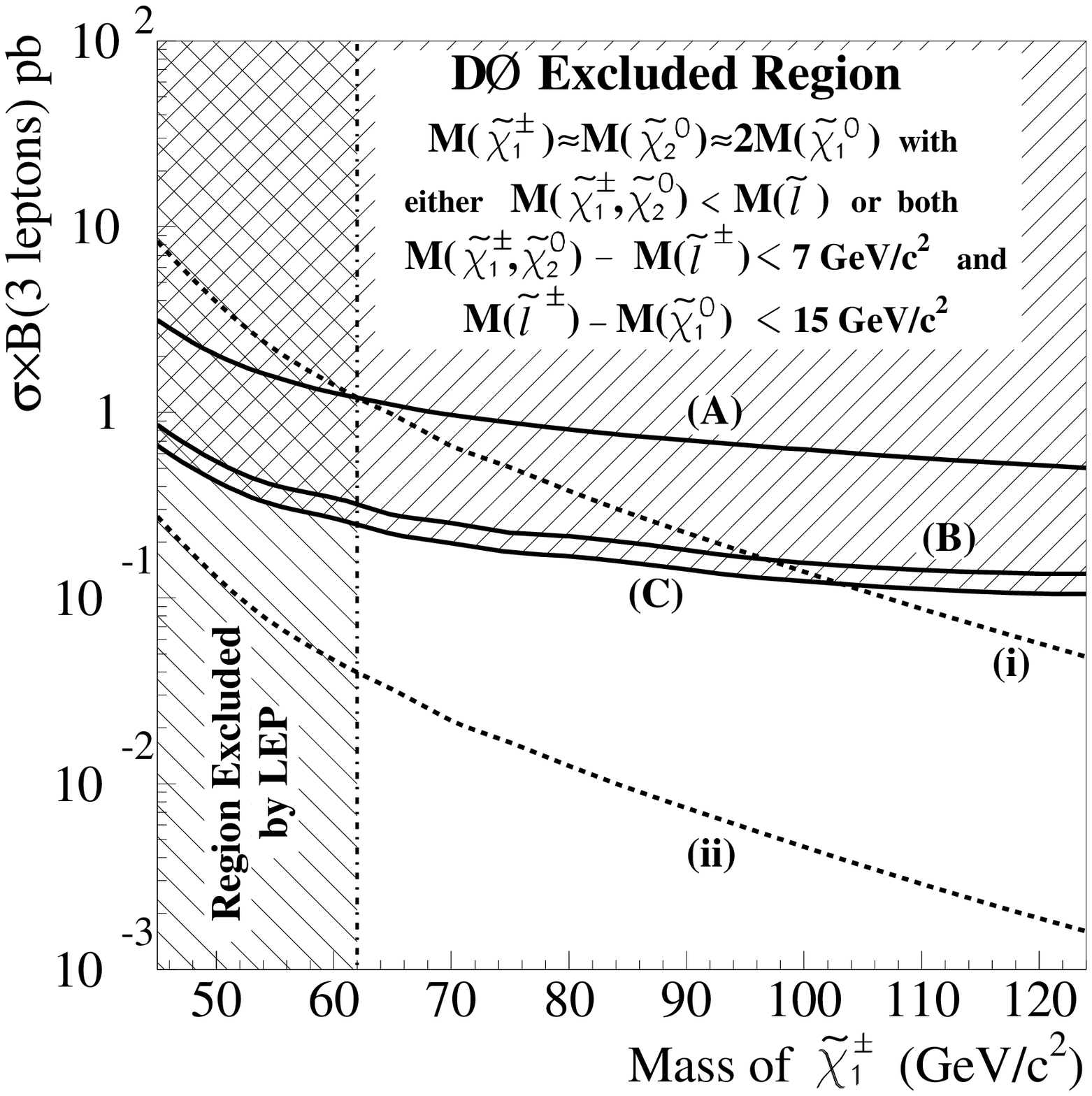,height=70mm}}
   }
}
\caption{
(Left) The CDF 95\% C.L. limits on cross section 
$\times$ branching ratio for $\winol \zinoh$ production
in 107 \ipb of data.  The limit is on the {\it sum} of the final states
$eee, ee\mu, \mu\mu\mu$ and $\mu\mu e$ when
$\winol\rightarrow \ell\nu\zinol$ and $\zinoh\rightarrow\ell\ell\zinol$.
The signals expected for three different RIPS scenarios are shown for
comparison.\protect\cite{hbaerslepton}
Typically, $M_{\zinoh} \simeq M_{\winol} \simeq 2 M_{\zinol}$.
(Right) Similar limits from \D0, but for the {\it average} of
all four channels.
The curves (A), (B), and (C) show the Run Ia, Run Ib, and combined limits.  
Curve (i) shows the predicted cross section $\times$ branching ratio 
assuming BR($\winol\to\ell\nu\zinol$)=
BR($\zinoh\to\ell^+\ell^-\zinol$)=1/3 ($\ell=e,\mu,\tau$).
Curve (ii) assumes
BR($\winol\to\ell\nu\zinol$)=$0.1$ and 
BR($\zinoh\to\ell^+\ell^-\zinol$)=$0.033$.
For both CDF and \D0,
kinematic efficiencies are calculated using the production cross 
section from {\tt ISAJET}.
}
\label{fig:cdfwino}
\end{figure}

The results of the CDF~\cite{CDF_trilepton_1a,CDF_trilepton_1ab} and
\D0~searches~\cite{d0_wino1a,d0_wino1b}  are shown in Fig.~\ref{fig:cdfwino}
analyzed using RIPS (see Sec.~\ref{sugra_inspired}). The searches include four
channels: $e^+e^-e^\pm$, $e^+e^-\mu^\pm$,  $e^\pm\mu^+\mu^-$ and
$\mu^\pm\mu^+\mu^-$.   The CDF analysis is based on the cuts listed in
Table~\ref{tab:cdftrilepton}, and requires one lepton with $E_T>$11~GeV, passing
tight identification cuts,  and two other leptons with $E_T>$5~GeV (electrons)
or  $p_T>$4~GeV (muons), passing loose identification cuts.  All leptons must be
isolated, meaning there is little excess $E_T$ in a cone of size $R=0.4$ in
$\eta-\phi$ space centered on the lepton. 
The event must have two leptons with  the same
flavor and opposite sign.  If two leptons of the same flavor and opposite charge
have a mass consistent with the $J/\Psi$, $\Upsilon$ or $Z$ boson, the event is
rejected.  After this selection, six events remain in the data set, while the
expected background, dominated by Drell--Yan pair 
production plus a fake lepton, is
8 events. After demanding  \met$>15$~GeV, no events remain, while 1.2 are 
expected from SM model sources.

\begin{table}[!ht]
\centering
\caption{Selection criteria and results
for the CDF trilepton gaugino search. The 'Very
Loose' muon category refers to isolated stiff tracks that leave only small
amounts of energy in the calorimeters, but do not have a corresponding
track in a muon chamber (this substantially increases the acceptance).}
\begin{tabular}{|l|l|}
\hline \hline
Quantity                        &  Criteria/Cut Value \\ \hline \hline
Lepton ID Categories            & 1 Tight + 2 Loose, $|\Sigma Q| \ne
3$\\ \hline
Lepton Isolation                & $\Sigma E_T < 2$ GeV in a cone
$R=0.4$ \\ \hline
Tight $E_T^e$, $\eta$ range     & $>11$ \gev, $|\eta|< 1.0$ \\ \hline
Loose $E_T^e$, $\eta$ range      & $>5$ \gev, $|\eta|< 2.4$ \\ \hline
Tight $p_T^\mu$, $\eta$ range    & $>11$ \gev, $|\eta|< 0.6$ \\ \hline
Loose $p_T^\mu$, $\eta$ range    & $>4$ \gev, $|\eta|< 1.0$ \\ \hline
Very Loose Muon $p_T^\mu$, $\eta$ range & $>10$ \gev, $|\eta|< 1.0$ \\
\hline
$Z$,$\Upsilon$,J/$\psi$ mass window cuts & $75-105, 9-11, 2.9-3.1$ 
\gev~ \\ \hline
$\Delta\phi$ between highest 2 $E_T$ leptons  & $< 170^\circ $ \\ \hline
$\Delta R$ between any 2 leptons                & $> 0.4 $ \\ \hline
\met & $>15$ \gev \\ 
\hline
\hline
\lumint                         & 107 \ipb \\
\hline
Expected Background  & 1.2 events \\
\hline
Observed Events              & 0 events \\
\hline\hline
\end{tabular}
\label{tab:cdftrilepton}
\end{table}

\begin{table}[!ht]
\centering
\caption{Selection Criteria and Results
for the \D0~Run Ib Trilepton Search.}
\hspace*{-1.7cm}
\begin{tabular}{|l|l|l|l|l|} \hline\hline
\multicolumn{1}{|c|}{} & \multicolumn{4}{c|}{ Channel and Trigger} \\ \cline{2-5}
  & $eee$\hfill Trigger  & $ee\mu$\hfill Trigger & 
    $e\mu\mu$\hfill Trigger & $\mu\mu\mu$\hfill Trigger \\ \hline
Energy ordered      &  $>22,5,5$\hfill$e$\met       
                    &  $>22($e$),5,5$~~\hfill$e$\met     
                    &  $>9($e$),10,5$\hfill$e\mu$ 
                    &  $>17,5,5$\hfill$\mu$       \\
$E_T$ (GeV)         & $>14,9,5$\hfill$2 e$\met
                    & $>14,9,5$($\mu$)~~\hfill$2 e$\met   
                    & $>17(\mu),5,5$\hfill$\mu$ 
                    & $>5,5,5$\hfill$\mu\mu$      \\
                    &                 
                    & $>9(e),10(\mu),5$~~\hfill$e\mu$
                    & $>5,5,5$\hfill$\mu\mu$
                    &                     \\ \hline
Mass window cut     & 80--100 GeV
                    & None 
                    & 0--5 GeV
                    & 0--5 GeV \\  \hline
\met      & $>15$ GeV& $>10$ GeV& $>10$ GeV& $>10$ GeV\\ \hline
$\Delta\phi$      &$|\pi - \Delta \phi_{e,e}|>$0.2 & None & $|\pi - \Delta \phi_{\mu , \mu}|>$0.1&  $|\pi - \Delta \phi_{\mu , \mu}|>$0.1 \\
cuts       &2 leading $e$'s & & & all combinations \\ \hline\hline
\lumint           & 94.9 \ipb& 94.9 \ipb& 89.5 \ipb& 75.3 \ipb \\ \hline
Background & 0.34 $\pm$0.07& 0.61$\pm$0.36& 0.11 $\pm$ 0.04& 0.20 $\pm$ 0.04 \\ \hline 
Observed & 0 & 0 & 0 & 0 \\ \hline \hline
\end{tabular}
\label{tab:tri_summary}
\end{table}

The \D0~analysis requires leptons with $E_T>5$~GeV satisfying the selection
criteria of Table~\ref{tab:tri_summary}.   However, several different triggers
are used, and  some lepton categories are required to have a larger $E_T$ to
pass the various trigger thresholds.   All leptons are required to be isolated. 
To reduce events with mismeasured \met, the \met~must not be along or opposite a
muon. Additional cuts are tuned for each topology.  For example,  the 
background from Drell--Yan pair production plus a fake lepton 
is highest in the $eee$ channel, so these
events are rejected if an electron pair is back--to--back. The \met~cut is
15~GeV for $eee$, and 10~GeV for the other three topologies. No events are
observed in any channel with a total of 1.26 events expected  from $(i)$
Drell-Yan production plus a fake lepton and $(ii)$ heavy--flavor
production.

To compare the \D0~and CDF 95\% C.L. results, note that the two experiments
present different quantities: the \D0~limit is on the ``average'' 
of the 4 modes
($eee,ee\mu,e\mu\mu$,and $\mu\mu\mu$), 
while the CDF limit is on the sum. After
accounting for this difference, the CDF limit is twice as
sensitive at a given $\winol$~mass.
The CDF limit shown is compared to three RIPS, which have different 
ratios of $m_{\squark}$ to $M_{\gluino}$.  The \D0~limit is compared to
a wide variation of possible branching ratios.  Curve $(i)$ assumes
BR($\winol\to\ell\nu\zinol$) = BR($\zinoh\to\ell\ell\zinol$) = 1/3 
(no hadronic decays), while curve $(ii)$ assumes
BR($\winol\to\ell\nu\zinol$) $=0.1$ and BR($\zinoh\to\ell\ell\zinol$) $= 0.033$ 
(gauge boson--like decays).
The \D0 theory curve assumes heavy squarks, suppressing the 
squark exchange diagram but the CDF curves do not.
The wide differences in the theory curves
in Fig.~\ref{fig:cdfwino} show the dangers of quoting a mass limit rather than
a cross section $\times$ branching ratio limit. 

\begin{figure}[!ht]
\center
\centerline{\psfig{figure=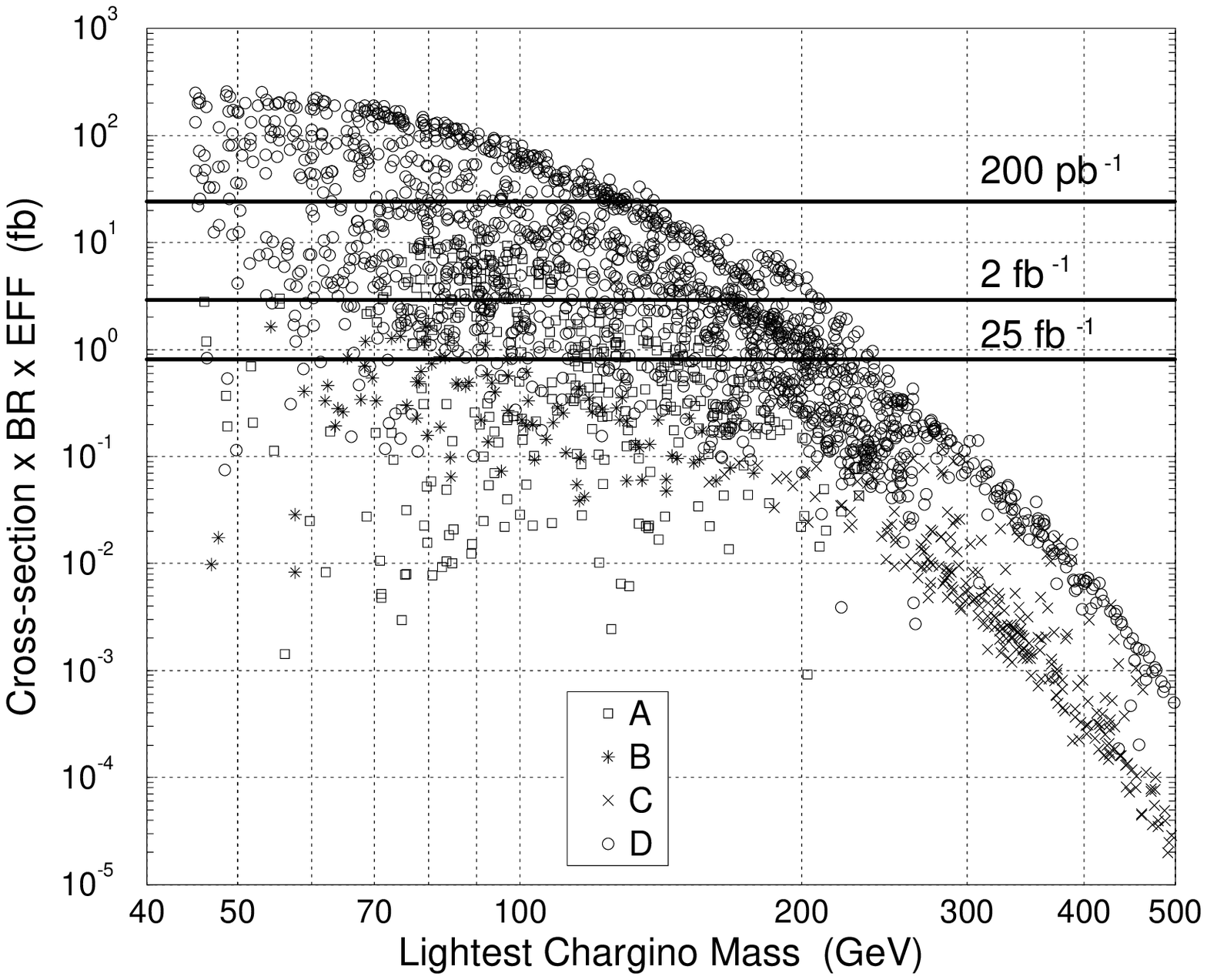,height=80mm}}
\caption{
The overall trilepton signal rate~\protect\cite{cmssm}
for an ensemble of
SUGRA models as a function of the lightest chargino
mass $M_{\winol}$.  The kinematics cuts are different
than those used in the present experimental analyses.
The different symbols refer to solutions showing
interesting behavior where $\zinoh$
has
(A) a neutral ``invisible'' branching ratio
    (generally $\zinoh\to\sneutrino\bar\nu$) $>90\%$,
(B) a large destructive interference in 3--body leptonic decays,
(C) a branching ratio to Higgs $> 50\%$ dominates, or
(D) all other solutions.
The horizontal lines represent the reach for various integrated
luminosities.
}
\label{fig:trilep_eff}
\end{figure}

The experimental limit on the cross section (times branching ratio) depends
on the kinematics of the decays, mostly through 
the mass splitting between $\winol,\zinoh$
and the LSP.   
For $\mu>0$, the mass splitting $M_{\winol}-M_{\zinol}$
is smaller, and the lepton \pt~cuts are less efficient.   As
long as the sleptons are heavier than $\winol$ and $\zinoh$,  
the leptonic decays of
$\winol$ and $\zinoh$
through virtual $W$ and $Z$ bosons and sleptons have similar
kinematics since the decays are dominated by phase space.
However, when the experimental result is presented as a limit
on the  mass of the lightest chargino rather than as a cross 
section limit, the 
result is highly model--dependent. The  theoretical cross section and branching
ratios are strongly affected by the SUSY parameters. If sleptons and
squarks are both heavy, the decays of $\winol$ and $\zinoh$ 
to leptons follow the pattern of the SM gauge
particles. If the sleptons are light and the squarks  are heavy, decays to
trileptons are enhanced.

Figure~\ref{fig:trilep_eff} shows the wide variation in total
efficiency$\times$cross section$\times$branching ratio
 at a given chargino mass
for the trilepton signature by sampling a large ensemble of 
SUGRA models.\cite{cmssm}  
The horizontal lines represent the
reach for various integrated luminosities, showing that the 
Tevatron reach can be quite good for sufficiently high luminosities.
However, even in the restrictive SUGRA framework, no absolute lower
limit on the chargino mass is possible.  

\setcounter{footnote}{0}
\subsection{Squarks and Gluinos}
\label{sec:squark_gluinos}
\begin{figure}[ht]
\centerline{
\hbox{
\hbox{\psfig{figure=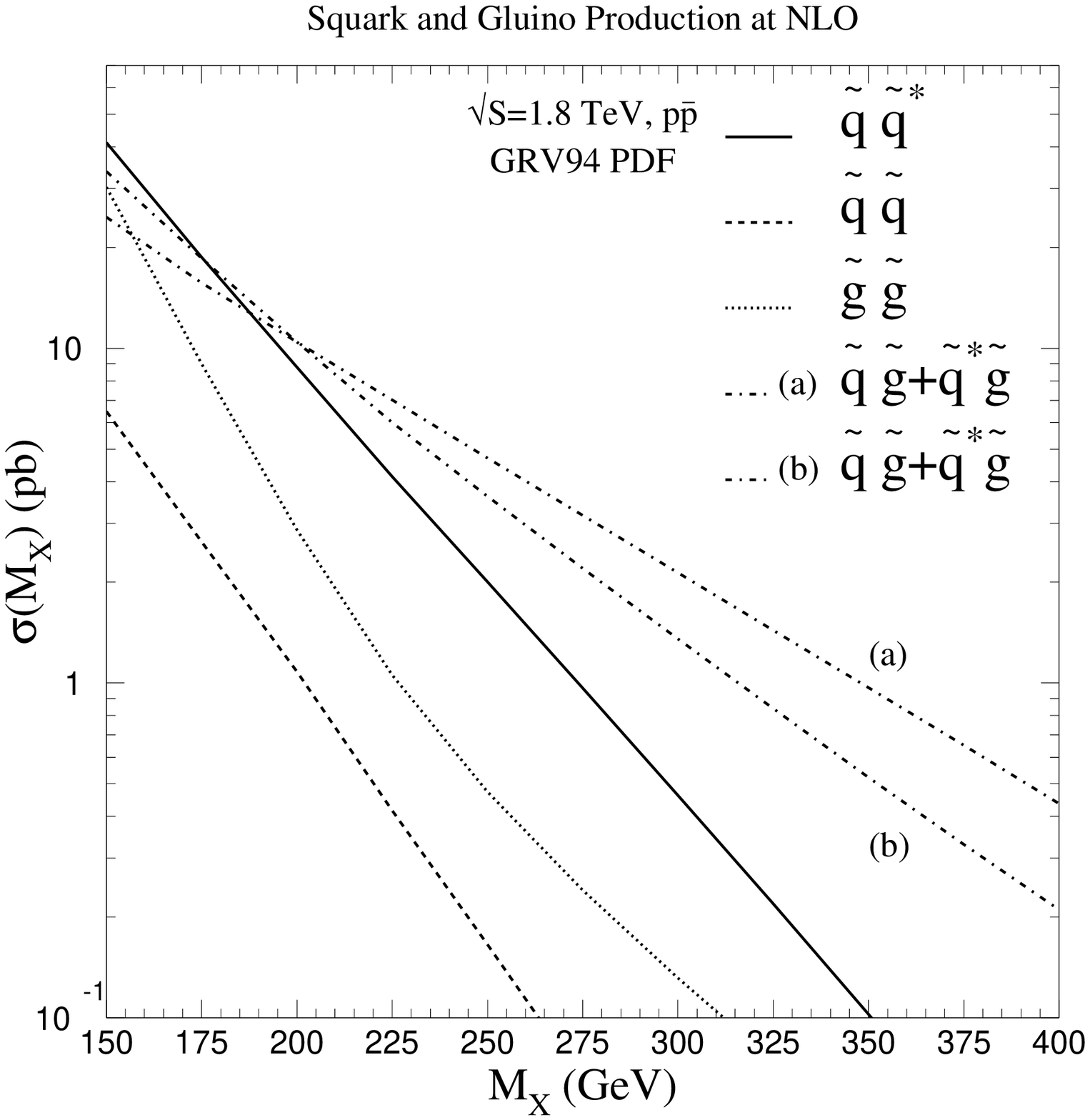,height=60mm}}
\hbox{\psfig{figure=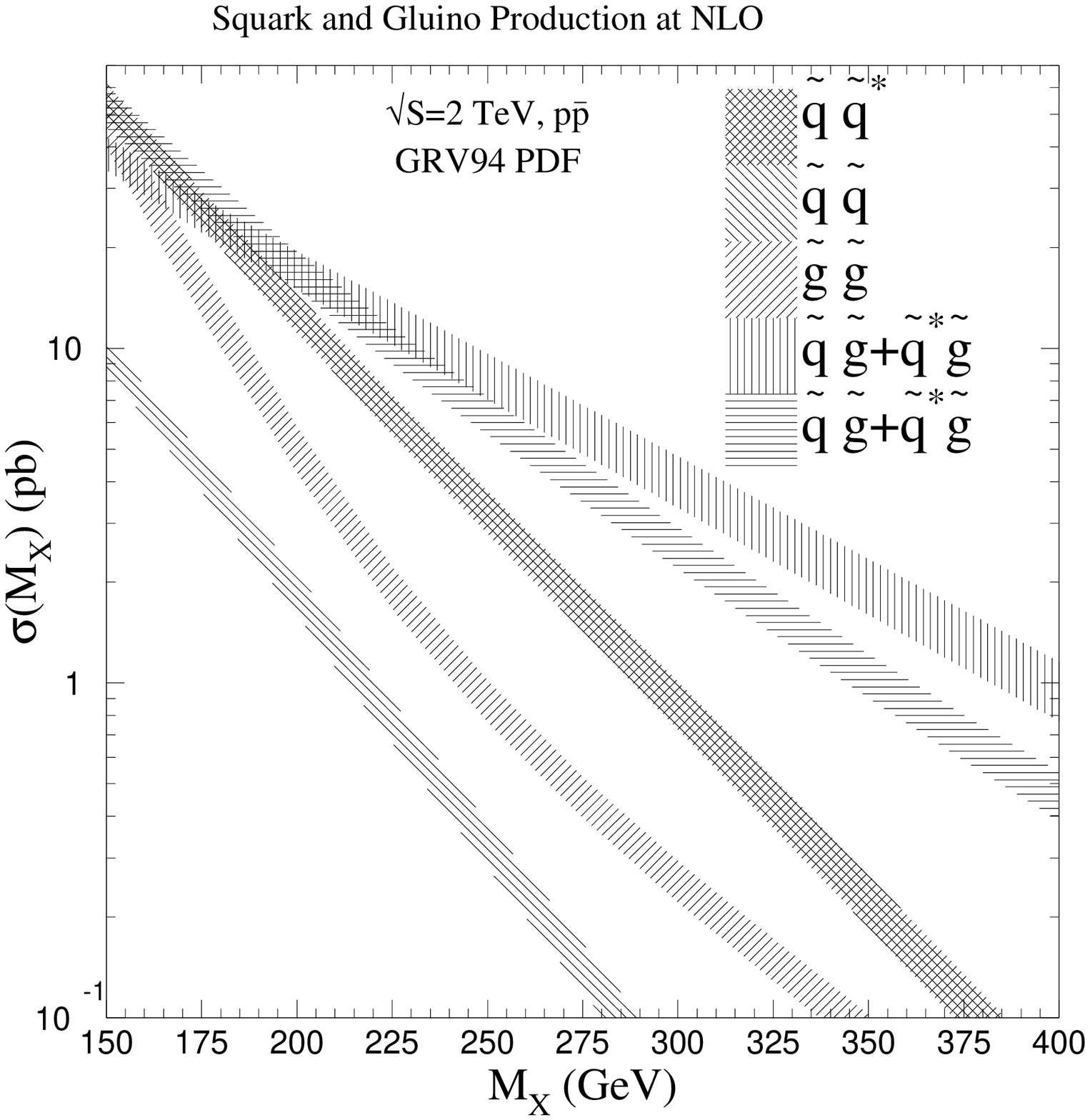,height=60mm}}}
}
\caption[0]
{(Left) Production cross sections for gluinos and squarks versus
sparticle mass $M_X$ at the Tevatron, $\sqrt{s}=1.8$ TeV, assuming
degenerate masses for 5 flavors of squarks.  For $\squark\squark^*$
and $\squark\squark$ production, $M_X$ is the squark mass and 
$M_{\gluino}=200$ GeV.  For $\gluino\gluino$ production, $M_X$ is
the gluino mass and $M_{\squark}=200$ GeV.  For $\squark\gluino$
production $(a)$, $M_X$ is the squark mass and $M_{\gluino}=200$ GeV;
for $(b)$, $M_X$ is the gluino mass and $M_{\squark}=200$ GeV.  The
scale is the average mass of the two produced sparticles.
(Right) The same curves with $\sqrt{s}=2$ TeV.
The bands show the change in rate from varying the scale
from $1/2$ to $2$ times the average mass of the produced particles.}
\label{fig:susyxseccol}
\end{figure}

Since the Tevatron is a hadron collider 
it can produce gluinos and squarks through their
$SU(3)_C$ couplings to quarks and gluons.  
The  dominant production mechanisms are 
$gg,q\bar q\to \gluino\gluino$ or $\squark\squark^*$,
$qq\to \squark\squark$ and $qg\to\squark\gluino,\bar q g\to 
\squark^*\gluino$. 
Because QCD is unbroken,\footnote{The strong couplings of gluinos and
squarks are the same as those of gluons and quarks, so that the production cross
sections are the usual strong interaction cross sections.} the production cross
sections of   gluinos and squarks can be calculated
as a function of only the squark and gluino masses
(ignoring EW radiative corrections).
Figure \ref{fig:susyxseccol} shows the production cross sections for
squarks and gluinos as a function of the sparticle masses 
at $\sqrt{s}=1.8$ TeV (left) and 2 TeV (right),
where NLO Supersymmetric QCD corrections  have been included.\,\cite{squarknlo}
The total cross sections can be of the order of a few picobarns for 
squark and gluino masses up to 400 GeV.
The NLO corrections are 
in general significant
and positive (evaluated at the scale\,\footnote{
In the perturbative calculation of scattering probabilities in field
theory, two scales appear:  a factorization scale, where the parton 
distribution functions are evaluated, and a renormalization scale, where the
running coupling is evaluated.  In practice, these scales are chosen
to be the same.  This scale should be representative of the typical
momentum flowing through a Feynman diagram.  In Drell--Yan production,
for example, the scale is the invariant mass of the Drell--Yan pair.
The higher the order of a perturbative calculation, the lesser the
dependence on this scale.
}  $Q = \bar m$, the average mass of the two 
produced particles), and much less sensitive to the choice of scale than
a LO calculation.\footnote{
The inclusion of NLO effects in the cross sections will typically raise the 
lower bounds  for squark and gluino masses by 10 to 30 GeV with respect to 
the LO cross sections evaluated at a scale equal to the
invariant mass of the produced particles.}
In Fig.~\ref{fig:susyxseccol}, five degenerate squark flavors are 
assumed,\,\footnote{Assuming 
the same mass for the bottom squark is a simplification that
becomes questionable in the large \tanb~region.}
which is one way to suppress FCNC's.

The ultimate
detectability of squarks and gluinos depends upon their decays,
which, in turn, depends on the electroweak
couplings of the squarks and the mixings in the   neutralino and
chargino sector.  
Since squarks and gluinos decay into charginos and neutralinos,
their signatures can be similar to $\ino\ino$ production,
but with accompanying jets.
If $m_{\squark} > M_{\gluino}$, then the squark has
the 2--body decay $\squark \rightarrow \gluino q$.
The gluino  has then the possible decays 
$\gluino\to q \bar{q}\zinog_i$ or
$\gluino\to q \bar{q}'\winog_i$,
 where $q$ can stand for $t$ or $b$ as well, or even
$\gluino\to t\stop^*$ or $\bar t \stop$ if kinematically
allowed.
The gluino can also decay via one--loop diagrams as 
 $\gluino\to g \zinog_i$.
If, instead, $m_{\squark} <  M_{\gluino}$, then the gluino
has the 2--body decay
$\gluino \to \squark q$.
The  squarks can then decay as
$\squark_{L,R} \rightarrow q \zinog_i$,      
$\widetilde{u}_{L } \rightarrow d \ino^+_i$,  and       
$\widetilde{d}_{L } \rightarrow u \ino^-_i$.
The final event signatures 
depend on the decay channels of the charginos and neutralinos, but,
typically involve \met~and 
a multiplicity of jets and/or leptons.  The 3--body
or top--stop decay modes of the gluino can produce a higher multiplicity of SM
particles in their decays than the squark 2--body decays to neutralinos
or charginos, particularly if $\squark\to q\zinol$ is the
dominant squark decay.
On the other hand, the LSP in gluino decays must share energy with 
more particles,
producing less \met~on the average (see Fig.~\ref{fig:gluinomet}).

Gluinos and squarks may also be produced at the Tevatron in association 
with charginos or neutralinos (analogous to $W$ and $Z$ + jet production).  
These processes can be more important than pair
production of gluinos and squarks if the latter are kinematically
limited. 
Event signatures are similar to $\squark$ and $\gluino$ production,
but possibly with fewer jets, 
though the events may still pass the selection criteria
for the squark and gluino searches.
For example, $\squark\winol$ production will have one less jet than
$\squark\squark$ production, assuming the decay $\squark\to q'\winol$
or $\squark\to q\zinol$

Promising signatures for squark and gluino production are $(i)$
multiple jets and \met\,\cite{ref1} and $(ii)$ isolated leptons and jets and 
\met.\,\cite{ref3}

\subsubsection{Jets + \met}

Both CDF and \D0 have performed searches for events with jets and \met.
This signature has significant physics
 and instrumental backgrounds.
The three  dominant physics backgrounds are 
$(i)$ $Z \to \nu\bar\nu$ plus jets,
$(ii)$ $W\to \tau \nu$ plus jets, where the $\tau$ decays hadronically, 
and $(iii)$ \ttbar$\to\tau$ plus jets, where the $\tau$ decays hadronically.  
The \met~in leptonic $W$ decays
peaks at $M_W/2\simeq 40$~GeV, 
with a long tail at high \met~due to off--shell or
high--\pt~$W$'s and energy mismeasurements, so a large \met~cut is needed 
to remove these events.
Instrumental backgrounds come from mismeasured vector boson, top, 
and QCD multijet events.  
Backgrounds from
vector boson production occur for  $W \to e\nu,\mu \nu$ plus jets  events
when the lepton is lost in a crack or is 
misidentified as a jet.  The same problem can occur when the $W$ 
is produced in a \ttbar~event.
QCD multijet production is a
background when jet energy mismeasurements cause false \met. 

The \D0~Run Ia analysis\,\cite{D0_metjets_1a} 
searches for events with 3 or more jets and \met~and 
 with 4 or more jets and \met.  The analysis is
described in Table \ref{tab:squarkcut3}. The resulting 
mass limits on squarks and gluinos
are shown in Fig.~\ref{fig:cdfd0squark} (right, the plot
containing the CDF results also shows the \D0 Ia results) and
were set using a RIPS model with the following parameters: 
$M_{H^{\pm}}$= 500 GeV,
\tanb = 2, $\mu=-250$ \gev, and  $M_{\slepton}=m_{\squark}$.
The efficiency and theoretical cross sections were calculated
using {\tt ISAJET}\,\cite{isajet} assuming 5 flavors
of mass degenerate squarks without top squark production
and a detector simulation.

\D0~also has a 3--jet analysis \cite{D0_metjets_1b} based on 79.2 \ipb~of 
Run Ib data. The basic requirements are three jets with $E_T>25$~GeV and a central
leading jet ($|\eta|<1.1$). The \met~may be significantly overestimated  if the
wrong interaction vertex is used;\,\footnote{The calculation of \met~uses 
the event vertex to calculate $E_T$ for all objects.} to reduce this 
effect, the
tracks  in the leading jet are required to point back to the primary vertex. 
The \met~is required to be uncorrelated in $\phi$ with  any jet.   A cut on 
the scalar sum of the $E_T$ of the  non--leading jets, 
called $H_T$, effectively reduces
events from vector boson backgrounds. The leading jet is also required to have
$E_T>115$~GeV because the only available unbiased sample to study the QCD
multijet background had this requirement. These cuts are summarized in
Table~\ref{tab:squarkcut3}. Vector boson backgrounds are estimated  using {\tt
VECBOS},\cite{vecbos} while the $t\bar t$ background  uses {\tt
HERWIG}\,\cite{herwig} normalized to the  \D0~measured $t\bar t$ cross section.  
The detector simulation is based on the {\tt GEANT}\,\cite{geant} program.  Two
techniques were used to calculate the QCD multijet background. One compares the
opening angle between the two leading jets and the \met~in the signal sample to
the distribution in a generic multi--jet sample.  The other selects events from
a single jet trigger which pass all the selection criteria except for 
the \met~requirement. The \met~distribution is fit in the low \met~region, and
extrapolated into the signal region. The complete set of background estimates
can be found in  Table~\ref{tab:squarkevts3}.

\begin{figure}[!ht]
\center
\centerline{\psfig{figure=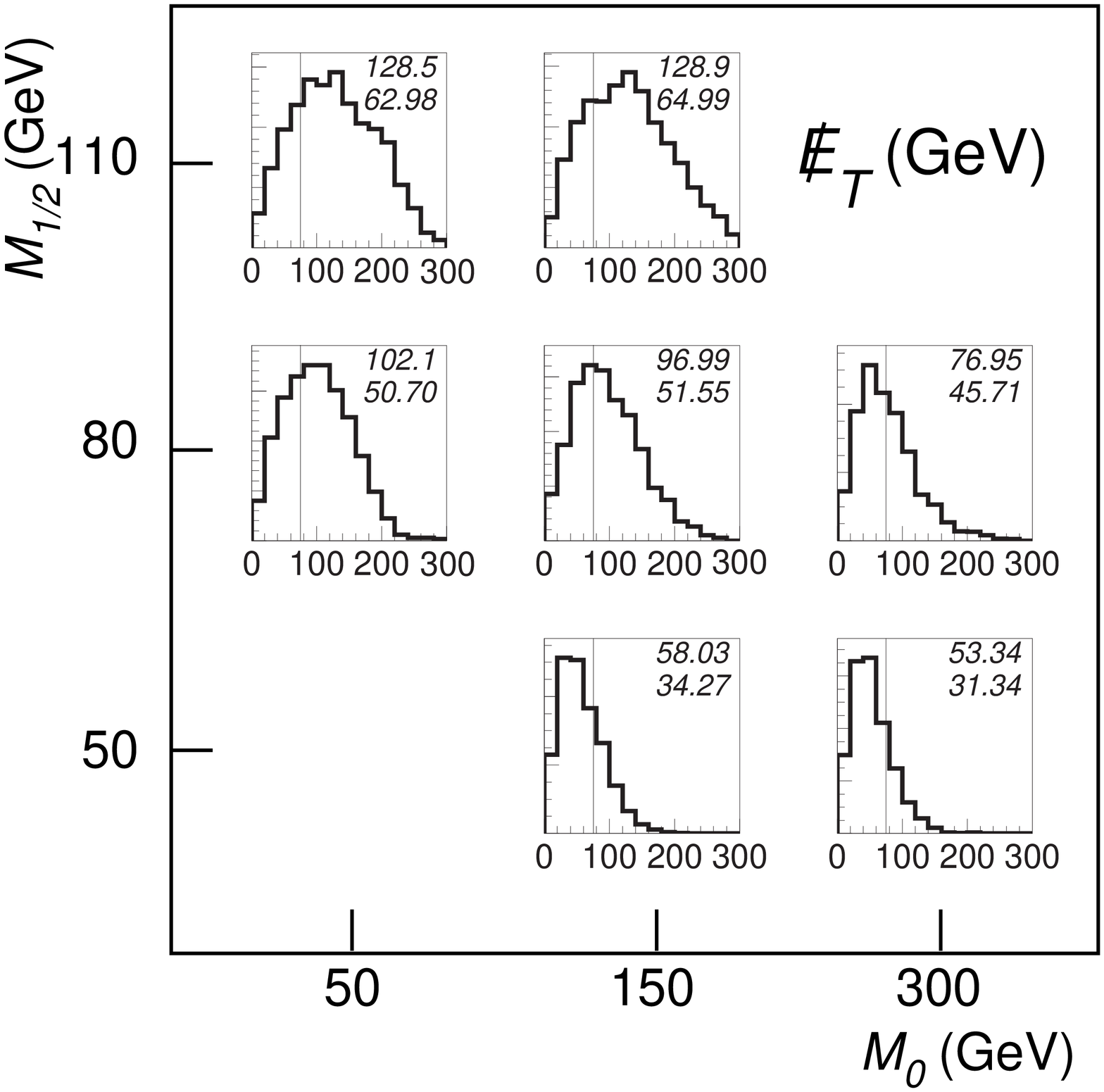,height=60mm}}
\caption{The \met~distribution from simulations of squark/gluino events
in the \D0~detector based on 
{\tt ISAJET}\,\protect\cite{isajet} and {\tt GEANT}.\protect\cite{geant}
The simulation used SUGRA mass relations assuming \tanb=2, \a0=0,
and $\mu<0$ and seven values of \m0~and~\mhalf.  
The numbers in the upper right-hand corner of each plot 
are the mean and RMS of the distribution.  The normalization is arbitrary.}
\label{fig:gluinomet}
\end{figure}

\begin{figure}[!ht]
\center                     
\centerline{
\psfig{figure=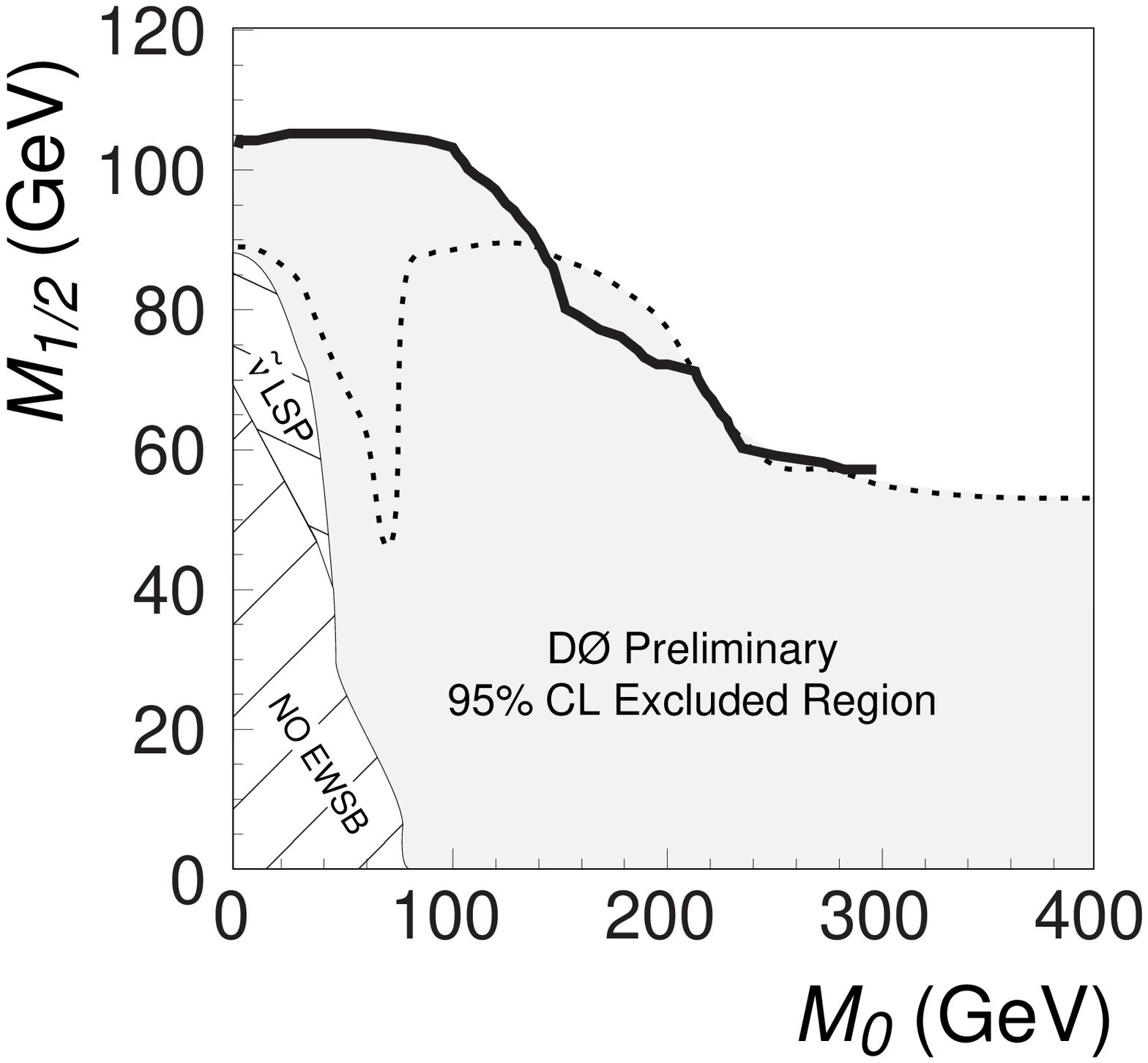,height=60mm}
\psfig{figure=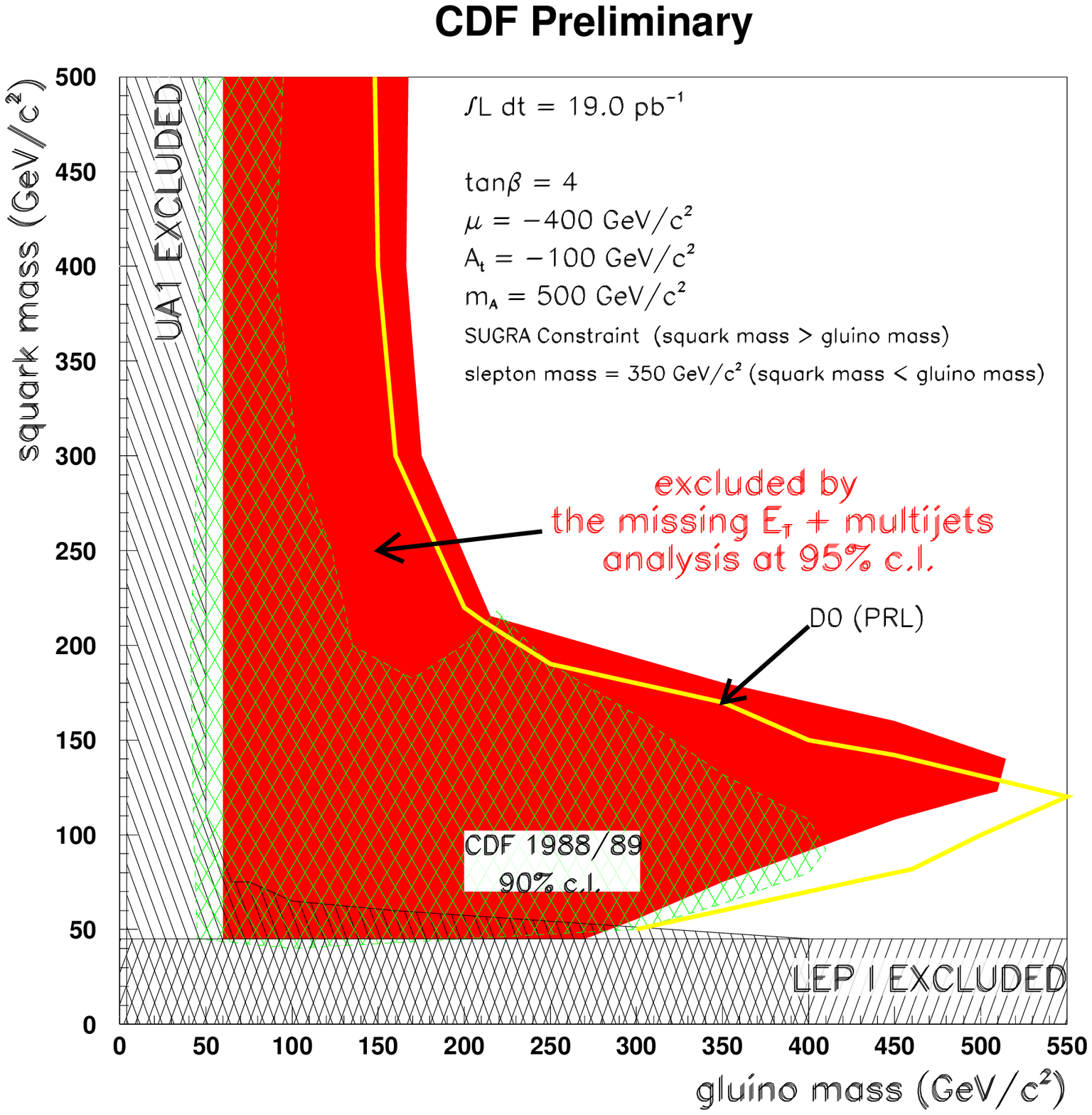,height=60mm}}
\caption{
(Left)  The \D0~Excluded region in the $m_0-m_{1/2}$ plane with fixed
parameters $\tan\beta = 2$, \a0 = 0, and $\mu<0$. The heavy solid line
is the limit contour of the \D0~jets and missing transverse energy analysis. 
The dashed line is
the limit contour of the \D0~dielectron analysis. The lower hashed 
area is a region where mSUGRA does not predict EWSB correctly.
The hashed region above is where the sneutrino is the LSP.
(Right) The CDF mass limits on squarks and gluinos from the search 
in jets and \met\,\protect\cite{CDF_metjets_1a}
using 19 \ipb~of data and the ISAJET 7.06 
Run I Parameter Set (RIPS) with the indicated values (solid area).
For $m_{\squark}<M_{\gluino}$,
the cross section used is leading order, and three or more jets are required.
For $M_{\gluino}<m_{\squark}$, 
the cross section is NLO,\,\protect\cite{squarknlo} 
and four jets are required.
The line labelled 
``\D0 PRL'' is the \D0~result from Run Ia using 13.5 \ipb~of 
data.\,\protect\cite{D0_metjets_1a}
}
\label{fig:cdfd0squark}
\end{figure}

The \D0 data have been analyzed in the context of a minimal SUGRA model.
For fixed \tanb, \a0, and sign of $\mu$, exclusion curves are plotted
in the $m_0-m_{1/2}$ plane, Fig.~\ref{fig:cdfd0squark} (left).
The limits are from the 3--jet, 79.2\ipb, analysis only.
Efficiencies are calculated using
{\tt ISAJET}\,\cite{isajet} for production of gluinos and five flavors of 
squarks without stop squark production. For each point in the  limit
plane, the \met~and $H_T$ cuts are reoptimized based on the predicted background
and SUSY signal. Figure~\ref{fig:gluinomet} shows the \met~as a function of
\m0~and  \mhalf~for  \tanb=2, ~\a0=0, $\mu<0$. When \m0$\gg$\mhalf, the 
\met~signature is degraded, because $m_{\squark}\gg M_{\gluino}$ and thus higher
multiplicity $\gluino\gluino$~events dominate.   Since higher multiplicity also
means higher $H_T$, varying the cuts can maintain sensitivity. These results are
robust within the SUGRA framework.\cite{D0_metjets_1b}

\begin{table}[!ht]
\centering
\caption{Selection criteria for Tevatron squark and gluino searches in
the 3 or 4 jets+\met~channels.  Cuts specific to the \D0 4--jet analysis are
in parentheses.} 
\hspace*{-1.7cm}
\begin{tabular}{|l|l|l|}
\hline \hline
Quantity                  & \multicolumn{2}{c|}{Experiment}        \\ \cline{2-3}
                          &  \D0                     & CDF        \\ \hline\hline
Trigger                   & \met $>$ 40 \gev         & \met $>$35~GeV \\
                          &                          & 1 jet with
  $E_T^j>50$ GeV \\ \hline
\met         &  $>75-100, (65)$ GeV  & $>60$ GeV with 
  $S>$2.2~GeV$^{1/2}$ \\ \hline
$E_T^j$            &  $>25, (20)$ GeV    & $>15$ GeV, $|\eta| <2.4$  \\ \hline
leading $E_T^j$    &  $>115$ GeV $|\eta|<1.1, (N.A.)$  & $>50$ GeV \\ 
                          &  \htran $> 100-160, (N.A.)$     &                          \\ \hline
$\Delta\phi_i$ between    & $5.7^\circ<\Delta\phi_i<174.3^\circ$ & $\Delta\phi_i>30^\circ$ \\ 
\met~and jet $i$; & $\sqrt{(\Delta\phi_1-180^\circ)^2+\Delta\phi_2^2}>28.6^\circ$ &
                            $\Delta\phi_1<160^\circ $ \\ 
$i$=1 is the leading jet  &   &  \\ \hline
Leptons                   &  Veto all     (N.A.)    &  Veto all \\
\hline
Vertices                  &  confirmed, (Only one)    & Any number \\
\hline\hline
\end{tabular}
\label{tab:squarkcut3}
\end{table}

The CDF analysis of the Run Ib data set is not yet complete,
but the Run Ia result based on 19 \ipb~has been published.\cite{CDF_metjets_1a}
The basic requirements are 3 or 4 jets and 60~GeV of \met.  
The full set of cuts is listed in Table~\ref{tab:squarkcut3}.
As in the \D0~search,
the direction of the \met~is not allowed to 
coincide with that of a jet, and 
events with leptons are rejected to 
reduce the background from $W$ and top events.  
The variable $S$, which indicates the significance of the \met,
is used to reduce fake \met~measurements; $S$ is calculated 
by dividing the \met~by the square root of the 
scalar sum of the $E_T$ in the calorimeters.  
The vector boson backgrounds are estimated using
{\tt VECBOS}\,\cite{vecbos} 
normalized to the CDF $Wjj$ data.
Top backgrounds are determined using {\tt ISAJET}\,\cite{isajet}
normalized to 
the CDF measured top cross section.
The QCD background is estimated using an independent data sample
based on a trigger that required one jet with $E_T>50$~GeV.
First all analysis cuts (Table \ref{tab:squarkcut3}) are 
applied to this sample 
except for the $S$ cut, the \met~cut, and the
3 or 4 jets cut.  Next the \met~distribution is fit and the number of events 
expected to pass the \met~cut is derived.  Finally the efficiency of the 
last three cuts is applied to arrive at the final background estimate,
shown in Table~\ref{tab:squarkevts3}.

The limits derived from the CDF analysis are shown in 
Fig.~\ref{fig:cdfd0squark} (right) within the RIPS framework 
(see Sec.~\ref{sugra_inspired}).
In RIPS, a heavy gluino implies a heavy
$\zinol$, so a light squark ($m_{\squark} \approx M_{\zinol}$)
decay will not produce much \met.  The
consequence is an apparent hole in the CDF limit for small $m_{\squark}$ and
large $M_{\gluino}$.  However, lighter gluinos always produce a large
\met~because of the enforced mass splitting between $M_{\gluino}$ 
and $M_{\zinol}$.
The results of this analysis do not change substantially  
as parameters are varied 
within the RIPS framework.\cite{CDF_metjets_1a} 

The results summarized in Fig.~6 are complemented by the dilepton+\met~analysis 
shown in Figs.~7~and~9.  
The limits on the gluino 
and squark masses in
each scenario (minimal SUGRA and RIPS) will be discussed below.

\begin{table}[!ht]                              
\centering
\caption{The number of expected and observed 
events for Tevatron squark and gluino searches in the jets+\met~channel after
performing the cuts in Table \ref{tab:squarkcut3}.}
\begin{tabular}{|l|c|c|c|c|}  \cline{2-5}
\multicolumn{1}{c|}{}   &  \multicolumn{2}{c|}{\D0} &
                 \multicolumn{2}{c|}{CDF} \\  \hline\hline
Analysis          &  3 jets    &  4 jets      &  3 or 4 jets      &  4 jets   \\ \hline
\lumint (\ipb)    & 79.2       &  13.5        &  19          & 19        \\ \hline
$W^{\pm}$         & $1.56\pm.67\pm.42$  & $4.2\pm1.2$  & $13.9\pm2.1\pm6.0$ & $2.6\pm0.9\pm1.7$ \\ 
$Z\to\ell\bar{\ell},\nu\bar{\nu}$ 
                  & $1.11\pm.83\pm.36$       & $1.0\pm0.4$  & $5.0\pm0.9\pm2.7$  & $0.4\pm0.2\pm0.4$ \\ 
\ttbar            & $3.11\pm.17\pm1.35$       & --           & $4.2\pm0.3\pm0.5$ & $2.2\pm0.2\pm0.4$ \\ \hline
QCD multijets     & $3.54\pm2.64$       & $1.6\pm0.9$  & $10.2\pm10.7\pm4.2$ & $3.2\pm3.8\pm1.3$ \\ \hline
Total Background  & $9.3\pm0.8\pm3.3$ & $6.8\pm2.4$ & $33.5\pm11\pm16$ & $8\pm4\pm4$     \\ \hline
Events Observed   & {\bf 15}   & {\bf 5}     &  {\bf 24}   & {\bf 6}   \\ 
\hline \hline
\end{tabular}
\label{tab:squarkevts3}
\end{table}

\subsubsection{Dileptons+\met}
\label{sec:dileptons}

If, in the cascade decay chain of the $\squark$'s and $\gluino$'s, 
two charginos decay $\winol\to\ell\nu\zinol$, or one neutralino
decays $\zinoh\to\ell^+\ell^-\zinol$, the final state can contain
2 leptons, jets, and \met.\,\cite{ref3}  This channel has the advantage of
being relatively clean experimentally. 
The requirement of two leptons significantly reduces jet
backgrounds and removes most of the $W$ backgrounds.
Requiring that the mass of the two leptons be inconsistent with the $Z$ mass
removes most of the rest of the vector boson backgrounds.
If the leptons are required to have
\pt$>$20~GeV, the major background from
physics processes
is \ttbar$\to bW^+\bar bW^- \to b\bar b\ell^+\ell^-$\met.
As the cut on lepton \pt~is lowered,
$Z \to \tau^+ \tau^-$, where the $\tau$'s decay semileptonically,
also becomes an important background.
The instrumental backgrounds are small.   The spectacular signature
of {\it like--sign}, isolated dileptons, which is difficult to
produce in the SM, can occur 
whenever a gluino is produced directly or in a cascade decay, since
the gluino is a Majorana particle.  This property is exploited in
the CDF dilepton searches.

\begin{table}[!ht]
\centering
\caption{Selection criteria for Tevatron searches for squarks or gluinos 
in the dileptons, 2 jets and \met~channel.}
\begin{tabular}{|l|c|c|}
\hline \hline
Quantity        & \multicolumn{2}{c|}{Experiment} \\ \cline{2-3}
                &  \D0          & CDF             \\ \hline \hline
\lumint~(\ipb)  &  92.9         & 81              \\ \hline
 $E_T^{e_1}, E_T^{e_2}$ & $>15, 15$ \gev & $>11, 5$ \gev  \\ \hline
 $p_T^{\mu_1},p_T^{\mu_2}$ & N.A. &  $>11,5$ \gev  \\ \hline
 $E_{T}^{j_1,j_2}$        & $>20$ \gev & $>15$ \gev \\ \hline
 Mass window cut & $M_Z\pm 12$ \gev & N.A.\\ \hline
$\Delta R$ between leptons and jets & N.A. & $>.7$ \\ \hline
\met            & $>25$ \gev  & $>25$ \gev \\ \hline
like sign dileptons &  no  & yes \\ \hline\hline
\end{tabular}
\label{tab:squarkll}
\end{table}

\begin{figure}[!ht]
\center
\centerline{\psfig{figure=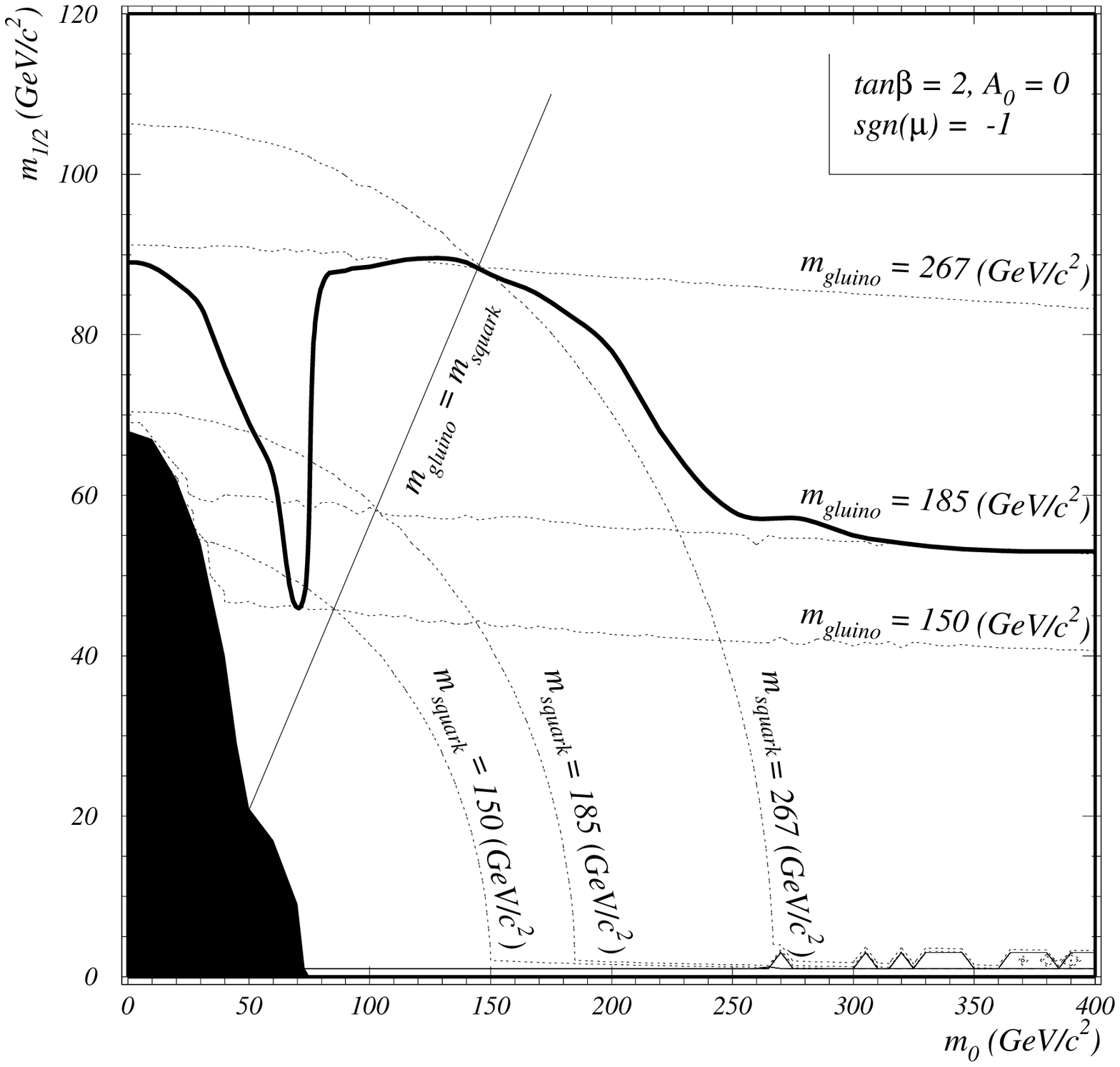,height=60mm}
\psfig{figure=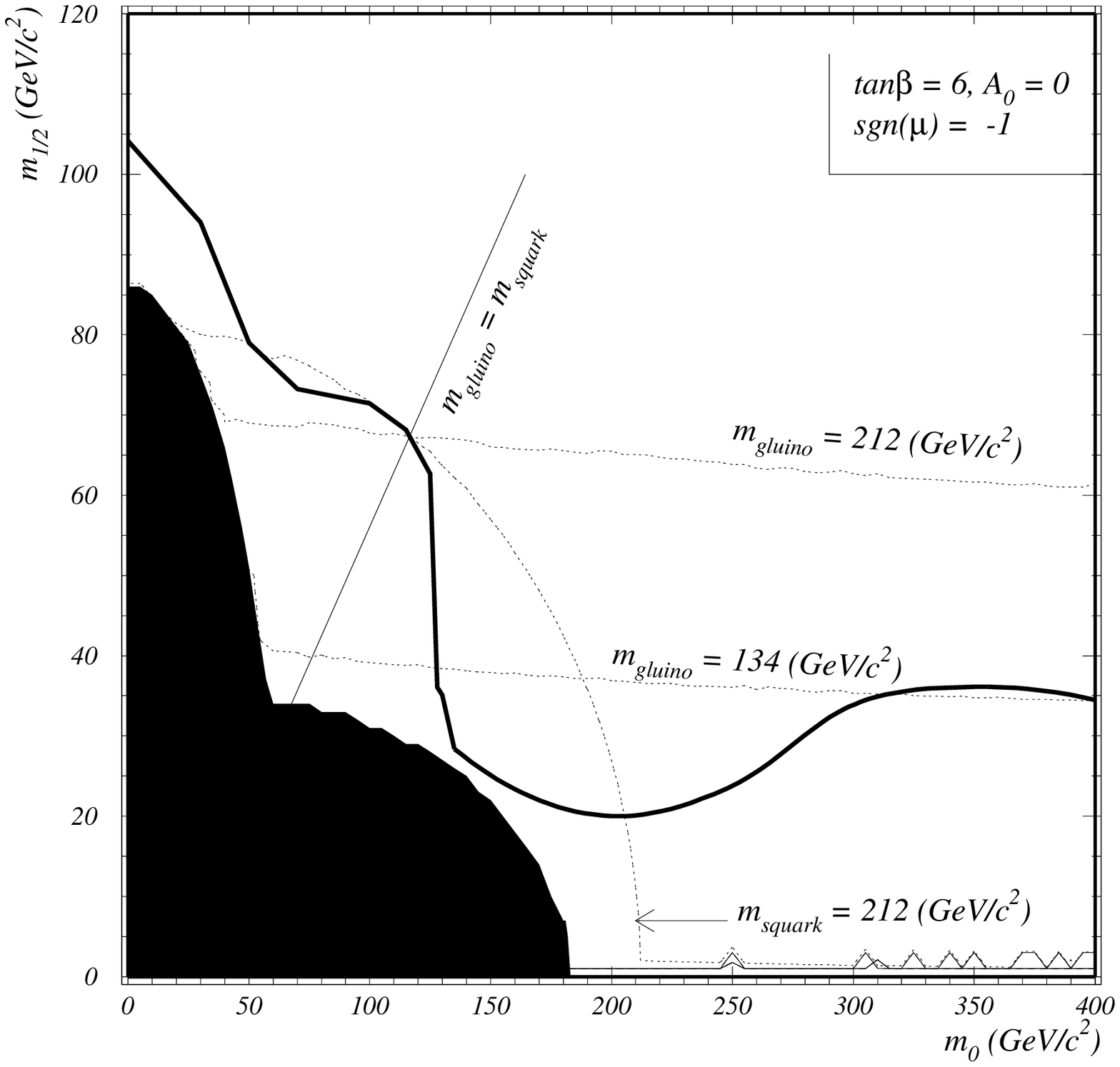,height=60mm}}
\caption{(Left) The \D0~limits on the SUGRA parameters \m0~and \mhalf~from the 
2 leptons, 2 jets, and \met~search\,\protect\cite{D0_metdilepton_1b}
for \tanb=2, \a0=0, and $\mu<0$.  (Right) The same
plot for \tanb=6, \a0=0, and $\mu<0$.  In both plots, the
dark shaded area is the region in which SUGRA does not produce
electroweak symmetry breaking.  Selected contours
of squark and gluino mass are also shown.
}
\label{fig:d0squarkee}
\end{figure}

\begin{figure}[!ht]
\center
\centerline{\psfig{figure=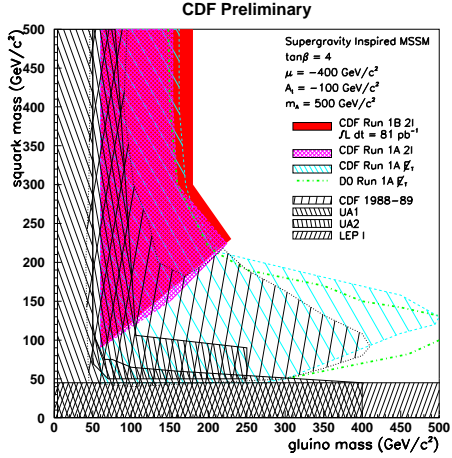,height=80mm}}
\caption{ CDF limits on the squark and gluino masses from the
2 like sign leptons, 2 jets, and \met~search in 81 \ipb.  
The limits were set using the ISAJET 7.06 Run I Parameter Set (RIPS)
with the indicated values.
}
\label{fig:cdfsquarkee}
\end{figure}

Figures~\ref{fig:d0squarkee} and ~\ref{fig:cdfsquarkee} show 
the \D0\,\cite{D0_metdilepton_1b}
and CDF\,\cite{CDF_metdilepton_1b,CDF_metdilepton_1a}
results from Run Ib, once again compared to SUGRA and RIPS, respectively.  
The CDF limit is based on NLO cross sections,\cite{squarknlo} and
the \D0~limit on LO cross sections.
The \D0~limits on \m0~and \mhalf~are calculated including contributions from
the production of  all sparticles (for instance, associated production of
neutralinos or charginos with squarks or gluinos),
while the CDF result only considers $\squark$ and  $\gluino$ production.
Table~\ref{tab:squarkll} gives the
selection criteria for the two analyses.  The experimental cuts are chosen
to identify two 
high--\pt~leptons, which come predominantly from 
$\squark$ and $\gluino$ decays into 
charginos or neutralinos which in turn decay 
into real or virtual $W$ or $Z$ bosons.

\begin{figure}[!ht]
\center                     
\centerline{
\psfig{figure=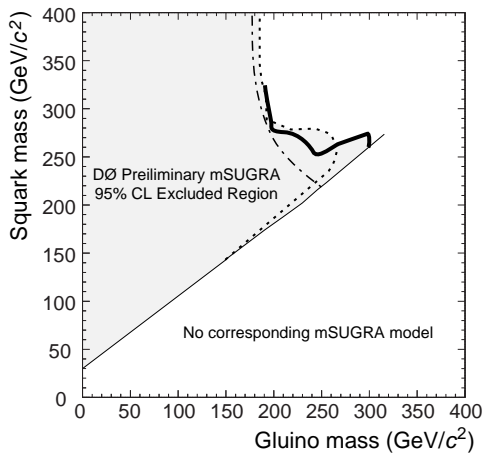,height=60mm}}
\caption{Excluded region from various \D0~analyses
in the $m_{\squark}-M_{\gluino}$ plane with fixed
mSUGRA parameters $\tan\beta=2$, \a0=0, and $\mu<0$. Note that 
there are no mSUGRA
models in the region to the right of the diagonal thin line. 
The heavy solid line is the limit
contour of the \D0~Run Ib 3 jets and missing transverse energy analysis. 
The dashed line is the limit
contour of the \D0~Run Ib dielectron analysis. 
The dot--dashed line is the limit contour of the \D0~Run Ia
3 and 4 jets and missing transverse energy analysis shown 
only in the region with valid mSUGRA models.}
\label{fig:onebdzcombined}
\end{figure}

\D0~has also presented an experimental limit in the $M_{\gluino}-m_{\squark}$ 
plane Fig.~\ref{fig:onebdzcombined}, which allows a comparison
with the CDF limit Fig.~\ref{fig:cdfsquarkee}.
For $m_{\squark}\gg M_{\gluino}$ or, equivalently,  for 
$m_0 \gg m_{1/2}$, $\gluino\gluino$ pair production
is the dominant SUSY process. 
As $m_0 (m_{\squark})$ is varied with the other parameters fixed, the 
branching ratios for the 3--body gluino decays to charginos or neutralinos
and jets become fairly constant, so the production rate of leptonic final
states becomes constant;
the experimental limit approaches a constant value asymptotically, as
can be seen in both the \D0~and CDF plots shown in 
Figs.~\ref{fig:cdfsquarkee} and \ref{fig:onebdzcombined}, respectively. 
Observe that, for large 
enough values of the gluino mass, the leptons easily
pass the experimental cuts, so the experimental efficiency also becomes
constant.

The relation $m_{\squark}\ll M_{\gluino}$ is not possible in SUGRA, and
is treated in an {\it ad hoc} manner in RIPS.  There is no limit in this region
for either opposite-- or like--sign dilepton pairs because the large, fixed slepton masses limit
the branching ratios to leptonic final states.  The possibility of 
like--sign dilepton pairs is further reduced because both
the $\gluino\gluino$ and $\gluino\squark$ cross sections 
(which produce like--sign leptons
because the gluino is a Majorana particle) and the $\squark\squark$
cross section (which produces like--sign leptons because the squarks
have the same charge) are small in this region.  It is very difficult
for $\squark\squark^*$ production to yield like--sign leptons in general.

When $m_{\squark}\simeq M_{\gluino}$, the $\gluino\gluino$ cross section
is supplemented by the $\gluino\squark$ cross section.
Just above the diagonal line at $M_{\gluino}=m_{\squark}$
({\it i.e.} $m_{\squark}$ just larger than $M_{\gluino}$) in 
Figs.~\ref{fig:cdfsquarkee} and~\ref{fig:onebdzcombined},
there are ``noses'' in the limit plots, with the limit becoming stronger close
to the diagonal.

The limits in Figs.~\ref{fig:cdfsquarkee} and~\ref{fig:onebdzcombined}
are for a specific choice of parameters within the RIPS or SUGRA framework.
If $\mu$, $A_t$ and \tanb~are varied, the branching ratios
into charginos or neutralinos can vary strongly.
The sensitive dependence on the parameters can be 
seen within minimal SUGRA models from the \D0~limits in 
Fig.~\ref{fig:d0squarkee}.
The dip in the \tanb=2 limit (left), around \m0$=70$ \gev, 
is a point where $m_{\slepton} > M_{\zinoh} > m_{\sneutrino}$ and
BR$(\zinoh\to\nu\bar\nu\zinol)\simeq 1$,
so the detection efficiency is very sensitive to the choice of high energy
parameters \m0~and \mhalf.
In Fig.~\ref{fig:d0squarkee} (right), with \tanb=6, 
the limits are severely reduced  compared to 
Fig.~\ref{fig:d0squarkee} (left), with
\tanb=2, in the region where the squark mass is large compared
to the gluino mass.  For large \tanb, the mass splitting 
$M_{\winol,\zinoh}-M_{\zinol}$ is reduced, so that the
leptons from the $\winol$ and $\zinoh$ decays are softer.
The non--trivial shape of the limit curves results from an interplay between
the cross section being larger when \m0~and \mhalf~are smaller
(sparticle masses are smaller) and the mass splittings being
smaller.
Consequently, although the dileptons+jets+\met~ signature 
is an excellent discovery 
channel with
little SM background, it is hard to set significant parameter limits
even using SUGRA models.

From the present analyses in the \met+jets 
and dileptons+\met~channels,
some preliminary conclusions can be drawn on the squark and gluino masses.
These depend, however,  on the assumed SUSY parameters.
The \D0~limit on the gluino mass effectively develops a plateau 
for large $m_0$ at 185 GeV for
\tanb$=2$, and at 134 GeV for \tanb$=6$.
The CDF limit on the gluino mass is 180 GeV for \tanb$=4$
for large $m_{\squark}$.
Instead, for equal squark and gluino masses,  
the \D0~mass limit for
\tanb=2 is 267~GeV, using all SUSY production and decay modes in the model.
From the CDF analyses and $m_{\squark}\simeq M_{\gluino}$,
the limit is about 220 GeV for 
$\tan\beta=4$.  
A direct comparison of all the above results is rather difficult since
\D0~and CDF have done analyses assuming different sets of MSSM parameters
(see Figs. 6--9). Moreover,
CDF considers only squark and gluino production, while \D0~ considers
all possible sparticle production, and the associated production of 
neutralinos or charginos with squarks or gluinos can have an impact on the experimental limits.

It would be very useful for purposes of comparing and combining the two
experimental limits to have both collaborations use at least one common 
model (such as SUGRA), and agree on several values of the parameters to do the 
searches.   
For example, the two collaborations could present their limits in the
$m_{1/2}-\tan\beta$ plane (for large \m0).
Secondly, they could move (partially)
towards more experimentally--based quantities by plotting contours of cross
section limit and also contours of acceptance $\times$ efficiency in the
\m0--\mhalf~plane.
This would eliminate the strong model dependence on the branching ratios.
The experimental acceptance for the signature of two
leptons+jets is much less model--dependent, since it simply reflects the
hard kinematics from the decays of two heavy objects.  A presentation
of cross section $\times$ branching ratio limits, in addition to the
mass limits, would be of more general use to model--builders.

\subsection{Top Squarks}

\begin{figure}[!ht]
\center
\centerline{\psfig{figure=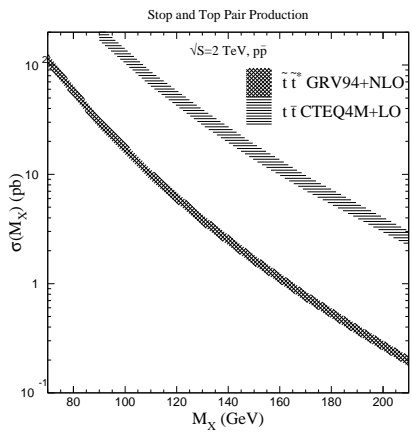,height=60mm}}
\caption{The cross section for pair production of one stop
quark versus the stop mass 
at the Tevatron, calculated to NLO.  The band represents
the variation of the scale from $1/2$ to $2$ times the
stop mass.
The top pair production cross section versus the top mass is
shown for reference, calculated only at LO.
The band represents
the variation of the scale from $1/2$ to $2$ times
the transverse mass $p_T\oplus m_t$,
where $p_T$ is the transverse momentum of the top quark.}
\label{fig:stopseccol}  
\end{figure}

The top squark (stop) is a special case worth a separate 
discussion.\cite{Godbole,Sender}
The mass degeneracy in the stop sector is expected 
to be strongly broken, and, for sufficiently large mixing, the lightest stop
can be expected to be rather light, possibly
lighter than the lightest chargino. 
The lightest stop has about 
a tenth the production cross section of a 
top quark
of the same mass, because the threshold behavior is $\beta^3$
(compared to $\beta$ for fermion pairs) and only half
the scalar partners are being considered.
At leading order, the cross section is independent of the 
gluino mass 
and depends only on the stop mass.\footnote{Since the 
incoming partons are not tops, then if flavor changing neutral
currents are suppressed by a Supersymmetric GIM mechanism,
stop production via gluino interchange is not allowed.
After including the NLO SUSY QCD corrections 
the dependence  on $m_{\tilde{g}}$ and stop mixing
becomes explicit but, in practice,
numerical results are insensitive to the exact gluino mass
and the mixing.}  
Due to the large left--right mixing,
the NLO SUSY QCD
corrections must deal with different left-- and right--handed couplings
of the quarks to squarks and gluinos.
The results for the stop--pair production 
cross section as a function of the stop mass are plotted in 
Fig.~\ref{fig:stopseccol}.\,\cite{spirastopx} 

The stop can be produced directly as $\stop\stop^*$ pairs or,
depending on the stop mass, indirectly\,\cite{Godbole}
in decays of the top $t\to \stop\zinol$, or sparticles, such as 
$\winog_i\to b\stop$.
Also depending on the stop mass, one of three decay modes is expected to
dominate.
If $(a)$ $m_{\tilde{t}_1} > m_{\winol}+ m_b$, then  
$\tilde t_1$ can decay into 
$b\winoop$, followed by the decay of the chargino.
This can look similar to the top decay $t\to bW$, but with different
kinematics and branching ratios for the final state.
Instead, if the stop is the lightest charged SUSY particle,  
it is expected to decay exclusively through a chargino--bottom loop as  
$(b)$ $\tilde{t}_1 \to c\zinol$, which looks quite different from
SM top decays.  Finally, the stop can decay
$(c)$ $\stop\to b W \zinol$ 
or if the stop is quite heavy into $\stop\to t\zinol$ 

The possible signals from $\stop\stop^*$ production, with
decay $(a)$  and
depending on the chargino decay modes, are:
  $(i)$ $b\bar b\ell^+\ell^-$\met,  
  $(ii)$ $b\bar b\ell^\pm jj$\met, or
  $(iii)$ $b\bar b jjjj$\met.
These are similar to $t\bar t$ final states, except $(iii)$ has real \met.
If decay $(b)$ dominates, 
this yields a signature of 2 acollinear charm jets
and \met.
Finally, if decay $(c)$ occurs, the events are similar to $t\bar t$
events, except that the kinematics of the individual $t$ and $\bar t$ are
altered and there can be much more \met.

If the stop is in the range $100-150$ GeV,
$\stop\stop^*$ production may be too small to observe and it might
be easier to observe a light stop in top quark decays $t\to \stop\zinol$.  
If $\zinol$ is Higgsino--like, 
then the BR($t \to \stop\zinol$) can be 50\%.  
If decay $(a)$ 
occurs, then top quark events have the same signatures as in the SM 
but they have more \met~and softer jets and leptons.  
In case $(b)$, fewer leptons and jets are produced and the \met~distribution is
affected.
If there is one SM top decay and one SUSY top decay, the final state can
be $b\ell^\pm c$\met, which would appear at a small rate in the $Wjj$
sample, but not in the SM $t\bar t$ event sample.

An indirect limit can also be set on the decay $(b)$.  Such decays
would not fall in the SM $t\bar t$ dilepton or lepton+jets samples,
but instead would deplete them.  Given a theoretical prediction for the
$t\bar t$ production cross section, the branching ratio for decays which
deplete the SM $t\bar{t}$ samples can be bounded in a straightforward manner.
If decay $(a)$ occurs, the analysis is more involved, since the
kinematic acceptance for the stop decays must be calculated for
many different choices of MSSM parameters.  Also in case $(a)$, 
some $\stop\stop^*$
events will feed into the top quark event samples.

\subsubsection{Direct Top Squark Pair Production}

\begin{figure}[!ht]
\center
\centerline{\psfig{figure=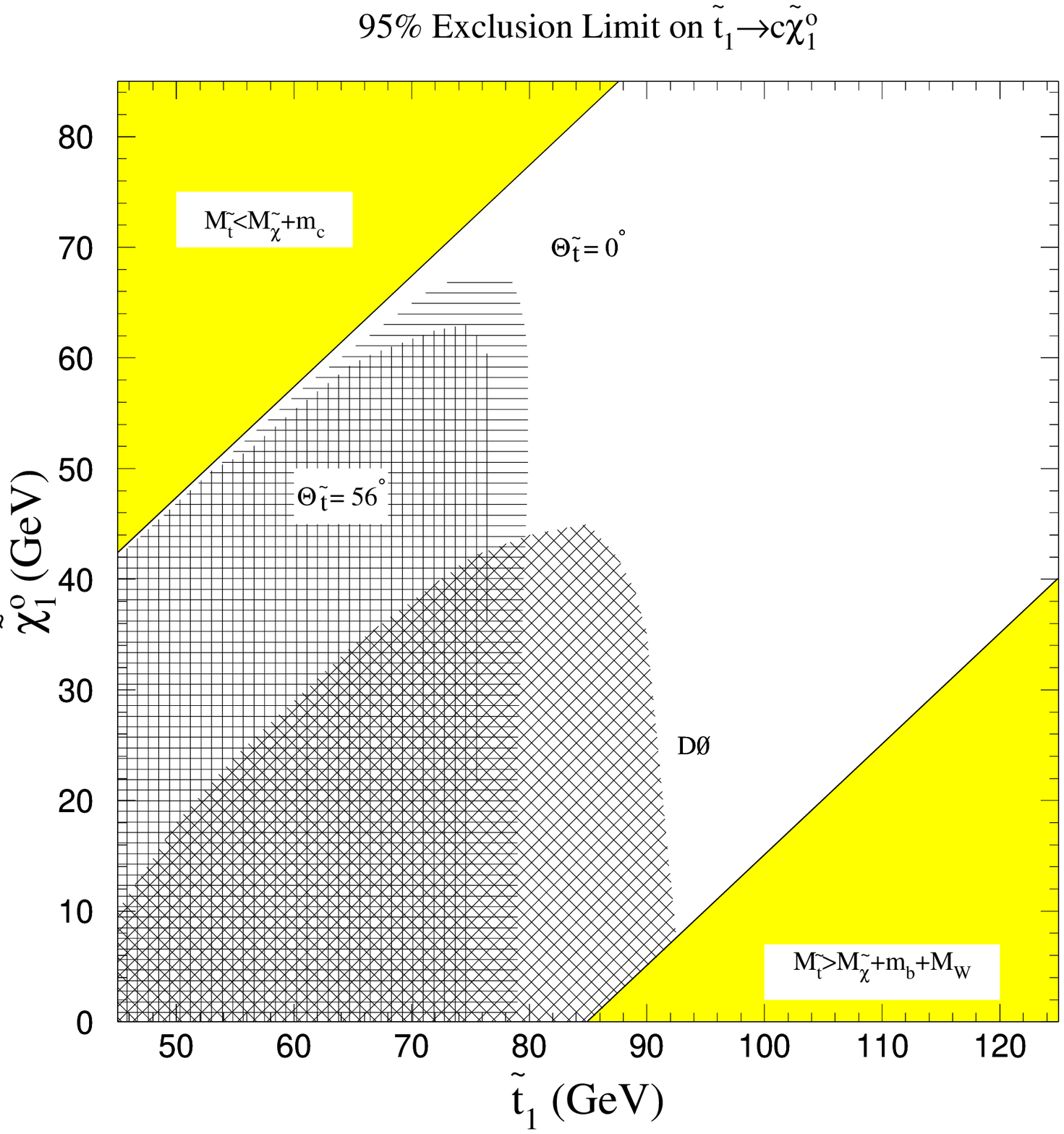,height=80mm}}
\caption{Mass limits from the \D0~search for $\stop\stop^*$ production
with the decay
$\stop\to c\zinol$ at the Tevatron.\protect\cite{D0_stopjj}   
The decay is kinematically forbidden in the two solid grey regions.
The hashed regions marked $\Theta_{\stop}$ show the 
LEP excluded regions as a function of the stop mixing angle, which
determines the strength of the stop coupling to the Z boson.
The mixing does not affect
the tree level process at hadron colliders.
}
\label{fig:d0hadstop}
\end{figure}

\D0~has searched for $\stop\stop^*$ production\,\cite{Sender} with
$\stop\to c\zinol$ using 7.4 \ipb~of Run Ia data.\,\cite{D0_stopjj}  
The signature is two acollinear jets and \met, satisfying
the selection criteria in Table~\ref{tab:claes}.
\begin{table}[!ht]
\centering
\caption{Selection criteria for the \D0~2 jet+\met~hadronic 
direct stop pair production search.  $j_1$ and $j_2$ are
the leading and sub--leading jets ranked by $E_T$.
}
\begin{tabular}{|l|l|}
\hline \hline
Quantity    &  \D0     \\ \hline \hline
$E_T^{j_1}, E_T^{j_2}$  & $>30$ GeV \\ \hline
$\Delta\phi$ $j_1$--$j_2$  & $90^\circ<\Delta \phi <165^\circ$ \\ \hline
$\Delta\phi$ $j_1$--\met & $10^\circ<\Delta \phi < 125^\circ $\\
\hline
$\Delta\phi$ $j_2$--\met & $10^\circ < \Delta \phi$ \\ \hline
Lepton veto                  & $e$ and $\mu$    \\ \hline
\met              & $>40$ GeV         \\ \hline\hline
\end{tabular}
\label{tab:claes}
\end{table}
The dijet cross section at the Tevatron
is large, and thus this signature has large instrumental
backgrounds.   It also has 
backgrounds from vector boson production.    
The QCD and vector boson backgrounds would, naively, be a factor of
1/$\alpha_s$ larger
than for the hadronic squark/gluino search, 
as this search requires only 2 jets while the
latter searches require at least 3 or 4 jets.
This is not the case since the multijet backgrounds can be controlled by
requiring $\Delta \phi> 45^\circ$
between the \met~and each jet,
and that the jets not be back--to--back.  The
vector boson backgrounds are controlled by requiring that 
the two leading jets are separated by at
least $\Delta \phi>90^\circ$.    After these cuts, the dominant backgrounds
are from $W$ and $Z$ boson production and decay, with the largest being 
$W \rightarrow \tau \nu$.
If the $\tau$ decays hadronically, only 1 additional jet
is necessary to fake the signature.
Top quark production is not as important a background for the light stop 
search as for the conventional hadronic squark search 
because of the lower jet multiplicity requirement.
As with the hadronic squark/gluino searches,
the cuts are not very efficient for signal events. The  efficiency is 
largest  when the stop is heavy compared to the
$\zinol$~(near the kinematic boundary for the decay $\stop_1\to t\zinol$),
reaching a 
maximum value of only 4\%.  
The mass difference $m_{\stop}-M_{\zinol}$ determines the 
$E_T$ of the charm jet and rapidly limits
this search mode as the charm jets 
become too soft (see Fig. \ref{fig:d0hadstop}).

With the assumption that BR($\stop_1\to c\zinol$)=1, the predicted SUSY
final state depends only on $M_{\zinol}$
and $m_{\stop_1}$.
The result of this search is a 95\% C.L. exclusion limit on a 
region in the $M_{\zinol}-m_{\stop_1}$ plane,
shown in Fig.~\ref{fig:d0hadstop}.
The production rate has been calculated using
only LO production cross sections evaluated at the scale
$Q^2=2stu/(s^2+t^2+u^2)$ 
from {\tt ISAJET},\cite{isajet} so 
the limit will change somewhat if re--evaluated 
with NLO production cross sections.

\begin{table}[!ht]
\centering
\caption{Selection criteria for the CDF lepton+jets+$b$--tag
direct stop pair production search.  $\ell=e$ or $\mu$.}
\begin{tabular}{|l|l|}
\hline \hline
Quantity    &  CDF cuts \\ \hline \hline
$E_T^\ell$    & $>20$ GeV   \\ \hline
\met          & $>20$ GeV  \\ \hline
$E_T^{j_1}$, $\eta$ range    &  $>15$~GeV,   $|\eta|<2.0$ \\ \hline
$E_T^{j_2}$, $\eta$ range    &  $>8$~GeV, $|\eta|<2.4$  \\ \hline
Jets                         &  $\le$4, $E_T^j>8$~GeV  \\ \hline
SVX $b$--tag       &  1, $E_T^j>$8~GeV  \\ \hline\hline
\end{tabular}
\label{tab:CDF_stop}
\end{table}

\begin{table}[!ht]
\centering
\caption{Selection criteria for the \D0~dielectron+jets+\met direct
stop pair production search.
} 
\begin{tabular}{|l|l|}
\hline \hline
Quantity    &  \D0~cuts \\ \hline \hline
$E_T^{e_1}, E_T^{e_2}$  &  $>16, 8$ GeV     \\ \hline
$E_T^{j}$               &  $>30$ GeV        \\ \hline
$E_T^{e_1}$+$E_T^{e_2}$+$|$\met$|$ & $<90$ \gev \\  \hline
$M_{ee}$ & $<60$ \gev \\ \hline
\met  & $>22$ \gev  \\ \hline\hline
\end{tabular}
\label{tab:amber}
\end{table}

\begin{figure}[!ht]
\centerline{
\hbox{\hbox{\psfig{figure=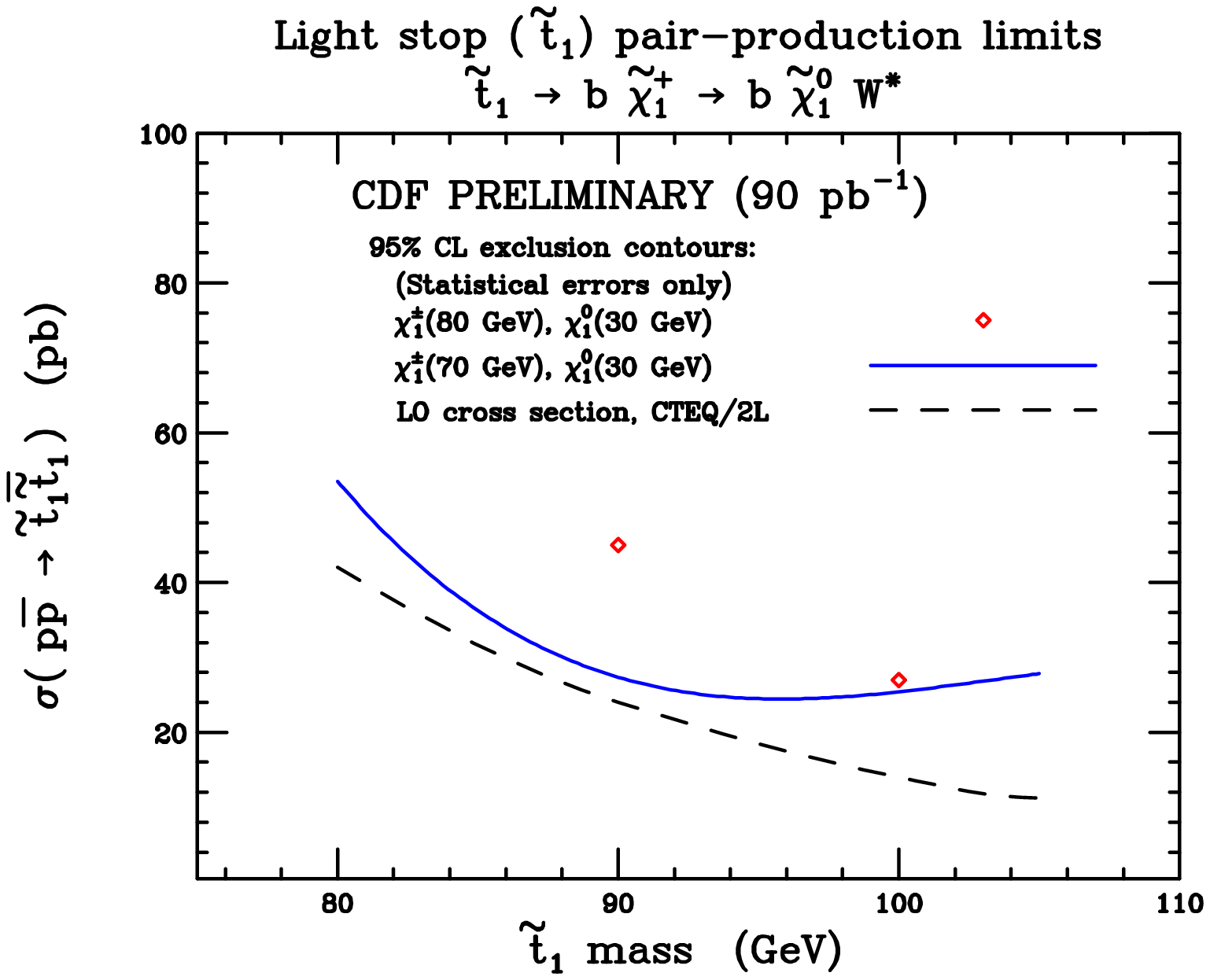,height=50mm}}
\hbox{\psfig{figure=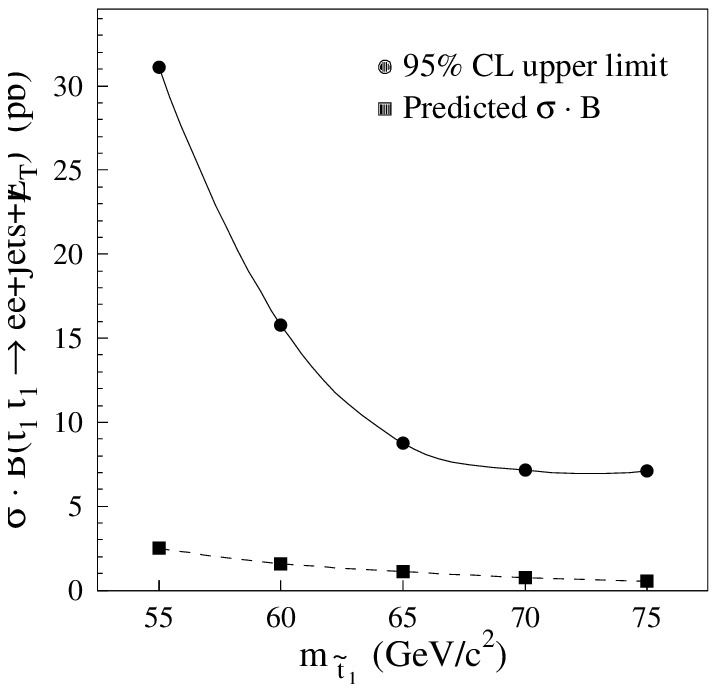,height=50mm}}}
}
\caption{(Left) The CDF cross section limit on direct production of
the top squark  using 90 \ipb~of data.  The decay mode is
$\stop\to~b\winoop(\to W^{*}\zinol)$.
One $W$ must decay semi--leptonically giving a signature of 
a lepton, \met, and jets.  
The theoretical cross section is from ISAJET 7.06.
(Right) The \D0~95\% confidence level cross section limit on  
the cross section for stop production times the branching ratio
to a final state containing 2 electrons as a function
of the mass of the $\stop$~is shown as a solid
line.\protect\cite{D0_stopee}    
The mass of the lightest
chargino is assumed to be 47~GeV.  The predicted cross section
times branching ratio from {\tt ISAJET}\,\protect\cite{isajet} is also shown as a
dashed line.}
\label{fig:cdfd0stopee}
\end{figure}

CDF and \D0~have also presented results from a search for $\stop_1\stop_1^*$
production, with $\stop_1\to b\winol$.\cite{Sender}
The CDF search is in the lepton+jets channel, and
uses a shape analysis of the transverse mass\,\footnote{
Transverse mass squared is 
$M_T^2=(|\vec{p}^{\hspace*{2pt} \ell}_T|+|\vec{\met}|)^2-
(\vec{p}^{\hspace*{2pt} \ell}_T+\vec{\met})^2$, and is a useful 
experimental quantity when information about the longitudinal component 
of momentum is missing.}
of the lepton and~\met.\,\cite{CDF_stop_gold}
The selection criteria are given in Table~\ref{tab:CDF_stop}. 
For both searches, the detection efficiency is smaller than for $t\bar t$ production
with a top quark of the same mass
because of the softer leptons from the 3--body decay $\winol\to
\ell\nu\zinol$.
The mass splitting $M_{\stop}-M_{\winol}$ sets the
efficiency for detecting the jets.
The results of the CDF search are shown in 
Fig.~\ref{fig:cdfd0stopee} (left).  The decay $\winol\to W^*\zinol$ is
assumed using the masses $(i)$ $M_{\winol}=80$ GeV and $M_{\zinol}=30$ GeV
and $(ii)$ $M_{\winol}=70$ GeV and $M_{\zinol}=30$ GeV.
Given these mass choices, there is little other parameter dependence.
Presently, 
the cross section limits are above the predicted cross sections 
due to the high $E_T$ cuts.

\D0~searches in the dilepton 
channel\,\cite{D0_stopee} using the cuts listed in 
Table~\ref{tab:amber}.  
The signature is similar to the squark and top dilepton searches.
The results are shown in Fig.~\ref{fig:cdfd0stopee} (right), assuming
$M_{\winol}=47$ GeV and $M_{\zinol}=28.5$ GeV.
A substantial background comes  from $Z\to\tau^+\tau^-$, 
again requiring a high threshold for the $E_T$ cuts, and no limit can be set.

The above analyses were done in regions of SUSY parameters that
have been excluded by LEP. They show, however, the 
procedures to be followed in redoing these studies for other regions of 
the MSSM parameter space.
\subsubsection{Top Squark Production From Top Decays}

CDF has presented another analysis using the  SVX--tagged
lepton+jets sample to search for the decay $t\to\stop_1\zinol$, with
$\stop_1\to b\winol$.\cite{CDF_stop_carmine}
If one of the top quarks in a $t\bar t $ event decays 
$t\to bW(\to \ell\nu)$ and the other $t\to\stop_1\zinol$ followed by
$\stop_1\to b\winol(\to jj\zinol)$
or $t\to bW(\to jj)$ and  $t\to\stop_1\zinol$ followed by
$\stop_1\to b\winol(\to \ell\nu\zinol)$,
the signature is $b\bar b\ell\nu jj$+\met, 
the same as in
the SM, but where the \met~includes the momentum of the $\zinol$.
The lepton+jets channel has a large number of events, so a
kinematic analysis can be performed on the event sample.
Due to the mass of the $\zinol$ and the intermediate sparticles 
in the decay chain, the jets from the SUSY decay are significantly 
softer.
This difference is exploited as the basis of the search.  

\begin{table}[!ht]
\centering
\caption{Selection criteria for the CDF search for top decaying 
into stop in the signature of 1 lepton ($\ell$), \met, and 
3 jets including at least one $b$ tag,
where $\ell=e$ or $\mu$.
The quantity
$|\cos\theta^*|$ is the polar angle of a jet in the rest frame of the 
$\ell$, \met~and jets.
$\Delta R_i$ is the distance between a jet $i$ and the next 
nearest jet in $\eta-\phi$ space.
The jets are ordered in $E_T$, so $E_T^1>E_T^2>E_T^3$.
} 
\begin{tabular}{|l|l|}
\hline \hline
Quantity    & CDF cuts  \\ \hline \hline
\lumint          & 110 \ipb \\ \hline
$E_T^\ell$       & $>20$~GeV \\ \hline
\met             & $>45$~GeV  \\ \hline
$M_T(\ell$\met)  & $>40$~GeV  \\ \hline
$p_T(\ell$\met)  & $>50$~GeV  \\ \hline
$E_T^{1,2}$      & $>20$~GeV, $|\eta|<2.0$   \\ \hline
$E_T^3$          & $>15$~GeV, $|\eta|<2.0$   \\ \hline
$|\cos\theta^*|_{1,2,3}$         & $<0.9,0.8,0.7$ \\ \hline
$\Delta R_{1,2,3}$  & $\ge 0.9$  \\ \hline
Number of SVX $b$--tags       & $\ge 1$ for $E_T^j>$15~GeV  \\ 
\hline\hline
\end{tabular}
\label{tab:CDF_top_stop}
\end{table}

The cuts listed in Table~\ref{tab:CDF_top_stop} are optimized for
acceptance of the SUSY decay and rejection of $W$+jets background.
A likelihood function is computed for each event
reflecting the probability that the jets with the $2^{\rm nd}$ and $3^{\rm rd}$ 
highest $E_T$ in the event are consistent with the stiffer 
SM distribution (as compared to the SUSY distribution).
The distribution of this likelihood function shows a significant 
separation of these two hypotheses.
After applying the cuts listed in
Table~\ref{tab:CDF_top_stop},
nine events remain, all of which fall outside of the SUSY signal region.
For stop masses between 80 and 150 GeV and chargino masses between
50 and 135 GeV, a BR($t\to \stop_1\zinol$)=50\% is excluded at the
95\% C.L., provided that $M_{\zinol}=20$ GeV.
Because $M_{\zinol}$ is fixed in this manner, it is not related to
$M_{\winol}$ as in SUGRA.  At present, 
only this one example is available (for $M_{\zinol}$ already excluded by LEP);
more statistics will significantly improve it.

\subsection{Sleptons}

\begin{figure}[!ht]
\center
\centerline{\psfig{figure=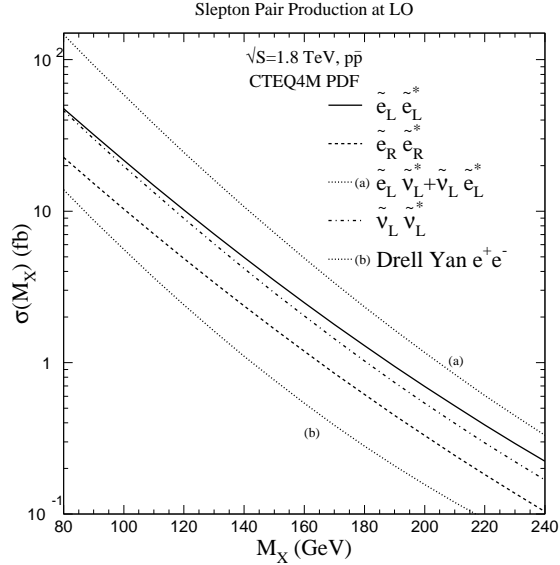,height=80mm}}
\caption{Production cross sections for sleptons versus
slepton mass at the Tevatron.  
The Drell--Yan cross section for producing $e^+e^-$
(curve b) is plotted as a function of $M_{e^+e^-}=2M_X$.
On this scale, the Drell--Yan peak ($Z\to e^+e^-$) would appear at $M_X=45$ GeV.}
\label{fig:susyxseclepton}
\end{figure}

At hadron colliders, sleptons can only be pair produced through their
electroweak couplings to the $\gamma, Z$ and $W$ bosons.
Figure~\ref{fig:susyxseclepton} shows the cross sections 
as a function 
of the corresponding slepton mass compared to the differential Drell--Yan
pair production cross section, $d\sigma_{\rm D-Y}/dQ$, where $Q=2M_X$.
The rate for slepton pair production is at most a few tens or hundreds
of fb at the Tevatron, and
so far neither collaboration has presented results on searches for sleptons in
the SUGRA or RIPS framework (we describe limits in gauge mediated 
models later).    

A (stable) charged slepton is not a viable LSP candidate, so the
decays $\tilde{\ell}_{L,R}^{\pm} \to \ell^{\pm} \tilde{\chi}^{0}_i $
or $\tilde{\ell}^{\pm}_L \to \nu\winog_i$ are expected.  The
sneutrino, instead, can be the LSP, or it can decay invisibly
$\sneutrino\to\nu\zinol$, or visibly $\sneutrino\to\winog_i\ell^\mp$.
If $m_{\sneutrino} < m_{\tilde\ell} < M_{\zinol}$, then the decay
$\tilde{\ell} \to \ell' \nu' \sneutrino $  (or $\tilde{\ell} \to q \bar{q}
 \sneutrino $) is possible.  Promising signatures are $(i)$
$e^+e^-,\mu^+\mu^-,\tau^+\tau^-$ plus \met,\,\cite{baer} $(ii)$
$e\mu,e\tau,\mu\tau$ plus \met, and $(iii)$ $e,\mu$ or $\tau$+jets plus
\met~(or jets plus \met).  
Although charged slepton production can lead to charged leptons
in the final state, there is no guarantee.

The major background to same--flavor lepton pairs 
is Drell--Yan pair
production, with
fake \met~from mismeasurement of the lepton or jets in the event.
Most of this background can be removed by vetoing on a dilepton
mass window around the $Z$ mass, by requiring significant \met($\ge 25$ GeV),
and by vetoing events
with the \met~pointing  in $\phi$ along one of the leptons or a jet.
Top quark production is also a major background to a slepton heavier than
the lighter gaugino, as it produces dilepton events
that have real \met.\cite{baer} 
Untangling a few heavy slepton events from top events would
be difficult at the present low level of statistics.

    Inclusive searches have the advantage of a larger
acceptance than searches in exclusive channels. A unique signature 
of slepton production directly or in cascade decays would be 
the apparent violation of lepton universality.  If the sleptons are 
not degenerate, both the production and decays of the sleptons will favor 
one or two leptons over the others, resulting in an imbalance in the detected 
$e/\mu/\tau$ ratios in SUSY--enhanced channels.
The dominant backgrounds to inclusive leptons come from heavy flavor 
production ({\it e.g.} $b$--quarks), and (single) $W$ and $Z$ boson
production.\,\cite{sacha}
Because sparticles are produced in
pairs, it may be possible to discriminate against SM backgrounds by
requiring the identification of a part of the decay of the second sparticle. 
Examples of channels that
may have enhanced SUSY contributions over SM backgrounds (and hence
possibly apparent lepton universality violation) are those that have, in
addition to the lepton, a  $\gamma$, $W$, $Z$, additional lepton, or a 
third--generation particle.

\subsection{Charged Higgs Bosons} 
\label{section:chargedhiggsresults}

Even though Higgs bosons are not sparticles,  the discovery of 
one or more would be considered {\it indirect} evidence
for SUSY.\,\footnote{If $M_{H^\pm}\lesssim 300$ GeV, then there must
exist extra light--matter fields beyond the SM to partially cancel the
$H^\pm$ contribution to BR($b\to s\gamma$).}
If it is light enough, the charged Higgs boson
$H^\pm$ can be produced 
in the decay of the top quark $t\to bH^+$.\,\cite{Roy}
The branching fraction for this decay depends on the 
charged Higgs boson mass and \tanb.  
When kinematically allowed, this branching ratio 
is larger than 50\% for \tanb~less than approximately 0.7 or greater than
approximately 50, and completely dominates for very small or 
very large values of \tanb.
However, as discussed in section 2.1, values of $\tan \beta \leq 0.6-0.7$
or above 60 would be associated with large top or bottom Yukawa couplings,
which will become infinite at scales not far above the TeV scale.
In general, at reasonably small values of $\tan\beta$, the charged 
Higgs boson decays $H^+\to c\bar s$; at large $\tan\beta$ it instead decays 
$H^+\to \tau^+\nu_\tau$.

CDF has searched for the decay $t\to bH^+$ using both
direct\,\cite{CDF_ch_higgs_jessop,CDF_ch_higgs_rutgers} and
indirect\,\cite{CDF_trilepton_1ab} methods. Direct searches look for an excess
over SM expectations of events with $\tau$ leptons from the charged Higgs 
boson decay
$H^+\to \tau^+\nu_\tau$ (dominant for large $\tan\beta$). 
On the other hand, indirect
searches are ``disappearance'' experiments, relying on the fact
that decays into the charged Higgs boson mode will deplete the SM decays
$t \to bW$, decreasing the number of events in the dilepton and 
lepton+jets channels.  

The CDF direct search at large \tanb~uses two sets of cuts,
listed in Table \ref{directchiggscuts}, to search for an excess of 
$\tau$'s in $t\bar{t}$ events.
The first set selects a sample containing a 
$\tau$ that decays hadronically, an SVX $b$--tagged jet,
\met~and objects indicating activity
from the second top decay: a second jet and a third jet or lepton.  
As $M_{H^\pm}$ approaches $m_t$, the $b$ produced in the
top decay $t \to bH$ becomes less energetic, causing a reduced
efficiency for the jet and $b$--tagging requirements.
To maintain efficiency in this region, a second set of cuts 
accepts events that have two high--$E_T$ $\tau$'s and \met.

The signature for hadronically decaying $\tau$'s is  a narrow jet associated
with one or three tracks  with no other tracks nearby.  Typically  the $\tau$
tracks are required  to be within a 
cone of $10^\circ$ with no other tracks within a
cone of $30^\circ$.  Fake rates, measured as the probability that a generic
jet is identified as a $\tau$, are approximately 1\% or less. These fake rates
are too high to identify $\tau$'s in a sample dominated by QCD.  However, if
another selection criterion is added that further purifies the sample,  $\tau$'s
can be identified with a 
good signal--to--background ratio.  For example, hadronic decays of
the $\tau$ are observed in 
$(i)$ monojet ($W\to\tau\nu$), 
$(ii)$ lepton, \met~and jet ($Z\to\tau\tau$) and
$(iii)$ the lepton+jets top quark 
samples.\,\cite{Marcus_PhD,Gallinaro_PhD}

\begin{figure}[!ht]
\centerline{\hbox{\hbox{\psfig{figure=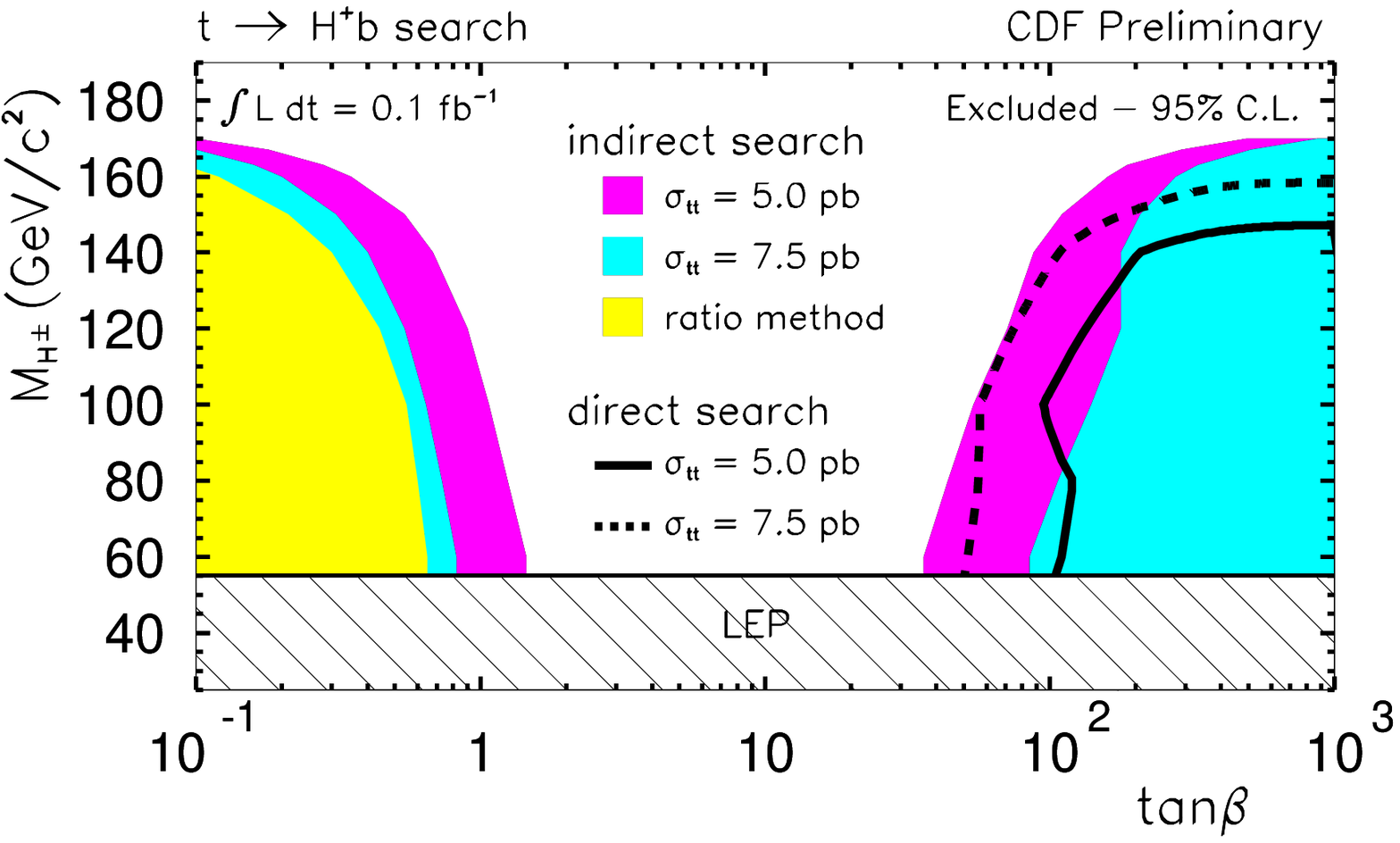,height=60mm}}
\hbox{\psfig{figure=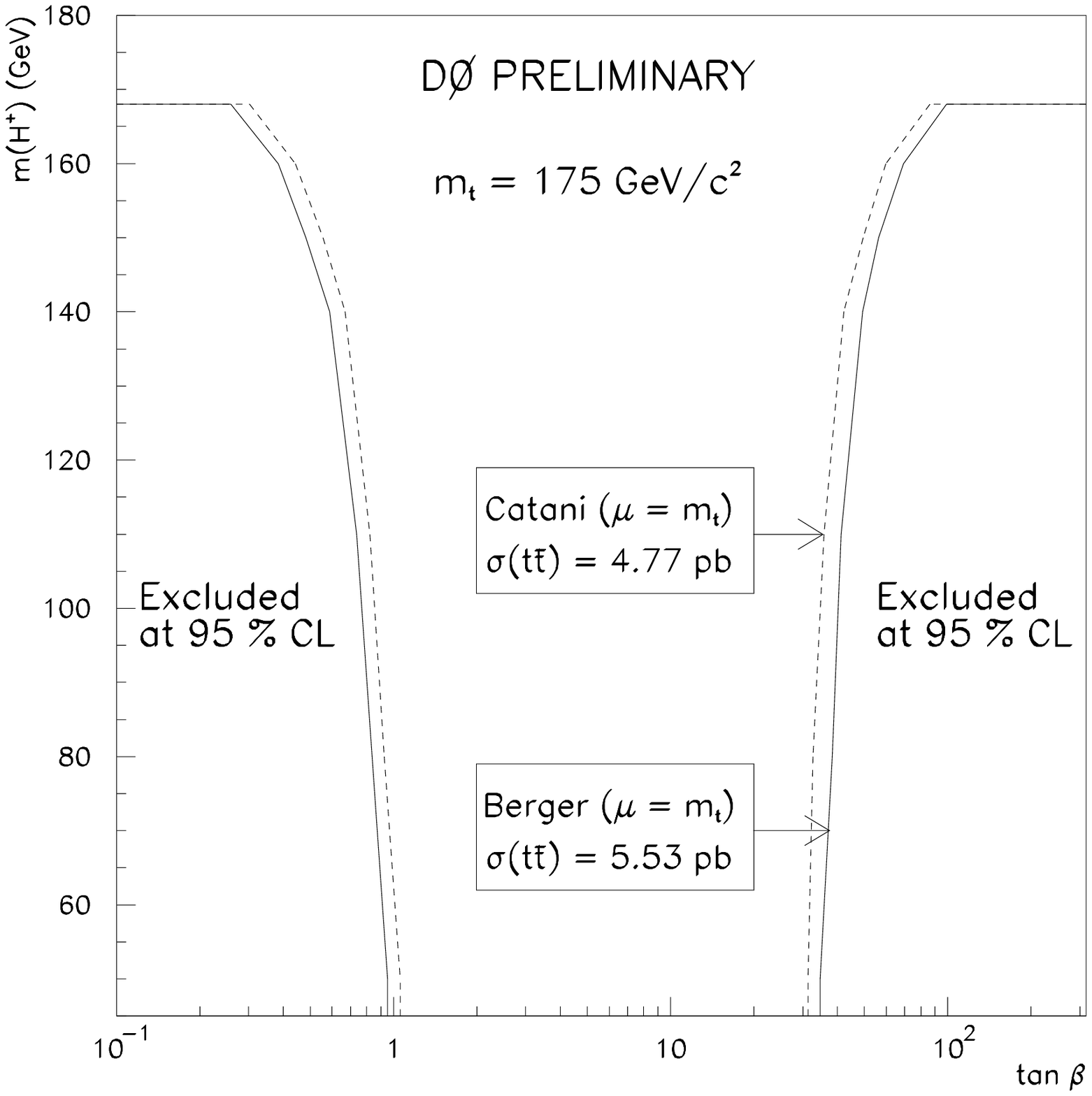,height=60mm}}}}
\caption{(Left) Exclusion space for the CDF searches for charged Higgs 
boson decays of the top quark in $t\bar{t}$ events.  
The shaded regions are from the indirect searches.
For the regions labeled $\sigma_{t\bar{t}}=5.0$ and 7.5~pb 
a top production cross section is assumed and points are excluded if
the predicted SUSY decays have depleted the SM channels to an extent that
they are inconsistent with the data.
The ``ratio method'' is an indirect method
comparing the number of lepton+jets events to the number 
of dilepton events and no top cross section is assumed.
The region excluded with solid lines at high
$\tan\beta$ is from a direct search for events where one or both 
top quarks in a $t\bar t$ event decay to $bH^+(\to\tau^+\nu)$
and information from the SM channels is ignored.
(Right) The results of a \D0~indirect search for a charged Higgs boson 
assuming $m_t=175$ GeV and a $t\bar t$ production cross section
of 5.53 pb and 4.77 pb.  This limit is based on the full Run I \D0~data sample.
}
\label{figure:chiggslimits}
\end{figure}

The CDF direct search for $t\bar{t}$ events with one or two  
charged Higgs bosons decaying to $\tau$ leptons
sets limits by two methods.
In the first method, a $t\bar{t}$ production cross section 
and a SUSY model ($M_{H^\pm}$ and \tanb) is assumed and 
the number of expected $\tau$ events is computed.  If the number 
expected is too large to be consistent with the observed number
at the 95\% C.L., the SUSY model is excluded for that $t\bar{t}$
cross section.  
This method excludes a charged Higgs with mass less  than 155~GeV 
(100~GeV) if \tanb~is
greater than approximately 100 (50) and the top cross section\,\footnote{
The reader should be aware that there are subtleties in analyses that assume
cross sections. The CDF experiment normalizes all cross sections
to its measured proton--antiproton cross sections, rather than to a hard
process such as $W$ production. The CDF total cross section, in the
judgement of one of the authors (HF), is most likely 10\% too high, and thus
all cross sections (in particular the top cross section) 
are too high. For analysis which compare different channels
internally this has no effect, but for analyses which compare to theoretical
predictions the reader should be careful. The D0 experiment normalizes to
a weighted mean of the CDF and E710 values for
the total cross section  which  is 2.4\% lower than
the CDF value.}
is 7.5~pb, as shown in Fig.~\ref{figure:chiggslimits} (left).

The second method combines the observation of $t\to bW$ decays 
into leptons ($e$ or $\mu$) and jets with the number of 
$\tau$ decays from the direct search.
This has the advantage that a top production cross section does not 
need to be assumed.  
The lepton+jets sample defines a top production cross section
which, in turn, through the SUSY model, predicts the number of 
$\tau$ events expected.  If the number is too large to be consistent
with the observation, the model is excluded.
The limits set by this method
are presented in detail elsewhere.\,\cite{CDF_ch_higgs_rutgers}
Qualitatively, the limits are similar to those set by the indirect 
method (discussed below).

\begin{table}[!ht]
\centering
\caption{Selection criteria for the CDF direct search for 
$t\rightarrow bH^\pm(\rightarrow \tau\nu)$ in 100~\ipb~of data.
The $\tau$'s are identified in their hadronic decay modes 
as one or three isolated, high--$p_T$ tracks.
Events are accepted if they pass the cuts in either analysis path.}
\begin{tabular}{|l|l|} \hline
Quantity & CDF \\ \hline\hline
\multicolumn{2}{|l|}{Analysis path 1:} \\ \hline
$E_T^\tau$ & $>20$~GeV \\ \hline
\met                     & $>30$~GeV \\ \hline
$E_T^{j_1}$, SVX tagged  & $>15$~GeV \\ \hline
$E_T^{j_2}$              & $>10$~GeV \\ \hline
Additional object & $e,\mu,\tau$ or \\
                  & 3$^{\rm rd}$ jet with $E_T>10$~GeV \\ 
   \hline\hline
\multicolumn{2}{|l|}{Analysis path 2:} \\ \hline
$E_T^{\tau_1}, E_T^{\tau_2}$   & $>30$~GeV     \\ \hline
$\Delta\phi(\tau_1,\tau_2)$    & $<160^\circ$  \\ \hline
\met                           & $>30$~GeV     \\ \hline
\end{tabular}
\label{directchiggscuts}
\end{table}

At small \tanb~a direct search is difficult since the 
charged Higgs decays into two jets.
Instead only the indirect method is applied.  

The indirect method can be applied to both small and large \tanb~searches.
The observed numbers of dilepton and lepton+jets events are consistent 
(at the 95\% C.L.) with a minimum production and decay rate in the SM.
For an assumed $t\bar t$ 
cross section (and a charged Higgs boson mass and \tanb), a  simulation of the 
expected mixture of SM and charged Higgs boson 
decays predicts a number of events 
in the SM channels.  At points in the parameter space where the decays
to charged Higgs bosons
are more pronounced, the SM contributions are diminished. 
If the number expected in a SM channel 
is not consistent with the rate defined by the data,
the assumed values are excluded. 
This method provides the limit displayed in Fig.~\ref{figure:chiggslimits}
(left).
The cross section of 5~pb is the expected cross section for a top mass of 
175~GeV; the curves using
7.5~pb show the sensitivity of the limit to the assumed top cross section.
Also shown in the figure is how the limit in the region of
large \tanb~can be extended using the assumptions of the indirect method.
In this region the possibility that a $\tau$ decay produces a high--$p_T$
lepton is included.

The area in Fig.~\ref{figure:chiggslimits} (left) labeled 
``ratio method'' is the exclusion 
region for an indirect search that does not make an assumption for the 
$t\bar{t}$ cross section.  If charged Higgs boson decays were competing with 
SM decays, the ratio of dilepton events to lepton+jets events would 
decrease, regardless of the $t\bar{t}$ cross section.  
This occurs because the lepton+jets yield is proportional to the SM
branching ratio while the dilepton yield is proportional 
to the SM branching ratio squared.
For each SUSY parameter point, the lepton+jets sample can
be used to infer a top cross section which, in turn, 
predicts a number of dilepton events.  The point is excluded
if the prediction is inconsistent with the dilepton data.
Although this method excludes less parameter space, it is important since 
the $t\bar{t}$ cross section may be enhanced
by SUSY mechanisms\,\cite{kanegtm} such as $\tilde{g}\to t\tilde{t}$.
At present, this method only excludes values of $\tan\beta \lesssim 0.7$,
which are not of much interest according to present theoretical bias.

\D0~has also searched for a charged Higgs boson lighter than the top quark
using the indirect method.\cite{D0_chhiggs}
The analysis compares the number of events observed in the lepton+jets 
channel to the number predicted assuming a theoretical
$t\bar t$ production cross section.  The limits depend on 
the mass of the charged Higgs, \tanb, and the top quark mass $m_t$.  
Table~\ref{tab:d0chargedhiggs} shows the selection criteria used in the search.
Fig.~\ref{figure:chiggslimits} (right) 
shows the excluded region.

\begin{table}[ht]
\centering
\caption{Selection criteria of the \D0~search for a charged Higgs
boson produced in top quark decays.  In addition, events are vetoed if the
\met~is aligned in $\phi$ within $25^\circ$ of a muon, or if 
the muons in a $\mu$ event with a $\mu$--tagged jet  
have a good fit to the decay $Z \to\mu\mu$.
}
\begin{tabular}{|l|l|l|}
\hline \hline
Quantity    &  \D0~ topological  & \D0~  tagged  \\ \hline \hline
 \et~ threshold on leptons& 20 \gev &  20 \gev \\ \hline
Max $\eta$ for leptons  & 2 (e) 1.7 ($\mu$) &  2 (e) 1.7 ($\mu$)\\ \hline
Number of jets  &  4 & 3 \\ \hline
jet \et~ threshold  & 15 \gev & 20 \gev \\ \hline
\met  & 25 \gev ~ $(e)$ 20 ($\mu$) & 20 \gev \\  \hline
$H_T$  & 180 \gev  & 110 \gev \\ \hline
Sum of lepton \et~ and \met~ & 60 \gev & N.A. \\ \hline
Aplanarity  & 0.065 & 0.04 \\ \hline
$|\eta_W|$  & 2.0 & N.A. \\ \hline
 $\mu$--tagged jets  & veto  & require  \\ \hline
\end{tabular}
\label{tab:d0chargedhiggs}
\end{table}

Recent studies have shown that quantum SUSY effects 
(SUSY QCD and electroweak radiative corrections) to the decay mode 
$t\rightarrow bH^+$ (with subsequent decays into $\tau$'s) 
may be important and should be considered 
in future analyses.\,\cite{ChHiggscorr}

\subsection{Neutral Higgs Bosons }
\label{section:neutralhiggsresults}

Within the MSSM, 
the main production channels for the lightest CP--even Higgs boson $h$ at the 
Tevatron are the same as for a SM Higgs boson, $Wh$
or $Zh$ production.\,\cite{stange2} 
The cross 
sections behave in such a way that these 
channels are relevant for large 
values of the CP--odd mass $M_A$ (the SM limit) or for small $M_A$ and small   
$\tan \beta$. 
The heavy CP--even Higgs boson $H$ could become marginally
relevant for searches at an upgraded Tevatron through $ZH$, $WH$ 
production,
in some restricted region of parameter space, complementary to 
the one relevant for the light CP--even Higgs boson searches.
In addition, the enhancement of the bottom Yukawa coupling in the 
large \tanb~regime can render the production processes
$hb\bar{b}$, $Ab\bar{b}$, and $Hb\bar{b}$ useful to perform searches in 
a large region of parameter space.\,\cite{Gunionhbb}

\begin{figure}[!ht]
\centerline{\psfig{figure=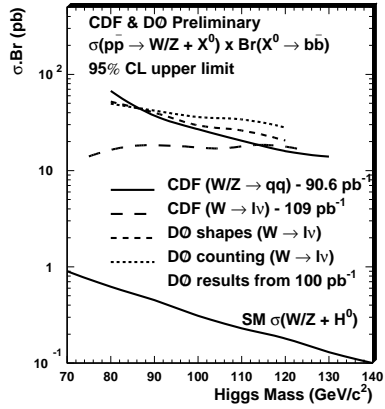,height=60mm}}
\caption{Limits from CDF and \D0~for the associated production
of a neutral Higgs boson and a $W$ or $Z$ boson.
The CDF limits are shown for the final
states of $\ell\nu$\bbbar~and $jj$\bbbar, and the \D0~limit is
for the final state $\ell\nu$\bbbar.  The limit is set using
a simple counting method and by fitting the \bbbar~spectrum (``shapes'').}
\label{figure:nhiggslimits}
\end{figure}

Both collaborations have searched for a neutral Higgs boson in the mode
$q\bar{q}'\rightarrow W^*\rightarrow 
W(\rightarrow e\nu,\mu\nu)h(\rightarrow b\bar{b})$.
\D0~has searched in 100~\ipb~of data using
a data sample containing a lepton, \met~and two jets.\cite{d0_neu_higgs}
One of the jets must have a muon associated with it 
for $b$--tagging.
The cuts are listed in Table \ref{nhiggscuts}.
Twenty--seven events pass the selection criteria; $25.5\pm 3$ events
are expected from $Wjj$ and $t\bar{t}$.  The limits shown in 
Fig.~\ref{figure:nhiggslimits} are set by a 
simple event--counting method and by 
fitting the $b\bar{b}$ dijet mass spectrum.

CDF has recently completed a similar search  for the same decay mode using
109 \ipb~of data.\cite{CDF_neu_higgs_weiming} 
All events must have one SVX $b$--tag.  These events 
are split into single--tagged (one SVX tag) and
double--tagged samples (two SVX tags or one SVX and one lepton 
($e$ or $\mu$) tag). 
The 36 (6) single--tagged (double--tagged) events  are consistent with the 
$30\pm 5$ ($3.0\pm 0.6$) expected from SM $W$+jets and $t\bar{t}$.
Both the single-- and double--tagged dijet mass distributions
are fit simultaneously to set the limits shown in 
Fig.~\ref{figure:nhiggslimits}.

The process $q\bar q\to Z^*\to Zh$ occurs at a comparable rate to the
$W^*$ process.  CDF has searched for both associated production processes
assuming $W/Z\to jj$.\cite{CDF_neu_higgs_valls}  
The event selection criteria are listed in Table~\ref{nhiggscuts}.
In 91 \ipb~of data, 589 events remain, consistent with the expectation 
from QCD heavy--flavor production and fake
tags.  To set limits, the $b\bar{b}$ dijet mass spectrum is fit.  
Also shown in Fig.~\ref{figure:nhiggslimits} is the SM production 
cross section 
for $Wh$ and $Zh$ as a function of the Higgs boson mass.  The present
experimental limits are roughly two orders of magnitude away from the
predicted cross section.  However, Run II will provide at least 20 times 
the data.  This plus the possibilities of looking at other
decay modes ({\it i.e.} $Zh\rightarrow \nu\nu b\bar{b}$) holds promise
for Higgs physics at the Tevatron.\cite{mrenna_higgs}

\begin{table}[!ht]
\centering
\caption{Selection criteria of Tevatron searches for the 
associated production of a neutral Higgs boson and a $W$ or $Z$,
and the Higgs boson decays to $b\bar{b}$.}
\begin{tabular}{|l|l|l|} \hline
Quantity & CDF & \D0 \\ \hline\hline
\multicolumn{3}{|c|}{$WH\rightarrow \ell\nu b\bar{b}$}   \\ \hline
$E_T^e (E_T^\mu)$     & $>25(20)$~GeV &  $>20(20)$~GeV   \\ \hline
\met $e$ ($\mu$)      & $>25(20)$~GeV &  $>20(20)$~GeV   \\ \hline
$E_T^{j_1},E_T^{j_2}$ & $>15$~GeV     &  $>15$~GeV       \\ \hline
$b$--tagging          & one SVX tag   &  one $\mu$ tag   \\ \hline\hline
\multicolumn{3}{|c|}{$(W,Z)H\rightarrow jj b\bar{b}$}    \\ \hline
Quantity &\multicolumn{2}{|c|}{CDF} \\ \hline
$E_T^{j_{1-4}}$   &\multicolumn{2}{|c|}{$>15$~GeV} \\ \hline
$b$--tagging      &\multicolumn{2}{|c|}{2 SVX tags} \\ \hline
$P_T(b\bar{b})$   &\multicolumn{2}{|c|}{$>50$~GeV} \\ \hline
\end{tabular}
\label{nhiggscuts}
\end{table}

\D0~has also searched for a fermio--phobic Higgs, {\it i.e.} one
with suppressed couplings to fermions.\cite{d0_gg_higgs}  For a light
neutral Higgs boson, the decay through a virtual $W$ loop
to a $\gamma\gamma$ final state can be dominant.\cite{stange}  Events are
selected containing two
photons with \et $>$ 20 and 15 \gev, and
two jets with \et $>$ 20 and 15 \gev. 
No evidence of a resonance is seen in 
the mass distribution of the 2 photons, and \D0~excludes,
at a 95\% C.L., such a Higgs with masses less than 81~GeV.
The branching fraction for $h\to\gamma\gamma$ is taken from
Ref.~\oldcite{stange}.
\begin{table}[!ht]
\centering
\caption{Selection criteria for the D0 search for a Higgs boson produced
in association with a hadronically-decaying W and which
decays to two photons.}
\begin{tabular}{|l|l|}
\hline \hline
Quantity    &  \D0  \\ \hline \hline
 \et~ threshold on photons& 1 above 20 \gev , 1 above 15 \\ \hline
$|\eta|$ on photons & $<$1.1 or 1.5$<|\eta|<$2.25 \\ \hline
\et~ threshold jets & 1 above 20 \gev , 1 above 15 \\ \hline
$|\eta|$ on jets  & $<$ 2 \\ \hline
vector sum of photon \et  & $<$ 10~GeV \\ \hline
vector sum of jet \et  & $<$ 10~GeV   \\ \hline
\end{tabular}
\label{tab:d0bosonichiggs}
\end{table}

\subsection{R--Parity Violation and a Short--Lived LSP}

 Allowing for R--parity violation  in the MSSM opens a host of
possibilities at the Tevatron.   
Both baryon--number--violating operators ($UDD$) 
and lepton--number--violating ones ($LLE$ and $LQD$) are possible.
There are many  resonant and non--resonant particle production mechanisms
and subsequent decay processes which have been analyzed in the 
literature.\,\cite{RPV} 
In this section, we restrict ourselves to the experimental analyses
performed so far.
 
The possible excess of HERA events at large $Q^2$ has triggered interest
in studying the consequences of the interaction of a light squark 
(preferably a top or charm squark) with an electron and a $d$ 
quark.\,\cite{CandR}
If the gluino were heavier than this squark, then gluino pair
production at the Tevatron and the decay 
$\tilde{g}\rightarrow \bar c{\tilde c}_L$ 
through R--conserving couplings, followed by the RPV decay 
$\tilde{c}_L\rightarrow e^+d$, would yield the signature of two electrons
and 4 jets.  
If the RPV decay ${\tilde c}_L\to e^+d$ is allowed through the
coupling $\lambda'_{121}$, then from the structure of the 
R--parity violating Lagrangian (Eq. (\ref{eq:RPVLLE})) it follows that 
${\tilde s}_L\to \nu_e d$,
${\tilde d}_R\to e^-c$, and
${\tilde d}_R\to \nu_e s$ are also allowed.
If $m_{{\tilde c}_L} \simeq m_{{\tilde s}_L}$ (which is guaranteed)
$\simeq m_{{\tilde d}_R}$ (which is probable), then the gluino decays
equally to ${\tilde c}_L\bar c$, ${\tilde s}_L\bar s$, and
${\tilde d}_R\bar d$ (+ h.c.) final states.
Assuming that only RPV decays occur, then 1/2 of gluino decays
produce a charged lepton.  Therefore, $\gluino\gluino$ production
produces like--sign dileptons 1/8 of the time.
The requirement of only RPV decays 
demands $M_{\zinol} > m_{\squark}$.

\begin{table}[!ht]
\centering
\begin{tabular}{|l|c|} \hline
Quantity & CDF \\ \hline
$E_T^{e_1}, E_T^{e_2}$    &  $>15$~GeV, $|\eta|<$1.1   \\ \hline
$Q_{e_1}+Q_{e_2}$         &  $\pm 2$                   \\ \hline
$E_T^{j_1}, E_T^{j_2}$    &  $>15$~GeV, $|\eta|<$2.4   \\ \hline
$S=$\met $/\sqrt{\Sigma E_t}$ & $<5$~GeV$^{1/2}$ \\ \hline
\end{tabular}
\caption{Selection criteria of the 
CDF search for R--parity violating processes using 105~\ipb~of data.}
\label{rpvcuts}
\end{table}

CDF has performed a search\,\cite{CDF_Rparity} considering the 
RPV squark decays with the signature of two like--sign electrons
and two jets.
In 105 \ipb~of Run Ia and Ib data, no events remain after all cuts 
are applied  (see Table~\ref{rpvcuts}).
Varying the masses of the SUSY particles does not 
alter the acceptance significantly since they are heavy enough for
the decay products
to easily pass the $E_T$ thresholds.  Because of this, the 
limit on the cross section times branching ratio is approximately constant 
at 0.19~pb.  For $m_{\tilde c_L}=200$~GeV, this excludes
$M_{\gluino} < 230$ GeV, assuming
${\rm BR}(\tilde{g}\tilde{g}\rightarrow e^\pm e^\pm X)=1/8$.

Allowing for possible 
R--parity conserving squark decays, the decay
$\squark\to q\zinol$ is possible, where $\zinol$ is the LSP.
Since the LSP has no R--parity conserving decays kinematically accessible,
the R--parity violating decay
$\zinol\to c \bar{d} e^-$ or $\bar c d e^+$ occurs through a virtual charm or down squark,
while $\zinol\to d \bar{s} \nu$ or $\bar d s \bar\nu$ 
occurs through a virtual strange or down squark.
The exact branching ratio for $\zinol\to e^\pm + X$
 depends on sparticle masses and
the mixing of the neutralinos.
For the analysis, five squark masses are assumed to be degenerate
and any squark pair can lead to like--sign dielectron events,
since $\zinol$ is a Majorana particle.
Squark masses less than 210~GeV are excluded if the 
mass of the $\zinol$ is more than half of the squark mass
and the gluino is heavy.  For lighter $\zinol$, the 3--body 
decay of the $\zinol$ can produce electrons that are too soft
to satisfy selection criteria.

\subsection{R--Parity Violation and Long--Lived Heavy Charged Sparticles} 

If R--parity is violated, and the LSP is charged, 
it can manifest itself
as a long--lived charged particle (see Sec.~\ref{sec:r_parity})
in a collider detector.
The particle can be identified by measuring the $dE/dx$ energy loss 
as it passes through the CDF SVX and CTC detectors.  For a 
given momentum, a heavy particle has a slower velocity and hence
a greater energy loss than a relativistic particle ($\beta\simeq 1$).  
If the particle is weakly interacting 
or massive enough to kinematically suppress showering, it will penetrate
the detectors and be triggered on and reconstructed 
as a muon with too much energy loss.
A result using part of the Run I data has been
presented by CDF \cite{CDF_stables} and is updated with the full data set here.
In 90 \ipb~of inclusive muon triggers ($p_T>$30 \gev), CDF 
searches for particles with ionization consistent with
$\beta\gamma < 0.6$ and finds 12 events 
depositing more than twice the energy expected from a minimum ionizing muon.
This is consistent with the number of events expected 
from muons which overlap with other tracks to fake a large $dE/dx$ signal.

The CDF result can be used to exclude some SUSY scenarios with
R--parity violation (RPV).
For example, the lightest tau slepton could be the LSP.  Its production
rate through R--parity--conserving couplings can be determined from 
Fig.~\ref{fig:susyxseclepton}.  If $\lambda_{333}$ is the only large 
RPV coupling,
the decay $\stau\to\tau\nu_\tau$ can occur with a lifetime fixed
by $\lambda_{333}$ and $m_{\stau}$ (see Sec.~\ref{sec:r_parity}).  
For small enough $\lambda_{333}$,
this decay can occur outside the tracker, leading to the desired signal
if the $\stau$ is travelling slowly enough.

\setcounter{footnote}{0}
\subsection{Photon and \met~Signatures}
\subsubsection{An Unexpected Turn: the CDF $ee\gamma\gamma$\met~Event} 
\label{section:introeeggm}

\begin{figure}[!ht]
\vspace*{0.5cm}
\centerline{\psfig{figure=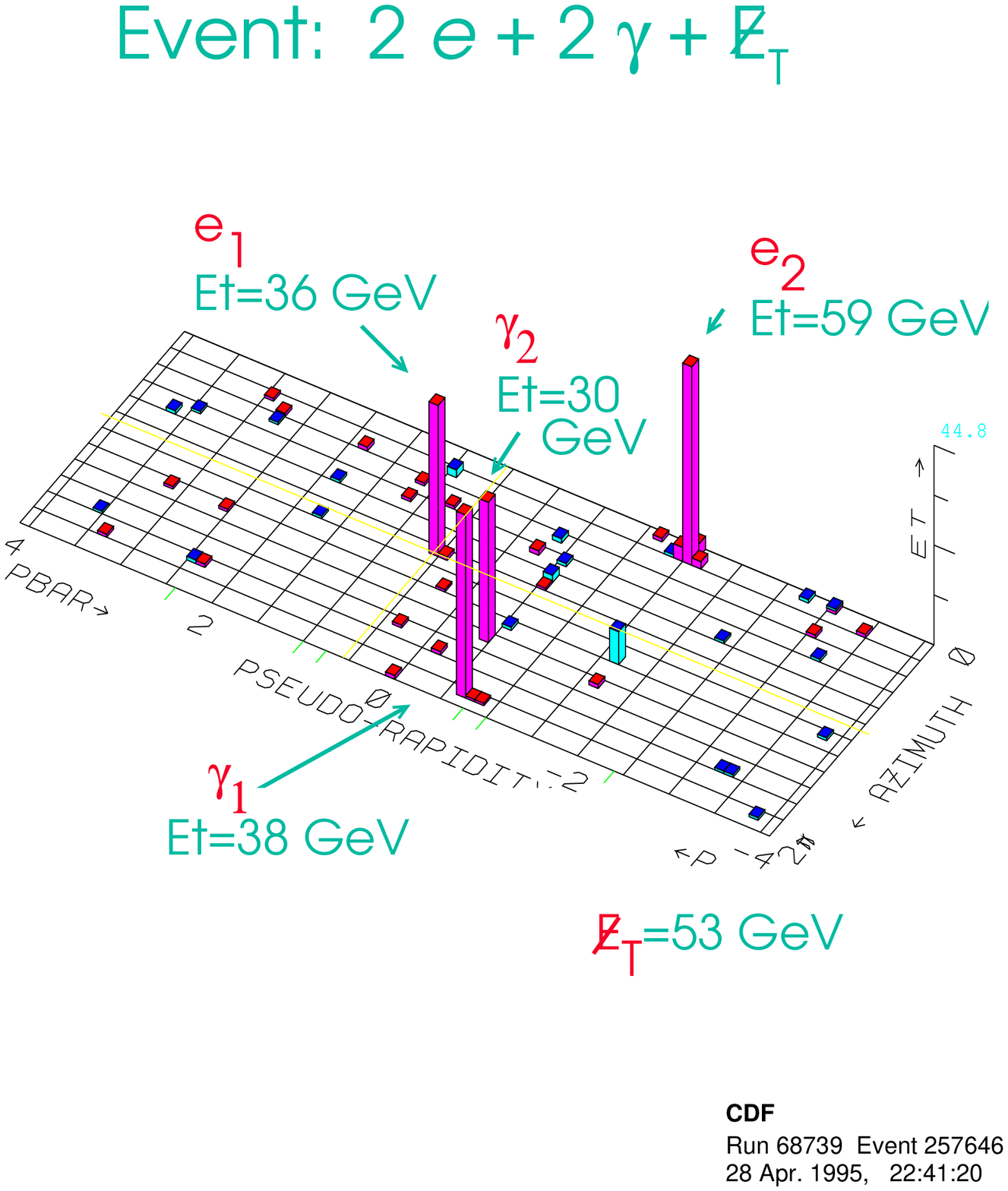,height=90mm}}
\caption{The very unusual CDF event containing two `electrons', two `photons'
and missing $E_T$. The display is the calorimeter cylinder unrolled 
into a plane.  The towers represent energy deposition, 
with the height of the tower proportional 
to $E_T$.}
\label{figure:cdfeeggmevent}
\end{figure}

Supersymmetry has so many parameters that
the full range of its allowed signatures may be hard to predict.
In April 1995, the CDF experiment recorded 
an event with a very unusual topology\,\cite{Park} which
may have SUSY interpretations.
It has four electromagnetic clusters, which pass the
typical cuts for two electrons and two photons, and \met.
A display of the event is shown in Fig.~\ref{figure:cdfeeggmevent}.  

The electron in the central region of the detector is
well--isolated and is associated with a  
track that has a $p_T$ in good agreement
with the $e^-$ hypothesis. The two photons are also well--isolated and have no
associated tracks.
The ``electron'' at large $\eta$ is more difficult to identify positively. The
associated track should only cross a part of the inner CTC where the occupancy
is too high to find the track.  Hence, its charge cannot be determined.  
The VTX, a wire
chamber surrounding the SVX but inside the CTC, measuring in the $r-z$ view,
has a track at the correct $\eta$ for the  electron hypothesis. The 
path through the cluster and
the event vertex can be
searched for tracks in the SVX, and this analysis is underway.
The probability that the event could be produced in the SM,
including the probability that one or more of the objects is fake, is being
estimated.  The preliminary results indicate that the number of expected
$\ell\ell\gamma\gamma$\met~events is many orders of magnitude 
less than one.
However, the data set was derived from over three trillion collisions,
and the probability of {\it all} signatures which would be 
considered ``rare'' must be
estimated  (an impossible task) to determine the significance of one event.

There have been two main proposals for a possible 
SUSY explanation of the event: the  Gravitino LSP and 
the Higgsino LSP model (for non--SUSY explanations, see 
Refs.~\oldcite{nonsusy}, for example).
Both proposals also 
suggest 
other signatures that should be expected within these models and which
are presented in the following.  The Tevatron
collaborations have completed some of these searches,
which are  also discussed below.

\subsubsection{Gauge--Mediated Low Energy SUSY--Breaking: Gravitino LSP}

The CDF analysis of the above one event reminded theorists of 
low-energy SUSY breaking models,\cite{thomas,n2ton1gprd,gaugetheory,ellis}  
which had long ago lost favor to SUGRA models.
In these models  the (usually ignored) 
gravitino ($\gravitino$) is very light and becomes the LSP.
The lightest
SM superpartner becomes the next--to--lightest supersymmetric particle
(NLSP), which is unstable and  decays  into its SM partner plus the
Goldstino component of the gravitino.\,\cite{Fayet}
In the simplest  gauge--mediated models, the squarks  are heavy and the
gauginos obey the unification relationship in Eq. (\ref{eq:gauginounif}).
Generically the NLSP can be a neutralino or a slepton (most plausibly a
right--handed slepton and, due to the larger Yukawa coupling,
a $\tilde{\tau}$).
If the scale of SUSY breaking is not far above the electroweak
scale ($\leq$ a few 1000 TeV), the NLSP will decay within the detector,
leading to distinctive signatures as displaced vertices or heavy charged
sleptons decaying into leptons, possibly with a kink to a minimum ionizing
track.\cite{thomas,gaugetheory,talkGM}

If a gaugino--like  neutralino is the NLSP, the only modification
to SUGRA phenomenology,
where all sparticles decay down to $\tilde{\chi}^0_1$, is that
$\tilde{\chi}^0_1$ then decays to a photon and \met. The production of
$\tilde{\chi}^0_2 \tilde{\chi}^{\pm}_1$,
$\tilde{\chi}^+_1 \tilde{\chi}^-_1$ and
$\tilde{\ell}^+_R \tilde{\ell}^-_R$ pairs, followed by 
cascade decays, leads to the final states
$WZ \gamma \gamma + \met$,  $W \ell^+ \ell^- \gamma \gamma + \met$,
$WW \gamma \gamma + \met$ and $\ell^+ \ell^- \gamma \gamma + \met$,  all with
comparable rates.\cite{gaugetheory}
A logical starting place for searches is in the inclusive
two photon and \met~channel.\cite{interest}
In particular, the CDF event can be interpreted as either
$\selectron\selectron^*$ production,\cite{thomas,ellis} followed by 
$\selectron\to e\zinol$  or
$\winoop\winoom$ production,\cite{ellis} followed by 
$\winoom\to e^-\bar\nu_e\zinol$.
If the coupling between the gravitino and matter is large enough,
then the lightest neutralino can
decay $\zinol\to\gamma\gravitino$ inside a collider detector, yielding
the desired signature of $e^+e^-\gamma\gamma$\met.
However, it follows that if one adjusts the parameters of the model to 
explain the multilepton plus photons CDF event, then a very large
rate of multijet plus multileptons plus photon(s) events is to be 
expected.\,\cite{interest,baertata}
The fact than none of these other signatures
has been detected makes the above LSP $\gravitino$ explanation of the CDF 
$ee\gamma\gamma$\met~event unlikely. Other possible explanations may, 
however, remain open.
 
Signatures of photons+\met~can point towards
models of low--energy SUSY breaking, but there are other 
possible signatures in these models.\,\cite {talkGM,Dicus}
If the NLSP is a neutralino which is mainly Higgsino--like, then
$\tilde{\chi}^0_1$ decays to the lightest Higgs boson
(or the heavy CP--even  or the CP--odd neutral Higgs bosons if they are
sufficiently light) plus a gravitino. The Higgs boson will subsequently decay into
$b \bar{b}$. Hence, the signature of 4 $b$--jets, which reconstruct 
the lightest Higgs boson
mass in pairs, plus \met~is possible.
If the NLSP is a right--handed slepton, then the decay $\tilde{\ell} \to \ell\tilde{G}$
occurs, yielding lepton pairs and \met~as final signature of slepton
pair production. The dilepton
signature will suffer from large irreducible backgrounds, but the production
and decay of heavier sparticles can give spectacular signals.
For example, the pair production of a left--handed slepton which
cascade decays into a right--handed slepton and a neutralino can yield
six leptons+\met~in the final state.
Also, since the NLSP slepton can be $\tilde{\tau}_R$,
signatures with many $\tau$ leptons are possible.

If any of the above  signatures were observed experimentally, a measurement
of the decay length of the NLSP would provide
information about the scale of supersymmetry breaking.
However, the scale of SUSY breaking might be sufficiently large
that an NLSP slepton would decay outside the detector.
In this case, heavy charged particle
pair production without missing energy could be a manifestation of
gauge--mediated low--energy SUSY--breaking models.

\subsubsection{Higgsino LSP}

The Higgsino LSP model\,\cite{n2ton1gprd} involves a region of MSSM parameter space in which the
$\zinoh$~is photino--like and the $\zinol$~is
Higgsino--like, so the
radiative decay $\zinoh\to\gamma\zinol$~dominates 
over other $\zinoh$ decay modes (see, for example,
Eq.~(\ref{eq:M2eqmu})).\,\cite{radiative_decay}
The event can be again interpreted as $(i)$ $\selectron\selectron^*$ production,
but with $\selectron\to e\zinoh$, or 
$(ii)$ $\winoop\winoom$~production, with $\winoom\to e^-\bar\nu\zinoh$,
and the subsequent radiative decay of the $\zinoh$~yielding 
the observed signature.  

In these models, photons only arise from the decay of $\zinoh$.
Other signatures involving two photons
might come from the process
$\sneutrino\sneutrino^*\to\nu\bar\nu\zinoh\zinoh$,
but there is no guarantee that the $\sneutrino$ is light enough
to produce a substantial signal.
Because the $\zinoh$ is photino--like, direct $\zinoh\zinoh$ production
is not large.  
In this model, the dominant neutralino and chargino
production processes are
$\winoop\winoom, \zinol\zinot,$ and $\winol\zinol, \winol\zinot$.
None of these involve
the direct production of $\zinoh$.
Typically, the decay $\zinot\to Z^*\zinol$
occurs, yielding no photon.
One of the next largest processes is $\winol\zinoh$,
which would produce a trilepton signature in SUGRA models,
but can produce $\ell\gamma$\met~or $jj\gamma$\met~signatures in
the Higgsino LSP model.

If the stop is light, this discussion changes, because 
the $\winol$~can
decay $\winol\to b\stop_1$, followed by $\stop_1\to c\zinol$.
The signature is then a rather distinct $\gamma bc$\met.
However, such a light stop would appear in top decays, depleting
the observed SM decays to an unacceptable level.  This is only
true, though, if there are no other sources of top quark production
from SUSY, which there obviously can be.  Surprisingly, such models
can be constructed that are in accord with the present SUSY 
limits.\cite{kanegtm}  If the gluinos are heavy enough so that
$\gluino\to t\stop_1^*$, or $\to \bar t\stop_1$, and gluino production
is further fed by squark decay $\squark\to q\gluino$, then one can
compensate for the lost top quarks in SUSY decay modes.
This leads to more sources of $\gamma bc$\met~events than just
$\winol\zinoh$ events, as well as other
signatures.

\subsubsection{Inclusive Two Photons and \met~Signatures}

The generic $\gamma\gamma$\met~$+X$ signature has 
no significant background from real photons.  The main backgrounds are 
caused by jets and electrons faking photons.   
The SM production of $W(\to e\nu)\gamma$ plus jets 
can fake some of the signatures if the electron is misidentified as a photon.
These events have a \met~spectrum typical of $W$ events, peaked
at about $M_W/2\simeq$ 40 GeV, with a long tail to high \met.    
The dominant instrumental background, however, is from di--jet and 
$\gamma+$jet production, where the large production cross section
overcomes the small probability 
($\simeq 10^{-4}-10^{-3}$) that a jet fakes
a photon. 

\begin{figure}[!ht]
\centerline{
\hbox{\hbox{\psfig{figure=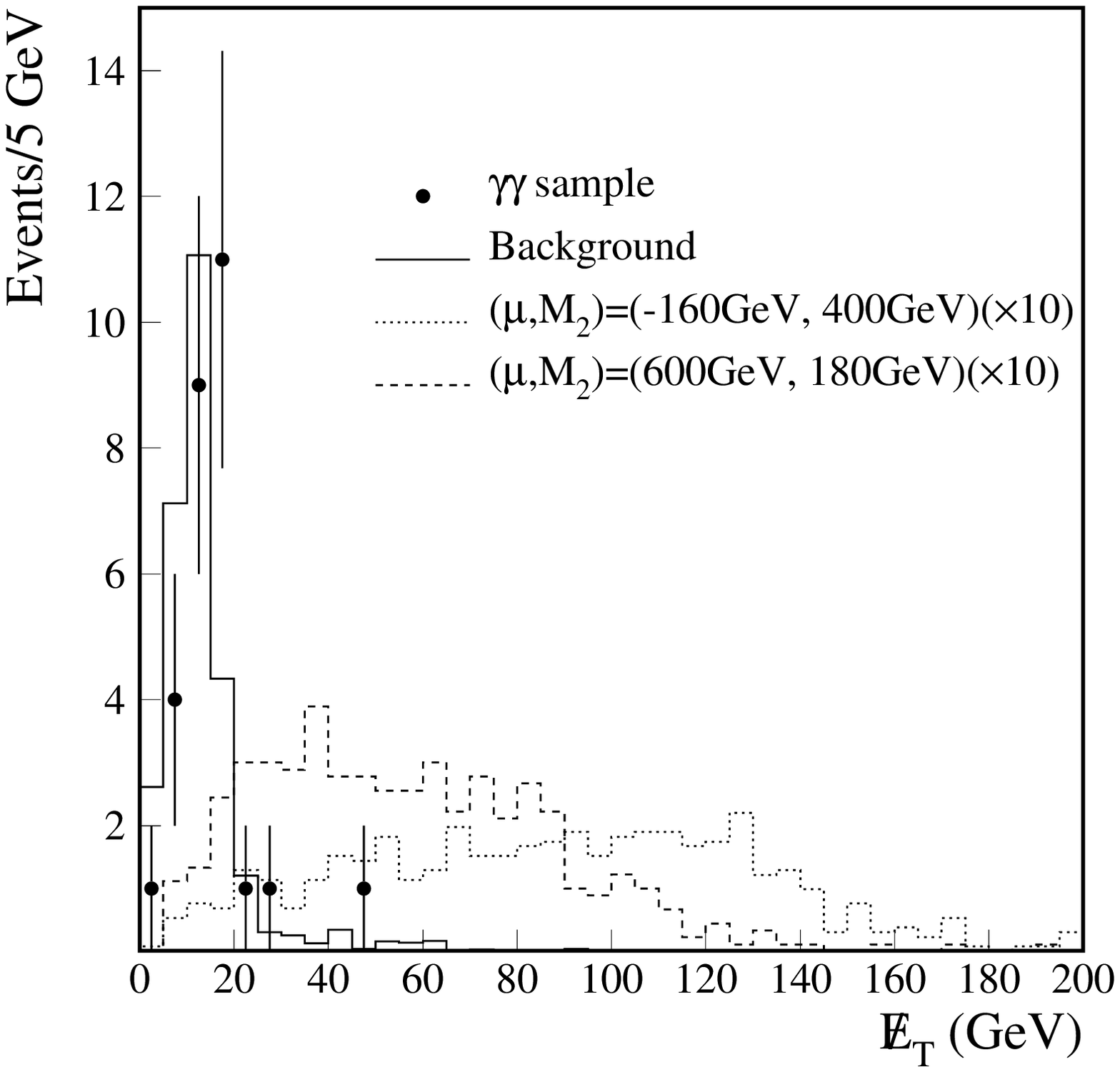,height=60mm}}
\hbox{\psfig{figure=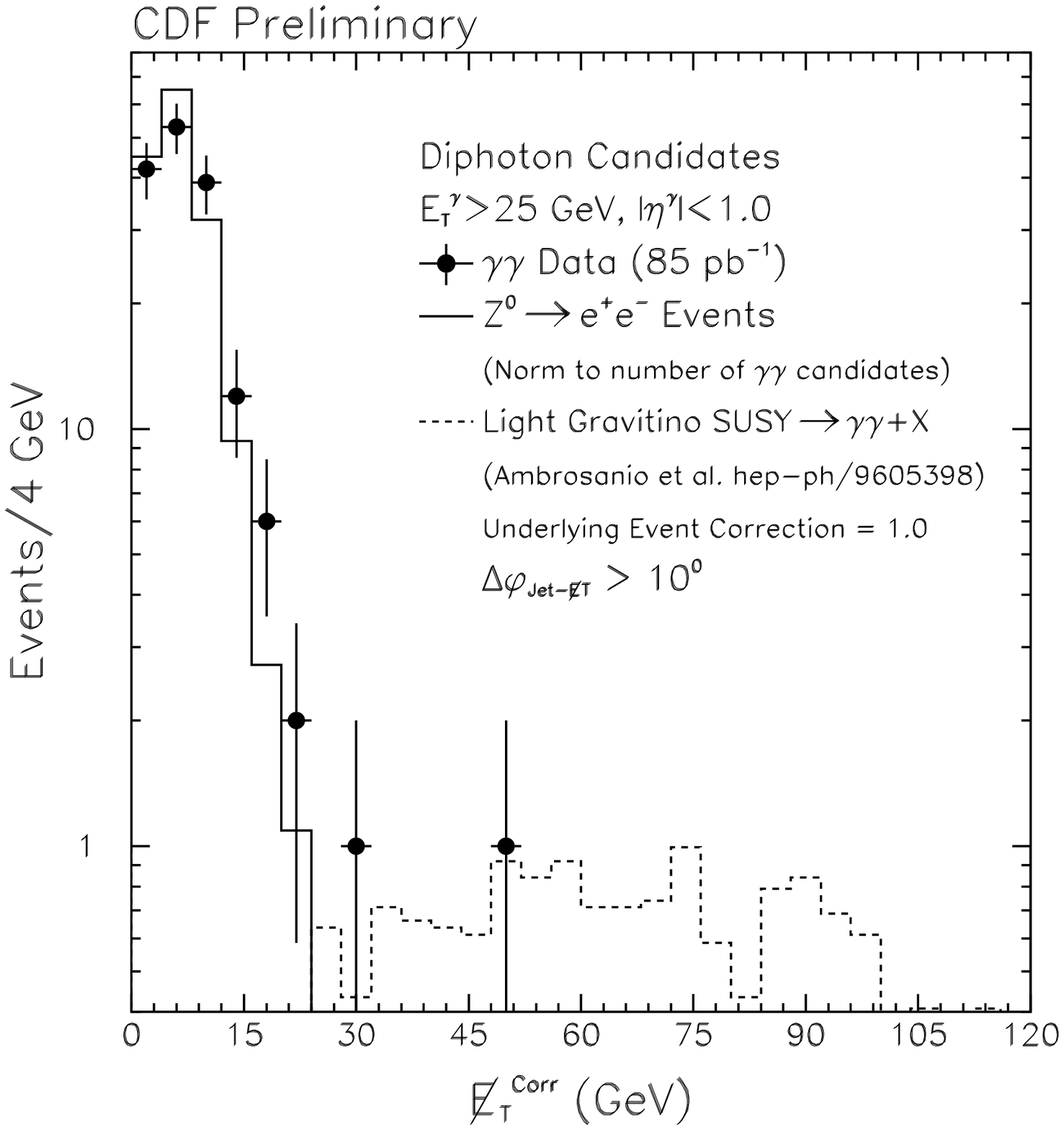,height=60mm}}
}
}
\caption{(Left) The 
\met~spectra in the \D0~search for events with 2 photons, one
with \et $>$ 25 \gev, 
the other with \et$>$ 12 \gev.\protect\cite{d0_diphoton}
The points are the data, the solid line is the estimated background
from di--jet events and direct photon events.  The dotted lines
are for gaugino production within gauge--mediated models using
the parameters listed and $M_1\simeq 2 M_2$.
(Right) The CDF \met~spectrum for events with two  central
photons with  $E_T > 25$~GeV.  
Events which have any jet with $E_T> 10$~GeV pointing 
within 10 degrees in azimuth
of the \met~are removed.
The solid histogram shows the resolution from
the $Z\to e^+e^-$ control sample.
The dashed line shows the expected distribution from 
all SUSY production in a model~\protect\cite{interest} 
with $M_2=225$ GeV, $\mu=300$ GeV, $\tan\beta=1.5$,
and $M_{\squark}=300$ GeV.
}
\label{fig:d0gamgam}
\end{figure}

Figure~\ref{fig:d0gamgam}
shows the \met~distributions from \D0~(left) and CDF (right)
diphoton events \cite{CDF_stop_rlc,d0_diphoton} after imposing the
selection criteria given in Table~\ref{tab:tab2gam}.
For the \D0~analysis, the shape of the \met~spectra agrees well 
with backgrounds containing
two electromagnetic--like clusters, where 
at least one of the two clusters fails the photon selection criteria.  
Two events satisfy all selection criteria, with a predicted
background, dominated by jets faking photons, of 2.3 $\pm$0.9 events.
For the CDF analysis,
the shape of the \met~distribution is in good agreement with the
resolution of the $Z\to e^+e^-$ control sample.
The event on the tail in \met~is the ``$ee\gamma\gamma$\met'' event.  
If the source of this event is an
anomalously large $WW\gamma\gamma$ production cross section that 
yields one event in $\ell\ell\gamma\gamma$\met, 
CDF would expect dozens of events with two photons and 
four jets.  However, the jet multiplicity spectrum in
diphoton events is well--modeled by 
an exponential, and there are no diphoton events with 3 or 4 jets.
As mentioned before, events 
with diphotons, jets and \met~can be signatures of 
gauge--mediated low--energy supersymmetry--breaking models.

\begin{table}[!ht]
\centering
\caption{Selection criteria for $\gamma\gamma+$\met$+X$ searches} 
\begin{tabular}{|l|l|l|} \hline\hline
Quantity    &  \D0  & CDF   \\ \hline \hline
$E_T^{\gamma_1}, E_T^{\gamma_2}$ & $>20, 12$ GeV & $>25, 25$ GeV \\
\hline
$|\eta^\gamma|$   & $<1.2$ or between 1.5 and 2.0 & $<1.1$ \\ \hline
\met              & $>25$ \gev  & $>35$ \gev \\  \hline 
$\Delta\phi$ between         & N.A. & $>10^\circ$ \\ 
\met~and nearest jet         &      &            \\ \hline \hline
\end{tabular}
\label{tab:tab2gam}
\end{table}

\begin{figure}[!ht]
\centerline{
\hbox{
\hbox{\psfig{figure=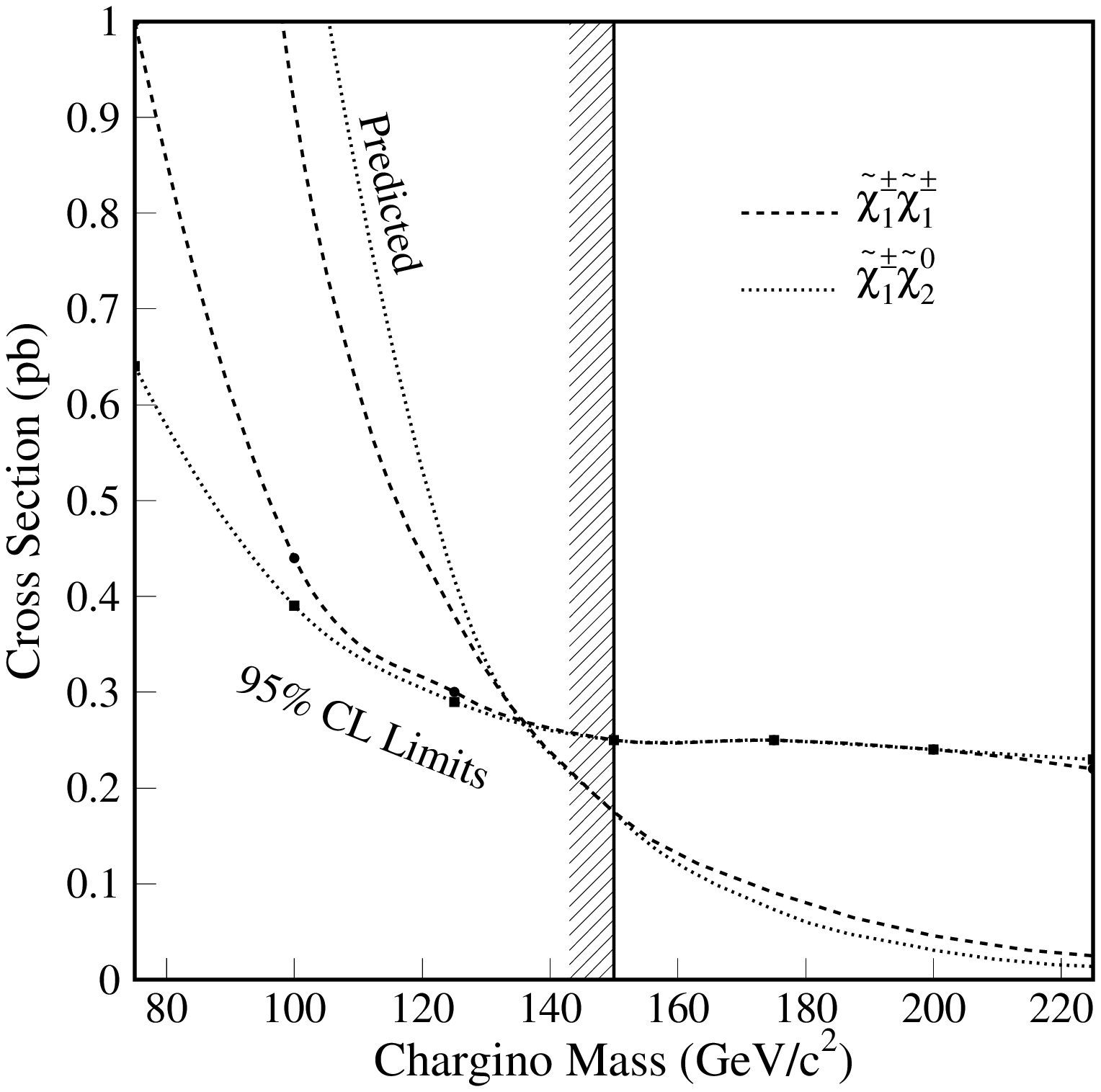,height=50mm}}
\hbox{\psfig{figure=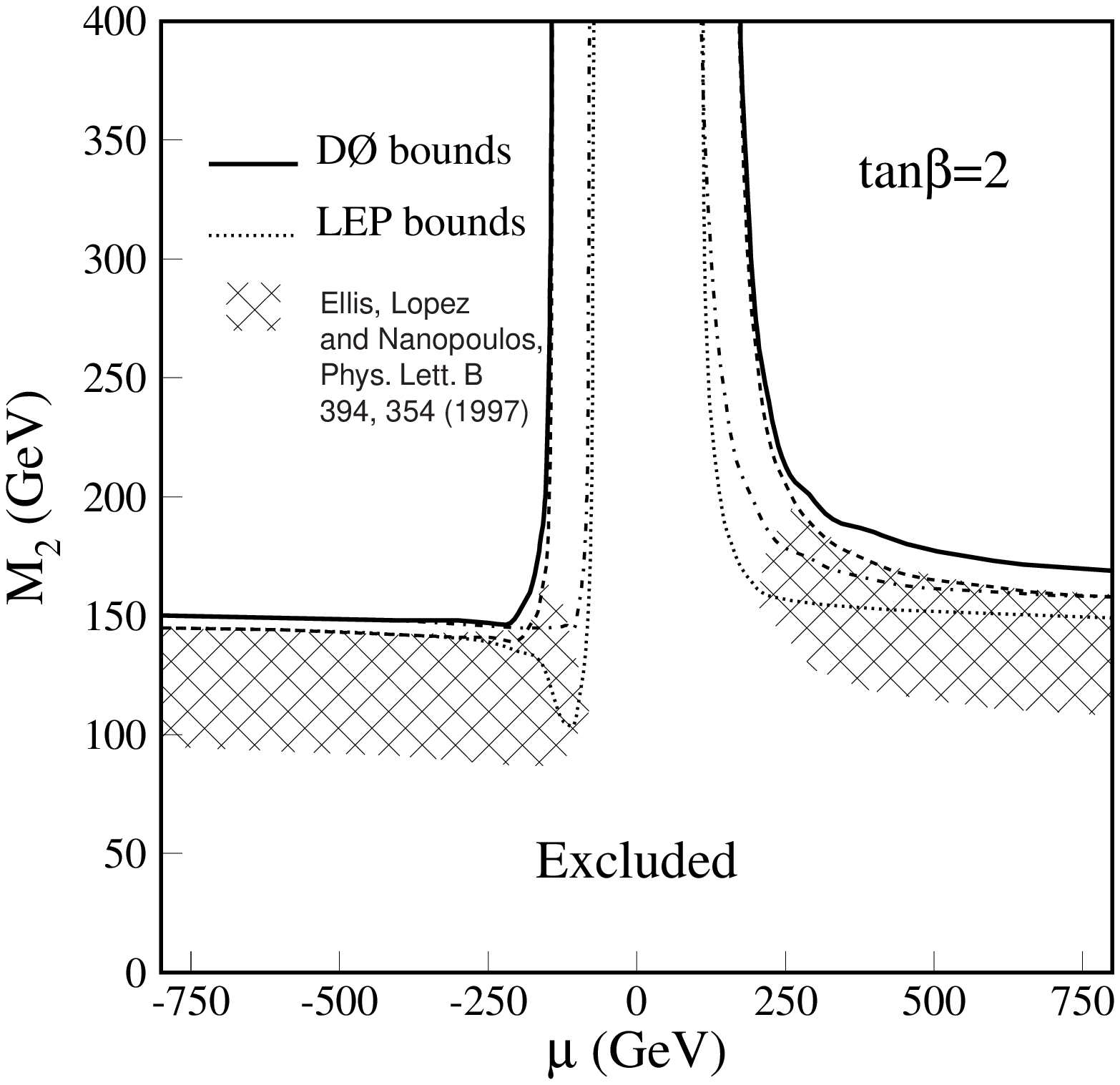,height=50mm}}
}
}
\caption{(Left) The
\D0~cross section limit on $\winol\winol$ and $\winol\zinoh$ 
production, assuming $M_{\winol}\approx 2 M_{\zinol}$ and
BR($\zinol \rightarrow \gamma \gravitino$) = 100\%.
The top dotted (dashed) curve is the cross section from {\tt PYTHIA} for
$\winol\zinoh$ ($\winoop\winoom$) production.   The bottom dotted
(dashed) curve is the cross section limit from the 
\D0~collaboration\,\protect\cite{d0_diphoton} 
on $\winol\zinoh$  ($\winoop\winoom$) production.
The vertical, hatched line marks the 95\% C.L. 
lower limit on the lightest chargino mass from considering
all chargino and neutralino pair production processes and all 
values of $\mu$.
(Right) The limits on the parameters $M_2$ and
$\mu$ in gauge--mediated models based on 
{\tt PYTHIA} for
\tanb=2 and $M_{\squark}$=800 \gev.\,\protect\cite{d0_diphoton}  
The hatched area is the region proposed\,\protect\cite{ellis} to explain
the CDF ee$\gamma\gamma$\met~event.
The solid line shows the \D0~bounds.
The long--dashed line shows a contour with $M_{\winol}=150$ \gev~and
the dash--dotted line shows a contour with $M_{\zinol}=75$ \gev.  
The dotted lines show an interpretation of preliminary LEP results 
at an energy of 161 GeV.}
\label{fig:d0gglimit}
\end{figure}

\D0~presents limits\,\cite{d0_diphoton}  
in the framework of the 
Gravitino LSP scenario by considering neutralino and chargino
pair production.  
Assuming $M_2 \simeq 2 M_1$ and large values of $m_{\squark}$,
the signatures are a function of only $M_2$, $\mu,$ and $\tan\beta$.
Event rates are predicted using {\tt PYTHIA}.\,\cite{spythia}
Figure~\ref{fig:d0gglimit} shows the limit
on the cross section for $\winol \winol$ and $\winol \zinoh$
production as a function of the $\winol$ mass when $|\mu|$ is large
and thus
the $\winol$ mass is approximately
twice the $\zinol$ mass.
The figure also shows,
more generally, the excluded region in the $M_2$--$\mu$ plane
($\mu<0$ gives larger $\winol,\zinoh-\zinol$ mass splittings, small $|\mu|$ 
means $\winol, \zinol,$ and $\zinoh$ are more Higgsino--like),
 along with a prediction
for the region that might explain the CDF 
$ee\gamma\gamma$\met~event as chargino pair production.
The latter explanation requires $100$~GeV $< M_{\winol} <150$~GeV 
with $M_{\zinol} < 0.6 M_{\winol}$ to produce one
event with a reasonable probability.\,\cite{ellis}

As can be seen from Fig.~\ref{fig:d0gglimit}, the cross section
limit is typically 0.24 pb for either $\winoop\winoom$ or
$\winol\zinoh$ production.  By combining all chargino and
neutralino pair production processes,
a $\winol$ with mass below 150~GeV is excluded.
Hence, to keep the
chargino interpretation of the $ee\gamma\gamma$\met~event,
it is necessary to expand on the analysis of Ref.~\oldcite{ellis}.
The chargino
mass limit is much higher than in SUGRA models, because of the 100\%
branching fraction for the decay $\zinol\to\gamma\gravitino$ and
the high detectability of the photon and \met.   
The result eliminates the possibility of
observing signatures of this particular model at LEP200.
The \met~cut needed to control QCD backgrounds makes the analysis
sensitive to the mass splittings between $\zinol$ and 
$\winol$ or $\zinoh$.   However, the simplest models predict unification
mass relations between the gauginos, which thus gives acceptable 
mass splittings.

\D0~also has a limit on the cross section for
$\selectron \selectron^* \rightarrow e^- e^+  \zinoh \zinoh$,
$\sneutrino \sneutrino^* \rightarrow \nu \bar\nu  \zinoh \zinoh$, and
$\zinoh \zinoh \rightarrow \gamma\gamma\zinol\zinol$ using the
same analysis as for the  Gravitino LSP search.  Such
signatures might also be expected in Higgsino LSP models.
The limit on the cross section for such processes is about
0.35~pb for $M_{\zinoh}-M_{\zinol}> 30$~GeV, which is close
to the maximum cross section predicted in these models.

\subsubsection{Single Photon, Heavy Flavor, and \met}
\begin{figure}[!ht]
\centerline{\psfig{figure=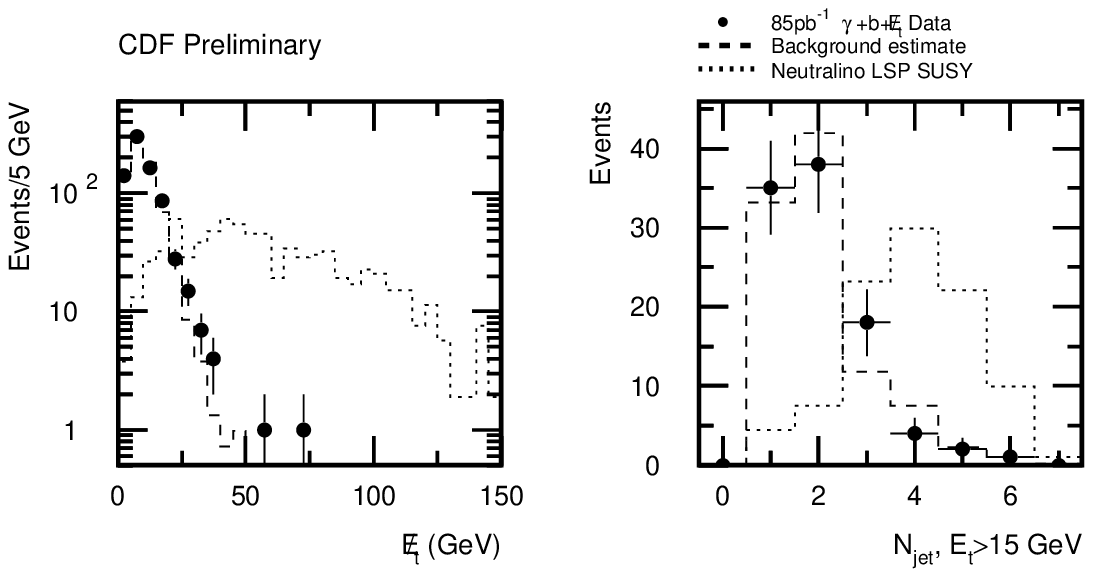,height=60mm}}
\caption{CDF results for the \met~(left) and the jet multiplicity 
with $E_T>15$~GeV (right) in events
with a photon and a SVX $b$--tag.  The search is for the signature
$b\gamma$\met~in a scenario where $\zinoh\to\gamma\zinol$
and the stop is light.
The jet multiplicity histogram
is made by requiring \met$>$20~GeV.  The SUSY model is normalized to
the area of the data histogram -- this is scaling by a factor of 100 for the
\met~histogram and a factor of 10 for the $n_{\rm jet}$ histogram.
The SUSY model has a Higgsino LSP \protect\cite{kanegtm}
generated with {\tt PYTHIA} 6.1.
}
\label{figure:cdfbcgmet}
\end{figure}

CDF has searched for the signature $\gamma bc\met$,
as predicted in Higgsino LSP models with a light stop.\cite{kanegtm}
The data sample of 85~pb$^{-1}$ contains events with 
an isolated photon with $E_T^\gamma>$ 25 GeV and a jet with an SVX $b$--tag.
The \met~spectrum of these events can be seen in  
Fig.~\ref{figure:cdfbcgmet}.  After requiring \met$>$20~GeV,
98 events remain.\cite{CDF_stop_rlc}

The estimated background to the 98 events is $77\pm 23\pm 20$ events.
The shape is consistent with background.
About 60\% of the background is due to jets faking photons, 13\% to 
real photons and fake $b$--tags, and the remainder to SM
$\gamma b\bar{b}$ and $\gamma c\bar{c}$ production; all of these
sources require fake \met.
When the \met~cut is increased 
to 40 GeV, 2 events remain.  More than 6.43
events of anomalous production in this topology is excluded.  

\begin{table}[!ht]
\centering
\caption{Summary of the 85~\ipb~data sample for the CDF 
$\gamma b$\met~search.  Limits are set using all cuts which results in two events.}
\begin{tabular}{|l|l|r|} \hline\hline
Quantity                  &  Cut        &  Cumulative Number of Events \\ \hline
$E_T^\gamma$, ID cuts     &  $>25$~GeV  & 511335 \\ \hline
SVX $b$--tag              &  $\ge 1$    &   1487 \\ \hline
$E_T^{b}$,  $|\eta|<2.0$ &  $>30$~GeV  & 1175 \\ \hline
\met                      &  $>20$~GeV  & 98 \\ \hline
\met, $\Delta\phi(\gamma-$\met$)<2.93$ & $>40$ GeV & 2 \\ \hline\hline
\end{tabular}
\label{gtmfilter}
\end{table}

The efficiency used in the limits is derived from 
a ``baseline'' model
with $M_{\zinol}=40$~GeV, $M_{\zinoh}=70$~GeV, $m_{\stop_1}=60$~GeV,
$m_{\squark}=250$~GeV, and $M_{\gluino}=225$~GeV.\footnote{This
analysis predates LEP results which exclude this example.}
The distribution of the number of jets in the data 
is shown in Fig.~\ref{figure:cdfbcgmet} compared to that
expected from backgrounds and the SUSY model (scaled $\times 10$).
There are more jets expected in the SUSY model than the data
indicates because of the hard kinematics of squark and gluino decays.
The baseline model
predicts 6.65 events, so this model is excluded (at
the 95\% C.L.).  This result does not rule out 
the Higgsino \LSP~model in general, only one version with a 
fairly light mass spectrum.
A more general limit can be set by 
holding the lighter sparticle masses constant and varying
the squark and gluino masses.  In this case squarks and gluinos
less than 200~GeV and 225~GeV, respectively, are excluded.

\setcounter{footnote}{0}
\subsection{Other Anomalies}
\label{anomalies}
There are other anomalies in the current data beyond the ``$ee\gamma\gamma$\met''
event. These are, so far, either single,  rare events  or
discrepancies on the tails of distributions where  statistics are low and
backgrounds difficult to calculate. In addition, there is the problem of
calculating probabilities for ``anomalies'' {\it a posteriori}.  
The expected number of events in any one channel from SUSY is usually
small with the present integrated luminosity.
New physics will most likely show up as a few events on the tails
of SM distributions.   Since there are many
potential SUSY signatures,
one can only
follow a strategy of systematically analyzing {\it all} 
high--mass channels and 
looking for discrepancies on the tails of distributions. The
single events  such as the ``$ee\gamma\gamma$\met'' event  have been useful as
``guideposts''  indicating promising new channels, 
such as the $\gamma bj$\met~channel described above.  
It is still possible that a sensible
picture of these events will emerge from the Run I data 
when a complete survey of all channels is 
completed using both detectors. At the very least, 
this is an important exercise
for preparing the Run II analyses.

\subsubsection{Top Dilepton Events}

As discussed earlier, the signature of dileptons+2 jets+\met~is a promising
SUSY search channel (see Sec.~\ref{sec:dileptons}).  However, such events
would also be a background to the SM top quark search using dileptons.
The consistency of this dilepton sample with that expected from $t\bar t$
production has been the subject of intense 
investigation.\cite{top_dilepton_disc} 
There are a number of
peculiarities, none by themselves statistically significant at a level required
to claim new physics. However, there are several events that have low
probabilities of being from top decay or any other SM 
process.\cite{barnett}  Such events should be
taken seriously as potential SUSY candidates. 

The most interesting of the anomalous CDF 
events\,\cite{Park,Marcus_PhD} is Event
129896 of Run 67581, which has  {\it three}, clean, isolated, 
high--\pt~leptons, large \met, and a high--\et~jet.  
In addition, the most energetic of the leptons
is a positron with $E_T\simeq$ 200 GeV, significantly larger than
is typical  for top events (0.06  $\pm$ 0.02 events are expected).  The 
corrected \met~is over 100 GeV, also large for  top decay ( 0.6 $\pm$ 0.1
events are expected). The event  contains a jet with $E_T\simeq$ 100 GeV; 
the total transverse energy
plus \met~is about 450 GeV. 
The  other two leptons are an electron with $E_T=27$ GeV
and a muon ($\mu^-$) with $p_T=27$ \gevc. 
The invariant mass of the $e^+e^-$ pair is 130 GeV, well
away from $M_Z$; the pair has very high \pt. 
In the SM top quark analysis, 
the event is classified
as a dilepton+2 jet event, because the lower \et~electron 
fails the fiducial cut by  4 mm and is thus defined to be a jet; 
however, the electron passes all other
standard electron criteria and is a ``golden'' electron in all other
ways.\footnote{In the top quark analysis the fiducial volume was 
conservatively chosen to be the same as for the 
precision  electroweak measurement of the ratio of $W$ to $Z$ cross sections.
The 4 mm miss does not affect the electron identification.}
The kinematics of the event are unusual: the invariant
mass $M_{e\mu j}$ is on the order of $m_t$,
while the other hemisphere contains only the
high--\et~positron.  The three isolated leptons and the kinematics  make the
event unlikely to come from SM top production and decay.  The event is a 
high--mass trilepton
+\met~event, and is consequently a good SUSY candidate.\cite{barnett} 

\begin{figure}[!ht]
\center
\centerline{\psfig{figure=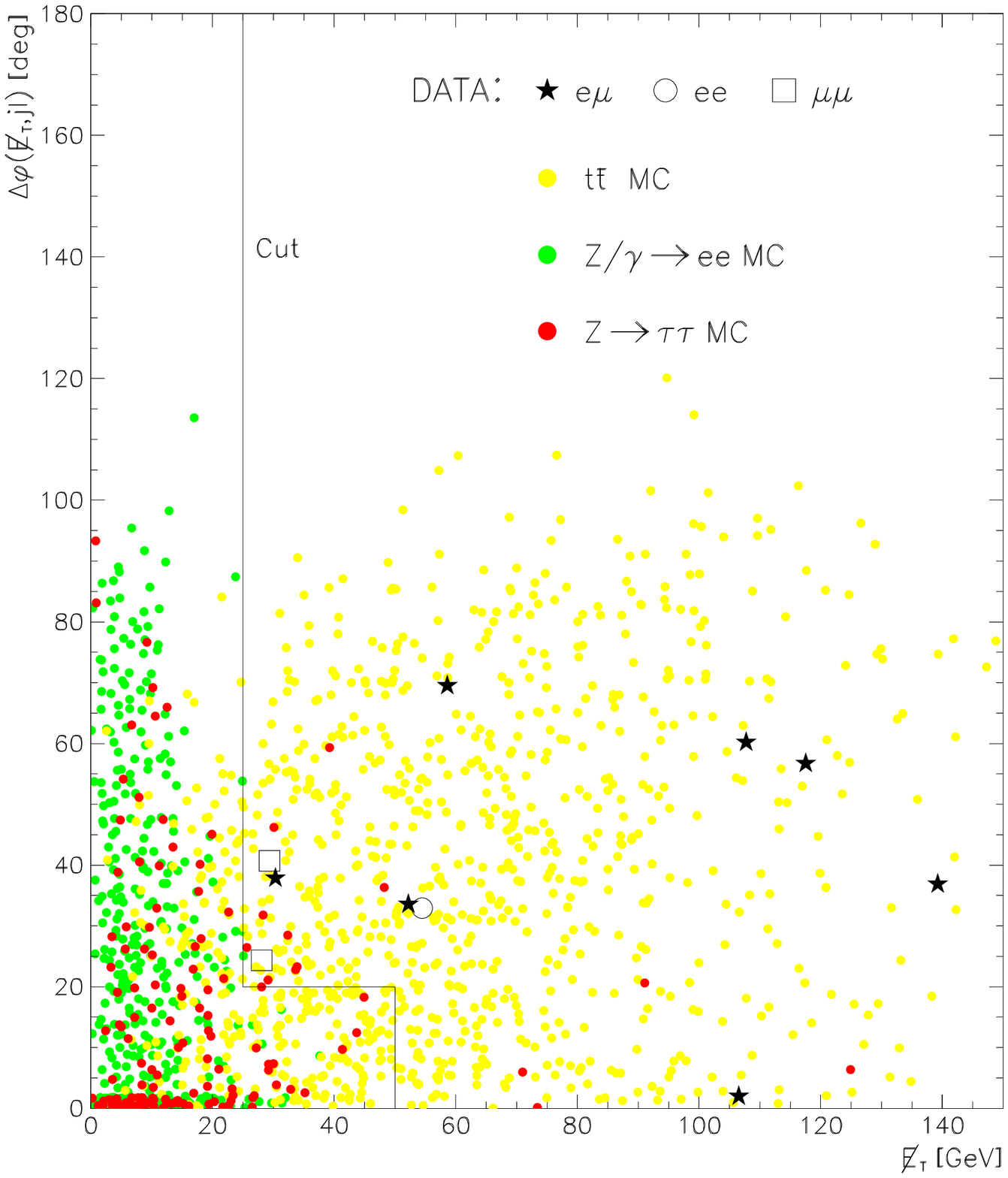,height=60mm}
\psfig{figure=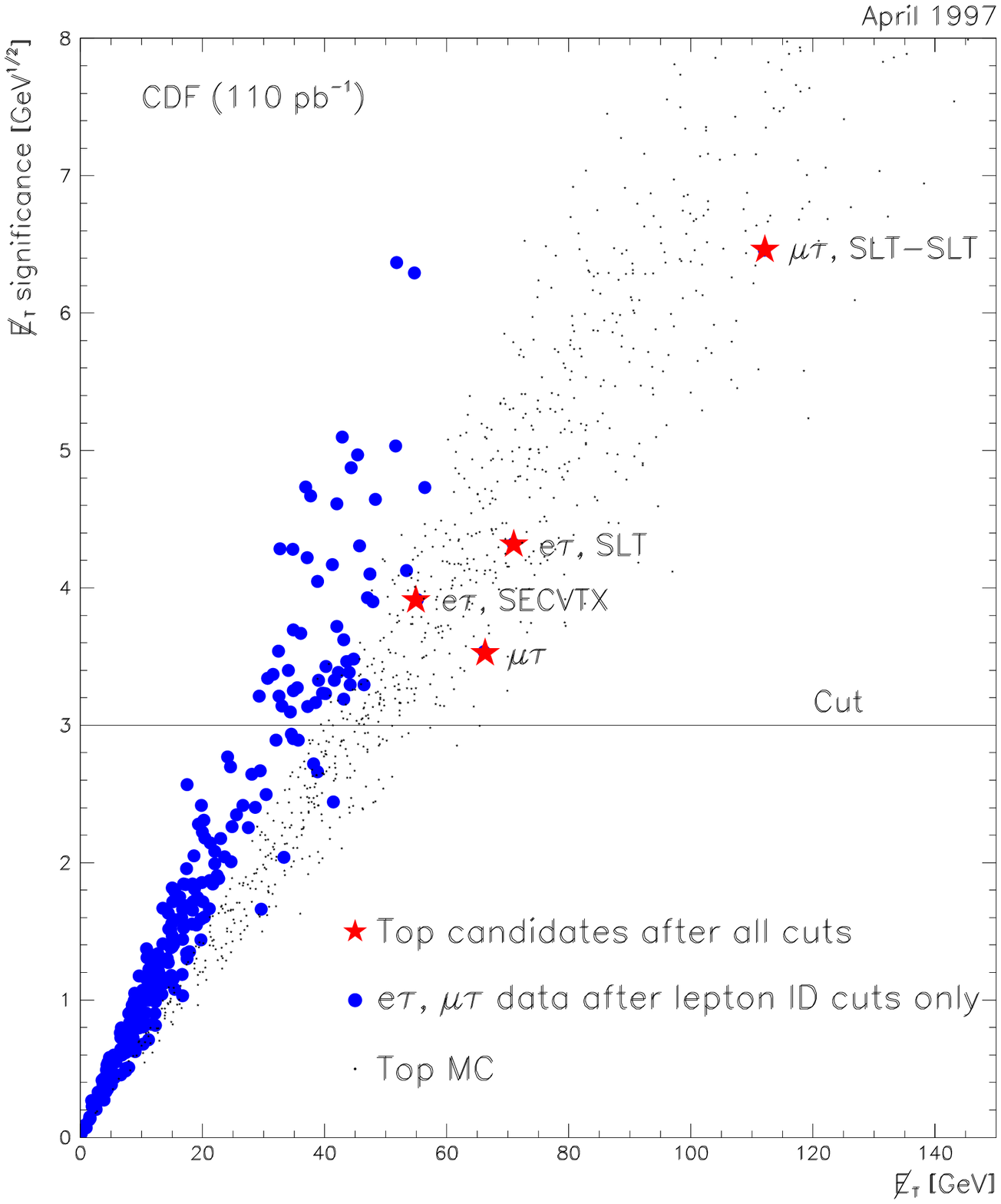,height=60mm}
}
\caption{
(Left) Scatterplot of the angle $\Delta\phi$(\met,$\ell j$) between the
corrected \met~and the closest lepton or jet versus corrected \met~for the
$ee$, $\mu\mu$, and $e\mu$ candidate events, compared with the expected
distributions for \ttbar~and background. Background and top contributions are
{\em not} normalized to the expected number of events.
(Right) The distribution of \met~significance versus 
\met~for events with a primary lepton and
a tau candidate (the slope of the data and background is
different, because the background is dominated by QCD)
in the CDF data compared with the \ttbar~Monte Carlo. Three of
the four final candidate events (stars) have $b$--tagged jets.
}
\label{fig:CDF_dil_kin}
\end{figure}

Other discrepancies in the top dilepton sample involve the
kinematics.  
Some of the
anomalous behavior in the kinematics can be seen in 
Fig.~\ref{fig:CDF_dil_kin} (left),
which shows 
\met~versus $\Delta \phi$ between the \met~and the
nearest jet or lepton.\cite{Marcus_PhD,tauprl}
Also shown is the distribution expected from
Monte Carlo $t\bar t$ events, but corresponding to 100 times
the luminosity.  There are
several events out in regions less populated by top quark events
(one is the trilepton event). 
Figure~\ref{fig:CDF_dil_kin} (right) shows the distribution in 
\met~significance~\cite{Marcus_PhD} 
vs \met~for 
the CDF tau--lepton top sample. None of these latter discrepancies is at a
statistical level that is significant; these will be channels of great interest
in Run II.

\section{Conclusions}
\label{Conclusions}
As can be seen from Table \ref{tab:summary}, there has been a large
effort in SUSY searches at the Tevatron.  
However, given the wide range of possible experimental signatures
in the minimal Supersymmetric extension of the Standard Model, there
is still work in progress and much to be done.  
Many Run I analyses are under way.

Our quantitative conclusions on Run I
are reflected in the Figures and Tables
of this review; here we will add a few more
general qualitative observations:
\begin{enumerate}

\item A systematic exploration of
signatures and channels is just starting.
In addition, the detectors have not yet
been exploited fully; for example, better $c$--tagging and 
dijet resolution to investigate final states with reconstructed $W$ and $Z$ 
bosons may be possible. 
These tools will allow the study of new channels.

\item There are some events involving leptons and/or photons 
that are provocative, and can be ``guideposts'' for Run II and further Run I 
analyses.

\item  There is a substantial need for theorists and experimentalists to
work together to understand better how to derive and present limits from
the Tevatron. 
We should move on parallel paths toward more 
``model--independent'' predictions and limits 
({\it e.g.} presenting 
plots of cross section versus experimentally measured quantities
like thresholds),
and confront specific models in ways
that allow the two experiments to compare their results.

\item The  analyses so far are luminosity limited: the reach of the
searches is just entering the interesting regions.  
\end{enumerate}

In Run II, two upgraded detectors at the Tevatron will collect more 
data at a higher energy of 2 TeV.  The nominal integrated luminosity
is 2 fb$^{-1}$, with a possible extension to 10 or even 30 fb$^{-1}$.
The production cross sections for heavy
sparticles will increase significantly with the higher energy.
Chargino and 
neutralino searches, as well as squark and gluino searches, will cover a 
wide range of SUSY parameter space in Run II.
Most importantly, by extending Run II up to an integrated luminosity
of about 20 fb$^{-1}$ and combining search channels, the Tevatron can
perform a crucial test of the MSSM Higgs boson sector.

The experience gained from Run I analyses will greatly increase
the quality of the Run II searches.\,\cite{tev2000,snowmass}
New triggering capabilities will open up previously inaccessible
channels, particularly those involving $\tau$'s and heavy flavor.
Increased $b$--tagging efficiency and \met~resolution 
will enhance many analyses.
A factor of 20 or more data combined with improved detector capabilities
makes the next Run at the Tevatron an exciting prospect.

\section*{Acknowledgments}
The authors would like to thank  the following people for useful
discussions and comments: 

Howard Baer,
Andy Beretvas,
Jeff Berryhill,
Brendan Bevensee,
Sue Blessing, 
Amber Boehnlein,
Dhiman Chakraborty,
Piotr Chankowski,
Max Chertok,
Dan Claes,
Regina Demina,
James Done,
Eric Flattum, 
Carla Grosso--Pilcher,
John Hobbs, 
Marcus Hohlmann,
Teruki Kamon,
Stefan Lammel,
Adam Lyon, 
Doug Norman, 
Marc Paterno,
Stefan Pokorski, 
Jianming Qian, 
Aurore Savoy-Navarro,
H.C. Shankar, 
Michael Spira,
David Stuart,
Benn Tannenbaum,
Xerxes Tata,
Dave Toback,
Carlos Wagner,
Noah Whiteman,
Peter Wilson
and 
Peter Zerwas.

Part of this manuscript
was completed at the Aspen Center for Physics.

\clearpage
\appendix
\section{Appendix A}
\subsection{Typical Decay Modes of Supersymmetric Particles. }
\begin{table}[!ht]
\centering
\caption{Typical final states from sparticle decay, assuming
$\zinog,\winog,\slepton,\sneutrino <
\squark(\ne \stop,\sbottom),\gluino$.
HLSP denotes models with a Higgsino LSP and GLSP denotes models with 
a Gravitino LSP.  Event signatures from sparticle pair production
can be constructed by combining two decays.}
\begin{tabular}{|l|l|l|c|} \hline\hline
Particle    &  Intermediate State   &  Final State & Comment \\ \hline
$\zinog_i$  &                       &  $\to$ \met           &    \\
            &                       &  $\to\ell\bar\ell$ \met & \\
            &                       &  $\to jj$ \met   &        \\
            &                       &  $\to \gamma$ \met & HLSP, GLSP \\ 
            &  $\to t\stop^*$       &  $\to bW\bar c$ \met  & \lstop \\
            &  $\cdots$                 &  $\to bW\bar b\ell$ \met & \\
            &  $\cdots$                 &  $\to bW\bar b jj$ \met &  \\
            &  $\to b\sbottom^*$    &  $\to b\bar b$ \met & \\ \hline
$\winog_i$  &                       &  $\to\ell$ \met  &     \\
            &                       &  $\to jj$ \met   &     \\
            &                       &  $\to\ell\gamma$ \met  & HLSP, GLSP \\
            &                       &  $\to jj\gamma$ \met   & HLSP, GLSP \\
            &  $\to b\stop^*$       &  $\to b\bar c$ \met  & \lstop \\
            &  $\cdots$                 &  $\to b\bar b\ell$ \met &   \\
            &  $\cdots$                 &  $\to b\bar b jj$ \met  &   \\
            &  $\to t\sbottom^*$    &  $\to bW\bar b$ \met    &   \\  \hline
$\slepton$  &  $\to\ell\zinol$      &  $\to\ell$ \met         &   \\
            &  $\to\ell\zinoh$      & $\to\ell$ \met  &  \\
            &  $\cdots$  &  $\to\ell\ell^{'}\bar\ell^{'}$ \met & \\
            &  $\cdots$                 &  $\to\ell jj$ \met   & \\
            &                 &  $\to\ell\gamma$ \met & HLSP, GLSP\\
            &  $\to \nu\winol$      &  $\to\ell$ \met      & \\
            &  $\cdots$                 &  $\to jj$ \met       & \\
            &  $\cdots$                 &  $\to b\bar c$ \met & \lstop \\ 
\hline\hline
\end{tabular}
\end{table}
\clearpage
\subsection{Typical Decay Modes of Supersymmetric Particles (continued,1).}
\begin{table}[!ht]
\centering
\caption{Typical final states from sparticle decays (continued,1).}
\begin{tabular}{|l|l|l|c|} \hline\hline
Particle    &  Intermediate State   &  Final State & Comment \\ \hline
$\sneutrino$ &  $\to\nu\zinol$       &  $\to$ \met  & \\
            &  $\to\nu\zinoh$       &  $\to$ \met  & \\
            &  $\cdots$                 &  $\to\ell^{'}\bar\ell^{'}$ \met & \\
            &  $\cdots$                 &  $\to jj$ \met & \\
            &                       &  $\to\gamma$ \met  & HLSP, GLSP \\ 
            &  $\to\ell\winol$      &  $\to\ell\bar\ell^{'}$ \met & \\
            &  $\cdots$                 &  $\to\ell jj$ \met & \\
            &  $\cdots$                 &  $\to\ell b\bar c$ \met & \lstop \\ \hline
$\stop$     &  $\to c\zinol$        &  $\to c$ \met & \lstop \\
            &  $\to b\winol$        &  $\to b\ell$ \met    &   \\
            &  $\cdots$                 &  $\to bjj$ \met &   \\
            &  $\to t\zinol$        &  $\to bW$ \met  &
$M_{\stop}>m_t+M_{\zinol}$  \\
            &  $\to \sbottom W$     &  $\to bW$ \met  &   \\ \hline
$\sbottom$  &  $\to b\zinol$        &  $\to b$ \met   &   \\
            &  $\to b\zinoh$        &  $\to b$ \met   &   \\
            &  $\cdots$                 &  $\to b\ell\bar\ell$ \met  &  \\
            &  $\cdots$                 &  $\to bjj$ \met  &  \\
            &  $\to t\winol$        &  $\to bW\ell$ \met & \\
            &  $\cdots$                 &  $\to bWjj$ \met   & \\
            &  $\cdots$                 &  $\to bW\bar b c$ \met & \lstop \\
            &  $\to \stop W$        &  $\to cW$ \met     & \lstop \\
            &  $\cdots$                 &  $\to b\ell W$ \met     &       \\
            &  $\cdots$                 &  $\to bjjW$ \met        &       \\ 
\hline\hline
\end{tabular}
\end{table}
\clearpage
\subsection{Typical Decay Modes of Supersymmetric Particles (continued,2).}
\begin{table}[!ht]
\centering
\caption{Typical final states from sparticle decays (continued,2).}
\begin{tabular}{|l|l|l|c|} \hline\hline
Particle    &  Intermediate State   &  Final State & Comment \\ \hline
$\squark$   &  $\to j\zinol$        &  $\to j$ \met  &  \\
            &  $\to j\zinoh$        &  $\to j$ \met  &  \\
            &  $\cdots$                 &  $\to j\ell\bar\ell$ \met &  \\
            &  $\cdots$                 &  $\to jjj$ \met & \\
            &  $\to j\winol$        &  $\to j\ell$ \met & \\
            &  $\cdots$                 &  $\to jjj$ \met &  \\
            &  $\cdots$                 &  $\to jb\bar c$ \met & \lstop \\
            &  $\to j\gluino$       &  $\to jjj$ \met &  \\
            &                  &  $\to j\gamma$ \met & HLSP, GLSP \\ \hline
$\gluino$   &  $\to j\squark$       &  $\to jj$ \met  &  \\
            &  $\to t\stop^*$       &  $\to bW\bar c$ \met & \lstop \\
            &  $\to b\sbottom^*$    &  $\to b\bar b$ \met &  \\
            &  $\to jj\zinol$       &  $\to jj$ \met & \\
            &  $\to jj\zinoh$       &  $\to jj$ \met & \\
            &  $\cdots$                 &  $\to jj\ell\bar\ell$ \met & \\
            &  $\cdots$                 &  $\to jjjj$ \met & \\
            &  $\to jj\winol$       &  $\to jj\ell$ \met & \\
            &  $\cdots$                 &  $\to jjjj$ \met & \\
            &  $\cdots$    & $\to jjb\bar c$ \met & \\
            &  $\to t\bar t\zinol$  &  $\to bW\bar bW$ \met & \\
            &  $\to t\bar b\winol$  &  $\to bW\bar b\ell$ \met & \\
            &  $\cdots$                 &  $\to bW\bar b jj$ \met  & \\
\hline\hline 
\end{tabular}
\end{table}
\clearpage
\section{Appendix B}
\subsection{Examples of Multijet Signatures for SUSY.}
\begin{table}[!htb]
\centering
\caption{Examples of R--Parity Conserving SUSY signatures at the 
Tevatron:  Jets+\met. 
Not all signatures are listed -- we
have (somewhat arbitrarily) restricted the list. 
We assume that the LSP is the $\zinol$. 
Note that $\squark$ decays give
one or 3 jets, $\gluino$ decays 
give 2,4, or 6 jets, and the $\winoi$, $\zinoi$,
$\slepton$, and $\sneutrino$ decays give an even number of jets.}
\begin{tabular}{|l|l|l|}
\hline \hline
\multicolumn{3}{|c|}{R--Parity Conserving Signatures: Jets +\met} \\
\hline
Signature & Production & Decay	\\
\hline
$j$\met & $\squark\zinol$ & $\squark \goes q\zinol$   \\
$jj$\met & $\squark\squark^*$ & $\squark \goes q\zinol$; 
		$\squark^* \goes \bar{q}\zinol$   \\
	      & $\winol \zinol$ &  $\winol \goes q\bar{q}\zinol$ \\ 
	      & $\stop\stop^*$ &  $\sTczinol$ \\ 
$jjj$\met & $\squark\zinol$ & $\squark \goes q\winol$,$\winol \goes
                                  q\bar{q}\zinol$ \\ 
	      & $\squark\gluino$ & $\squark \goes q\zinol;\gluino \goes
                                         q\bar{q}\zinol$  \\ 
$jjjj$\met 	& $\sQsQ$ & $\sQjone$; $\sQsjthree$ 	 \\ 
		& $\glgl$ & $\gljtwo$        \\ 
$5j$\met 	& $\sQgl$ & $\sQjthree$; $\gljtwo$    \\
$6j$\met 	& $\sQsQ$ & $\sQjthree$   \\
		& $\glgl$ & $\gljtwo$; $\gljfour$     \\
$>6j$\met 	& $\sQgl$ & $\sQjthree$; $\gljfour$   \\
		& $\glgl$ & $\gljsix$; $\gljtwo$  \\
\hline \hline
\end{tabular}
\label{tab:signatures_jet}
\end{table}
\clearpage
\subsection{Examples of SUSY Signatures that Include b-quarks.}
\begin{table}[!ht]
\centering
\caption{Examples of R--Parity Conserving SUSY signatures at the 
Tevatron: $b$--tags + jets +\met.  We have shown only a few modes. 
The signatures change depending on the relative
masses of the $\sbottom, \stop$, $\winol$ and $\zinol$.}
\begin{tabular}{|l|l|l|}
\hline \hline
\multicolumn{3}{|c|}{R-Parity Conserving Signatures: b quarks} \\
\hline
Signature & Production & Decay \\
\hline
$b$\met            & $\sbottom\zinol$ & $\sBbzinol$       \\
$bj$\met    & $\sbottom\stop$  & $\sBbzinol;\sTczinol$     \\
$bjj$\met     & $\sbottom\gluino$ & $\sBbzinol;\gljtwo$    \\
		      & $\stop\zinol$    & $\sTbwinol$      \\
$bjjj$\met    & $\sbottom\sbottom$ & $\sBbzinol;~\sBstW,\sTczinol$ \\
	      & $\sbottom\stop$   & $\sBbgluino,\gljtwo;~\sTczinol$ \\
$bjjj\ldots$\met  & $\sbottom\sbottom$ & $\sBbgluino\gljtwo;~\sBstW\sTczinol$  \\
\hline 
$bbj\ldots$\met 	& $\sbottom\sbottom$  & $\sBbzinol$  \\
		& $\stop\stop^*$  & $\sTbwinol,\sWqq$     \\
$bbbj\ldots$\met    & $\gluino\sbottom$   & $\glbbbar;~\sBbzinol$    \\
		& $\gluino\stop$      & $\glbbbar;~\sTbwinol,\sWqq$   \\
$bbbbj\ldots$\met    & $\gluino\gluino$    & $\glbbbar$     \\
\hline
$bb \ell jj$\met    & $\stop\stop$ & $\sTbwinol,\sWenlsp,\sWqq$ \\ 
$bb\ell\ell$\met    & $\stop\stop$ & $\sTbwinol,\sWenlsp$ \\ 
\hline \hline
\end{tabular}
\label{tab:signatures_b}
\end{table}
\clearpage
\subsection{Examples of SUSY Signatures that Include Leptons.}
\begin{table}[!ht]
\centering
\caption{Examples of R--Parity Conserving SUSY signatures at the 
Tevatron: Leptons.  We have shown only a few modes.
Note that (for example) $\selectron$ decays can
give 0, 1, or 3 (charged) leptons, and $\sneutrino$ decays can  give 1 or 2
leptons. The signatures will change depending on the relative masses of
the sneutrinos and sleptons in the different generations. 
We have not shown explicitly the differences in the 
``left'' and ``right'' slepton
decays. The decays that single out the $\tau$ (for example, from the $H^+$)
are omitted here.  
Decay modes involving neutralinos and charginos
can be created by feed--down
from squark and gluino decays. 
Gluino decays can lead to leptons with uncorrelated charges.}
\begin{tabular}{|l|l|l|}
\hline \hline
\multicolumn{3}{|c|}{R--Parity Conserving Signatures -- ``Generic'' Leptons} \\
\hline
Signature & Production & Decay	 \\
\hline
$\ell$\met & $\sEsN$ & $\sEezinol;~\sNnzinol$  \\
			  & $\sWsZh$ & $\sWenlsp;~\sZnn$    \\
$\ell\ell$\met 	& $\sWsW$ & $\sWenlsp$       \\
				& $\sZsZ$ & $\sZee;~\sZnn$   \\
				& $\sEsE$ & $\sEezinol$   \\
$\ell\ell\ell\ldots$\met  
				& $\sWsZh$ & $\sWenlsp;~\sZee$   \\
				& $\sZsZ$ & $\sZee;~\sZee$   \\
				& $\sEsE$ & $\sEezinoh,\sZee$   \\
				& $\sEsN$ & $\sEezinoh,\sZee;~\sNewinol$  \\
\hline 
$\ell j\ldots$\met & $\sWsZh$ & $\sWenlsp;~\sZqq$  \\
		       & $\glgl$ &$\glqqwino;~\gljtwo$ \\
$\ell\ell j\ldots$\met & $\sEsE$ & $\sEezinol;~\sEenzinol$  \\
			& $\sWsZh$ & $\sWqq;~\sZee$  \\
		       & $\glgl$ &$\glqqwino,\sWenlsp$ \\
		       & $\stop\stop^*$ &$\sTbwinol,\sWenlsp$ \\
$\ell\ell\ell\ldots j\ldots$\met 
		       & $\sWsZt$ &$\sWenlsp;~\sZtqq$ \\
		       & $\glgl$ &$\glqqwino;~\glqqzinoh$ \\
\hline 
\end{tabular}
\label{tab:signatures_lep}
\end{table}
\clearpage
\subsection{Examples of SUSY Signatures that Include Photons.}
\begin{table}[!ht]
\centering                               
\caption{Examples of R--Parity Conserving SUSY signatures at the 
Tevatron: $\gamma$'s+\met. 
In the light Gravitino scenario (GLSP) $\sZgG$ always occurs. 
If the decay has a long lifetime, one of the two $\zinol$ may decay
outside the detector (LLG).
In the Higgsino LSP scenario (HLSP), $\sZgZ$ often occurs.}
\begin{tabular}{|l|l|l|l|}
\hline \hline
\multicolumn{4}{|c|}{R-Parity Conserving Signatures: Photons} \\
\hline
Signature & Production & Decay	& Comment  \\
\hline
$\gamma$\met          & $\sZlsZl$  & $\sZgG,\sZgG$   & LLG \\
$\gamma j$\met  & $\sQsZ$  & $\sQjone,\sZgG$ & LLG \\
$\gamma jj$\met   & $\sWsZ$  & $\sWwsZ,\sZgG$  & LLG \\
		& $\sWsZh$  & $\sWqq,\sZgZ$  & HLSP \\
$\gamma jjj\ldots$\met & $\glgl$ & $\gljfour$ & LLG \\
		& $\sQgl$ & $\sQjone;~\gljtwoh,\sZgZ$ & HLSP \\
\hline
$\gamma b$\met  & $\winol\zinoh$  & $\sWstb;~\sZgsz$ & HLSP\\
\hline 
$\gamma \ell$\met & $\winol\zinol$  & $\sWenlsp, \sZgG $  & LLG \\
	 & $\winol\zinoh$  & $\sWenlsp;~ \sZgsz$  & HLSP \\
	 & $\sEsN$  & $\sEezinoh,\sZgZ;~\sNnzinol$  & HLSP \\
$\gamma \ell jj\ldots$\met & $\glgl$  & $\glqqwinoe;~ \glqqzinoh,\sZgZ $  & HLSP\\
\hline 
$\gamma\ell\ell$\met        & $\zinoh\zinoh$  & $\sZee;~\sZgsz$  & HLSP \\
$\gamma\ell\ell jj\ldots$\met & $\glgl$  & 
                $\glqqzinoh,\sZee,\sZgZ$ & HLSP    \\
\hline 
$\gamma\gamma jj\ldots$\met & $\winol\zinoh$  & $\sWqq;~\sZqq,\sZgG$  & GLSP    \\
$\gamma\gamma\ell jj\dots$\met & $\winol\zinoh$  
                                       & $\sWenlsp;~\sZqq,\sZgG$ & GLSP    \\
$\gamma\gamma\ell\ell jj\ldots$\met & $\sEsE$  & $\sEezinol,\sZgG$ & GLSP\\
 & $\sWpsWm$  & $\sWenZh,\sZgsz$ & HLSP\\
\hline \hline
\end{tabular}
\label{tab:signatures_gamma}
\end{table}

\end{document}